\documentclass[review=false]{jfp-epi}

\newcommand{\jfpversion}{}
\usepackage{xcolor}
\usepackage{listings}
\usepackage{fancyvrb}
\usepackage{booktabs}
\usepackage{colortbl}
\usepackage{array}
\usepackage{pifont}
\usepackage{subcaption}
\usepackage{tikz}
\usetikzlibrary{arrows.meta,calc,positioning,fit}
\usepackage{enumitem}
\ifdefined\jfpversion\usepackage{placeins}\else\usepackage[section]{placeins}\fi
\definecolor{Navy}{HTML}{1F3C6F}
\definecolor{AlertRed}{HTML}{D8433A}
\definecolor{BlockRed}{HTML}{E76B67}
\definecolor{BlockRedBody}{HTML}{FCEEED}
\definecolor{BlockTeal}{HTML}{66B2A5}
\definecolor{BlockTealBody}{HTML}{EBF5F2}
\definecolor{BlockBlue}{HTML}{196183}
\definecolor{BlockBlueBody}{HTML}{EDF0F7}
\definecolor{TextGray}{HTML}{868686}
\definecolor{PortFill}{HTML}{EDF0F7}
\definecolor{FBoxFill}{HTML}{FBE3E3}   % Alice's box
\definecolor{GBoxFill}{HTML}{DCD9F4}   % Bob's box

\definecolor{HsKeyword}{HTML}{A626A4}   % keywords: purple, bold
\definecolor{HsType}{HTML}{B45309}      % types, classes, constructors: amber
\definecolor{HsFunction}{HTML}{2563EB}  % functions: blue
\definecolor{HsOperator}{HTML}{0E7490}  % operators: cyan
\definecolor{HsString}{HTML}{15803D}    % strings: green
\definecolor{HsComment}{HTML}{8B8B93}   % comments: grey, italic

\definecolor{CExp}{HTML}{1D4ED8}     % the experiment layer, Exp e, the step i
\definecolor{CMem}{HTML}{0F766E}     % the membrane, RunTrial e t
\definecolor{CTrial}{HTML}{A21CAF}   % the trial layer, TrialModel t
\definecolor{CSet}{HTML}{C2410C}     % the settings (a_i, b_i)
\definecolor{CLam}{HTML}{15803D}     % the strategy lambda_i
\definecolor{CRes}{HTML}{D8433A}     % the answers (x_i, y_i)

\newcommand{\codesize}{\fontsize{8}{9.6}\selectfont}
\newcommand{\codebasewidth}{0.525em}
\newcommand{\outputfont}{\ttfamily}
\newcommand{\reportfamily}{tt}

\lstdefinelanguage{Hask}{
  morekeywords=[1]{
    case,class,data,deriving,do,else,forall,if,import,in,infix,infixl,infixr,
    instance,let,module,newtype,of,qualified,then,type,where},
  morekeywords=[2]{
    A0,A1,ASetting,Applicative,B0,B1,BSetting,BadSampler,BellHiddenState,
    Bool,Bounded,Clocked,Collapsing,Comparison,Context,Contextual,Double,
    Empirical,Enum,Eq,Event,Exp,False,Functor,IO,Identity,Int,Just,LHV,
    Lambda,Matrix,Maybe,Minus,Monad,MonadIO,MonadReader,MonadState,Nothing,
    Num,Outcome,PR,PRHiddenState,Plus,Quantum,ReadOnlyBell,ReadOnlyPR,Reader,
    ReaderT,Record,Report,RunTrial,Sampler,Scheduler,Shared,Source,State,
    StateT,String,TrialModel,True,Vector},
  morekeywords=[3]{
    aBit,abs,add,allLambdas,altExtrema,antiCorr,ask,asks,bBit,
    biasPlus2Blocks,biasPlus2Loophole,biasedPlus2IO,bitOut,blind,cell,
    cheatLam,cheatNS,chsh,chshId,chshOf,chshReport,chshST,chshSuperdet,coin,
    collapse,comparisons,concat,const,contexts,contextualityReport,
    correlation,correlator,cos,dependent,empirical,entangled,evalState,
    evalStateT,even,every,experiment,filter,fixedSchedule,flipOutcome,
    flipWhen,freq,fresh,fromIntegral,get,gets,globalAssignmentExists,
    identity,independent,jointAB,lambdaAllPlus,lambdaAntiAll,legalAllPlus,
    length,lhvCHSHIO,lhvCHSHId,liftA2,liftIO,liftIdentity,localA,localB,mapM,
    maxBound,maximallyEntangled,measA,measB,minBound,minus2,mkStdGen,modify,
    modify',noSignalingIn,nonlocalA,nonlocalB,normalize,not,null,numTrials,
    observableA,observableB,outBit,outcomePairs,pAPlus,pBPlus,pEq,
    pPlusOverlapL,pages,pauliX,pauliZ,peek,pi,plus2,prConstraint,
    prettyReport,printReport,prob,probability,projector,pure,randomOutcome,
    randomRIO,randomScheduleIO,rare,readA,readB,renderComparisons,
    renderReport,renderTable,report,report3_loophole_closed,
    report3_loophole_open,report3_random_loophole_closed,
    report3_random_loophole_open,reportBellHiddenState_blocks_fixed,
    reportBellHiddenState_uniform_fixed,reportBellHiddenState_uniform_random,
    reportLHV_bias_blocks_7of8,reportLHV_bias_loophole_7of8,
    reportLHV_uniform_fixed,reportLHV_uniform_random,
    reportPRHiddenState_plus2_fixed,reportPRHiddenState_uniform_fixed,
    reportPRHiddenState_uniform_random,reportPR_fixed,reportPR_random,
    reportPeek_fixed,reportPeek_random,reportQuantum_fixed,
    reportQuantum_partial,reportQuantum_random,reportShared_fixed,
    reportSuperdet,reportSuperdet_cheatNS,reportSuperdet_control,
    reportSuperdet_fixed,reportSuperdet_random,runIdentity,runReader,
    runReaderT,runTrialBellHiddenState,runTrialCI,runTrialCZ,
    runTrialClockLoose,runTrialClockTight,runTrialCollapsing,
    runTrialContextual,runTrialLHV,runTrialPR,runTrialPRHiddenState,
    runTrialQuantum,runTrialReadOnlyBell,runTrialReadOnlyPR,runTrialSource,
    runTrialState,runTrialSuperdet,scale,seeded,setStdGen,show,showA,showB,
    sqrt,sub,suchThat,sum,test2_fixed,test3CI,test3CZ,test3RCI,test3RCZ,
    test3_loophole_closed,test3_loophole_open,test3_random_loophole_closed,
    test3_random_loophole_open,testContextuality,testIdentity_fixed,
    testIdentity_random,testLHV_biasIO,testLHV_bias_blocks_7of8,
    testLHV_bias_loophole_7of8,testLHV_minus2,testLHV_plus2,
    testLHV_uniform_fixed,testLHV_uniform_random,testPRHiddenState,
    testPRHiddenState_fixed,testPRHiddenState_random,
    testPRHiddenState_uniform_fixed,testPRHiddenState_uniform_random,
    testPR_fixed,testPR_random,testQuantum_fixed,testQuantum_random,
    testSuperdet_control,testSuperdet_fixed,testSuperdet_random,trial,
    trialFrom,trialValue,uniformAll,uniformBlocks,uniformCycle,uniformIO,
    unlines,violated,xor},
  sensitive=true,
  morecomment=[l]{--},
  morestring=[b]",
}

\newcommand{\opc}[1]{{\color{HsOperator}#1}}
\lstdefinestyle{hask}{
  language=Hask,
  basicstyle=\ttfamily\codesize,
  keywordstyle=[1]{\color{HsKeyword}\bfseries},
  keywordstyle=[2]{\color{HsType}},
  keywordstyle=[3]{\color{HsFunction}},
  stringstyle=\color{HsString},
  commentstyle=\color{HsComment}\itshape,
  columns=fixed, basewidth=\codebasewidth,
  keepspaces=true, showstringspaces=false, upquote=true,
  frame=leftline, framerule=1.2pt, rulecolor=\color{BlockTeal},
  xleftmargin=12pt, framexleftmargin=7pt,
  aboveskip=4pt, belowskip=4pt,
  literate=*
    {=}{{\opc{=}}}1 {|}{{\opc{|}}}1 {\$}{{\opc{\$}}}1
    {\\}{{\opc{\textbackslash}}}1 {.}{{\opc{.}}}1 {*}{{\opc{*}}}1
    {+}{{\opc{+}}}1 {/}{{\opc{/}}}1 {<}{{\opc{<}}}1 {>}{{\opc{>}}}1
    {->}{{\opc{->}}}2 {<-}{{\opc{<-}}}2 {=>}{{\opc{=>}}}2 {::}{{\opc{::}}}2
    {==}{{\opc{==}}}2 {/=}{{\opc{/=}}}2 {>>=}{{\opc{>>=}}}3 {>>}{{\opc{>>}}}2
    {<\$>}{{\opc{<\$>}}}3 {++}{{\opc{++}}}2 {||}{{\opc{||}}}2 {..}{{\opc{..}}}2
    {&&}{{\opc{\&\&}}}2 {><}{{\opc{$\otimes$}}}2,
}
\makeatletter
\newcounter{codelisting}
\renewcommand{\thecodelisting}{L\arabic{codelisting}}
\newcount\bf@lines
\def\bf@addrange#1-#2\@nil{\advance\bf@lines by \numexpr #2-#1+1\relax}
\newcommand{\bf@countlines}[1]{\bf@lines=0
  \@for\bf@r:=#1\do{\expandafter\bf@addrange\bf@r\@nil}}
\newif\ifbf@tagged
\newcommand{\bf@codetag}[1]{\ifbf@tagged\else\global\bf@taggedtrue
  {\normalfont\sffamily\scriptsize\textcolor{TextGray}{\thecodelisting}}\fi}
\lstdefinestyle{tagged}{style=hask, numbers=right, numbersep=8pt, numberstyle=\bf@codetag}
\NewDocumentCommand{\code}{O{} m m o}{\par\bf@countlines{#2}%
  \refstepcounter{codelisting}\IfValueT{#4}{\label{#4}}\global\bf@taggedfalse
  \ifnum\bf@lines>10
    \lstinputlisting[style=tagged,linerange={#2},#1]{src-ascii/#3}\par
  \else
    \vspace{2pt}\noindent\begin{minipage}{\linewidth}%
    \lstinputlisting[style=tagged,linerange={#2},#1]{src-ascii/#3}\end{minipage}\par\vspace{2pt}%
  \fi}
\makeatother
\newcommand{\hs}{\lstinline[style=hask,basicstyle=\ttfamily,frame=none]}
\newcommand{\printedreport}[1]{\par\vspace{3pt}\noindent\begin{minipage}{\linewidth}%
  \VerbatimInput[fontfamily=\reportfamily, fontsize=\codesize, frame=leftline, framerule=1.2pt,
    rulecolor=\color{BlockTeal}, xleftmargin=12pt, framesep=7pt]{generated/report-#1.txt}%
  \end{minipage}\par\vspace{3pt}}

\makeatletter
\newcommand{\setval}[2]{\expandafter\def\csname bf@val@#1\endcsname{#2}}
\newcommand{\val}[1]{\ifcsname bf@val@#1\endcsname\csname bf@val@#1\endcsname
  \else\PackageError{paper}{No value #1 in generated/values.tex}{}\fi}
\makeatother
\setval{nTrials}{80000}
\setval{nPerContext}{20000}
\setval{nSep}{20000}
\setval{eps}{0.02}
\setval{nRuns}{31}
\setval{nMissed}{4}
\setval{tsirelson}{2.828}
\setval{chshId}{2}
\setval{partial/chsh}{2.414}
\setval{partial/AliceA0}{0.854}
\setval{partial/AliceA1}{0.500}
\setval{partial/Bob}{0.750}
\setval{pr/biasedB0}{0.900}
\setval{pr/biasedB1}{0.500}
\setval{theta/zero}{1.414}
\setval{theta/classical}{0.214}
\setval{theta/classicalDeg}{12.2}
\setval{chshST}{4}
\setval{order/bell-uniform-fixed/chsh}{2.000}
\setval{pure-fixed/chsh}{2.000}
\setval{pure-fixed/ns}{\yes}
\setval{pure-fixed/sep}{\yes}
\setval{pure-fixed/run/AliceA0/1}{1.000}
\setval{pure-fixed/run/AliceA0/2}{1.000}
\setval{pure-fixed/run/AliceA1/1}{0.000}
\setval{pure-fixed/run/AliceA1/2}{0.000}
\setval{pure-fixed/run/BobB0/1}{0.000}
\setval{pure-fixed/run/BobB0/2}{0.000}
\setval{pure-fixed/run/BobB1/1}{1.000}
\setval{pure-fixed/run/BobB1/2}{1.000}
\setval{pure-fixed/sep/AliceA0/1}{1.000}
\setval{pure-fixed/sep/AliceA0/2}{1.000}
\setval{pure-fixed/sep/AliceA1/1}{0.000}
\setval{pure-fixed/sep/AliceA1/2}{0.000}
\setval{pure-fixed/sep/BobB0/1}{0.000}
\setval{pure-fixed/sep/BobB0/2}{0.000}
\setval{pure-fixed/sep/BobB1/1}{1.000}
\setval{pure-fixed/sep/BobB1/2}{1.000}
\setval{pure-random/chsh}{2.000}
\setval{pure-random/ns}{\yes}
\setval{pure-random/sep}{\yes}
\setval{pure-random/run/AliceA0/1}{1.000}
\setval{pure-random/run/AliceA0/2}{1.000}
\setval{pure-random/run/AliceA1/1}{0.000}
\setval{pure-random/run/AliceA1/2}{0.000}
\setval{pure-random/run/BobB0/1}{0.000}
\setval{pure-random/run/BobB0/2}{0.000}
\setval{pure-random/run/BobB1/1}{1.000}
\setval{pure-random/run/BobB1/2}{1.000}
\setval{pure-random/sep/AliceA0/1}{1.000}
\setval{pure-random/sep/AliceA0/2}{1.000}
\setval{pure-random/sep/AliceA1/1}{0.000}
\setval{pure-random/sep/AliceA1/2}{0.000}
\setval{pure-random/sep/BobB0/1}{0.000}
\setval{pure-random/sep/BobB0/2}{0.000}
\setval{pure-random/sep/BobB1/1}{1.000}
\setval{pure-random/sep/BobB1/2}{1.000}
\setval{shared/chsh}{4.000}
\setval{shared/ns}{\no\ Bob at \texttt{B1}}
\setval{shared/sep}{\no}
\setval{shared/run/AliceA0/1}{0.000}
\setval{shared/run/AliceA0/2}{0.000}
\setval{shared/run/AliceA1/1}{0.000}
\setval{shared/run/AliceA1/2}{0.000}
\setval{shared/run/BobB0/1}{0.000}
\setval{shared/run/BobB0/2}{0.000}
\setval{shared/run/BobB1/1}{0.000}
\setval{shared/run/BobB1/2}{1.000}
\setval{shared/sep/AliceA0/1}{0.000}
\setval{shared/sep/AliceA0/2}{0.000}
\setval{shared/sep/AliceA1/1}{0.000}
\setval{shared/sep/AliceA1/2}{0.000}
\setval{shared/sep/BobB0/1}{0.000}
\setval{shared/sep/BobB0/2}{0.000}
\setval{shared/sep/BobB1/1}{0.000}
\setval{shared/sep/BobB1/2}{1.000}
\setval{clock-visible-fixed/chsh}{4.000}
\setval{clock-visible-fixed/ns}{\no\ Bob at \texttt{B1}}
\setval{clock-visible-fixed/sep}{\yes}
\setval{clock-visible-fixed/run/AliceA0/1}{1.000}
\setval{clock-visible-fixed/run/AliceA0/2}{1.000}
\setval{clock-visible-fixed/run/AliceA1/1}{1.000}
\setval{clock-visible-fixed/run/AliceA1/2}{1.000}
\setval{clock-visible-fixed/run/BobB0/1}{1.000}
\setval{clock-visible-fixed/run/BobB0/2}{1.000}
\setval{clock-visible-fixed/run/BobB1/1}{1.000}
\setval{clock-visible-fixed/run/BobB1/2}{0.000}
\setval{clock-visible-fixed/sep/AliceA0/1}{1.000}
\setval{clock-visible-fixed/sep/AliceA0/2}{1.000}
\setval{clock-visible-fixed/sep/AliceA1/1}{1.000}
\setval{clock-visible-fixed/sep/AliceA1/2}{1.000}
\setval{clock-visible-fixed/sep/BobB0/1}{0.750}
\setval{clock-visible-fixed/sep/BobB0/2}{0.750}
\setval{clock-visible-fixed/sep/BobB1/1}{0.750}
\setval{clock-visible-fixed/sep/BobB1/2}{0.750}
\setval{clock-visible-random/chsh}{0.990}
\setval{clock-visible-random/ns}{\yes}
\setval{clock-visible-random/sep}{\yes}
\setval{clock-visible-random/run/AliceA0/1}{1.000}
\setval{clock-visible-random/run/AliceA0/2}{1.000}
\setval{clock-visible-random/run/AliceA1/1}{1.000}
\setval{clock-visible-random/run/AliceA1/2}{1.000}
\setval{clock-visible-random/run/BobB0/1}{0.750}
\setval{clock-visible-random/run/BobB0/2}{0.749}
\setval{clock-visible-random/run/BobB1/1}{0.748}
\setval{clock-visible-random/run/BobB1/2}{0.753}
\setval{clock-visible-random/sep/AliceA0/1}{1.000}
\setval{clock-visible-random/sep/AliceA0/2}{1.000}
\setval{clock-visible-random/sep/AliceA1/1}{1.000}
\setval{clock-visible-random/sep/AliceA1/2}{1.000}
\setval{clock-visible-random/sep/BobB0/1}{0.750}
\setval{clock-visible-random/sep/BobB0/2}{0.750}
\setval{clock-visible-random/sep/BobB1/1}{0.750}
\setval{clock-visible-random/sep/BobB1/2}{0.750}
\setval{clock-hidden-fixed/chsh}{2.000}
\setval{clock-hidden-fixed/ns}{\yes}
\setval{clock-hidden-fixed/sep}{\yes}
\setval{clock-hidden-fixed/run/AliceA0/1}{1.000}
\setval{clock-hidden-fixed/run/AliceA0/2}{1.000}
\setval{clock-hidden-fixed/run/AliceA1/1}{1.000}
\setval{clock-hidden-fixed/run/AliceA1/2}{1.000}
\setval{clock-hidden-fixed/run/BobB0/1}{1.000}
\setval{clock-hidden-fixed/run/BobB0/2}{1.000}
\setval{clock-hidden-fixed/run/BobB1/1}{1.000}
\setval{clock-hidden-fixed/run/BobB1/2}{1.000}
\setval{clock-hidden-fixed/sep/AliceA0/1}{1.000}
\setval{clock-hidden-fixed/sep/AliceA0/2}{1.000}
\setval{clock-hidden-fixed/sep/AliceA1/1}{1.000}
\setval{clock-hidden-fixed/sep/AliceA1/2}{1.000}
\setval{clock-hidden-fixed/sep/BobB0/1}{1.000}
\setval{clock-hidden-fixed/sep/BobB0/2}{1.000}
\setval{clock-hidden-fixed/sep/BobB1/1}{1.000}
\setval{clock-hidden-fixed/sep/BobB1/2}{1.000}
\setval{clock-hidden-random/chsh}{2.000}
\setval{clock-hidden-random/ns}{\yes}
\setval{clock-hidden-random/sep}{\yes}
\setval{clock-hidden-random/run/AliceA0/1}{1.000}
\setval{clock-hidden-random/run/AliceA0/2}{1.000}
\setval{clock-hidden-random/run/AliceA1/1}{1.000}
\setval{clock-hidden-random/run/AliceA1/2}{1.000}
\setval{clock-hidden-random/run/BobB0/1}{1.000}
\setval{clock-hidden-random/run/BobB0/2}{1.000}
\setval{clock-hidden-random/run/BobB1/1}{1.000}
\setval{clock-hidden-random/run/BobB1/2}{1.000}
\setval{clock-hidden-random/sep/AliceA0/1}{1.000}
\setval{clock-hidden-random/sep/AliceA0/2}{1.000}
\setval{clock-hidden-random/sep/AliceA1/1}{1.000}
\setval{clock-hidden-random/sep/AliceA1/2}{1.000}
\setval{clock-hidden-random/sep/BobB0/1}{1.000}
\setval{clock-hidden-random/sep/BobB0/2}{1.000}
\setval{clock-hidden-random/sep/BobB1/1}{1.000}
\setval{clock-hidden-random/sep/BobB1/2}{1.000}
\setval{lhv-plus2/chsh}{2.000}
\setval{lhv-plus2/ns}{\yes}
\setval{lhv-plus2/sep}{\yes}
\setval{lhv-plus2/run/AliceA0/1}{1.000}
\setval{lhv-plus2/run/AliceA0/2}{1.000}
\setval{lhv-plus2/run/AliceA1/1}{1.000}
\setval{lhv-plus2/run/AliceA1/2}{1.000}
\setval{lhv-plus2/run/BobB0/1}{1.000}
\setval{lhv-plus2/run/BobB0/2}{1.000}
\setval{lhv-plus2/run/BobB1/1}{1.000}
\setval{lhv-plus2/run/BobB1/2}{1.000}
\setval{lhv-plus2/sep/AliceA0/1}{1.000}
\setval{lhv-plus2/sep/AliceA0/2}{1.000}
\setval{lhv-plus2/sep/AliceA1/1}{1.000}
\setval{lhv-plus2/sep/AliceA1/2}{1.000}
\setval{lhv-plus2/sep/BobB0/1}{1.000}
\setval{lhv-plus2/sep/BobB0/2}{1.000}
\setval{lhv-plus2/sep/BobB1/1}{1.000}
\setval{lhv-plus2/sep/BobB1/2}{1.000}
\setval{lhv-minus2/chsh}{-2.000}
\setval{lhv-minus2/ns}{\yes}
\setval{lhv-minus2/sep}{\yes}
\setval{lhv-minus2/run/AliceA0/1}{1.000}
\setval{lhv-minus2/run/AliceA0/2}{1.000}
\setval{lhv-minus2/run/AliceA1/1}{1.000}
\setval{lhv-minus2/run/AliceA1/2}{1.000}
\setval{lhv-minus2/run/BobB0/1}{0.000}
\setval{lhv-minus2/run/BobB0/2}{0.000}
\setval{lhv-minus2/run/BobB1/1}{0.000}
\setval{lhv-minus2/run/BobB1/2}{0.000}
\setval{lhv-minus2/sep/AliceA0/1}{1.000}
\setval{lhv-minus2/sep/AliceA0/2}{1.000}
\setval{lhv-minus2/sep/AliceA1/1}{1.000}
\setval{lhv-minus2/sep/AliceA1/2}{1.000}
\setval{lhv-minus2/sep/BobB0/1}{0.000}
\setval{lhv-minus2/sep/BobB0/2}{0.000}
\setval{lhv-minus2/sep/BobB1/1}{0.000}
\setval{lhv-minus2/sep/BobB1/2}{0.000}
\setval{lhv-uniform-fixed/chsh}{0.000}
\setval{lhv-uniform-fixed/ns}{\no\ Bob at \texttt{B0}}
\setval{lhv-uniform-fixed/sep}{\yes}
\setval{lhv-uniform-fixed/run/AliceA0/1}{0.500}
\setval{lhv-uniform-fixed/run/AliceA0/2}{0.500}
\setval{lhv-uniform-fixed/run/AliceA1/1}{0.500}
\setval{lhv-uniform-fixed/run/AliceA1/2}{0.500}
\setval{lhv-uniform-fixed/run/BobB0/1}{0.000}
\setval{lhv-uniform-fixed/run/BobB0/2}{1.000}
\setval{lhv-uniform-fixed/run/BobB1/1}{1.000}
\setval{lhv-uniform-fixed/run/BobB1/2}{1.000}
\setval{lhv-uniform-fixed/sep/AliceA0/1}{0.500}
\setval{lhv-uniform-fixed/sep/AliceA0/2}{0.500}
\setval{lhv-uniform-fixed/sep/AliceA1/1}{0.500}
\setval{lhv-uniform-fixed/sep/AliceA1/2}{0.500}
\setval{lhv-uniform-fixed/sep/BobB0/1}{0.500}
\setval{lhv-uniform-fixed/sep/BobB0/2}{0.500}
\setval{lhv-uniform-fixed/sep/BobB1/1}{0.500}
\setval{lhv-uniform-fixed/sep/BobB1/2}{0.500}
\setval{lhv-uniform-random/chsh}{0.008}
\setval{lhv-uniform-random/ns}{\yes}
\setval{lhv-uniform-random/sep}{\yes}
\setval{lhv-uniform-random/run/AliceA0/1}{0.501}
\setval{lhv-uniform-random/run/AliceA0/2}{0.499}
\setval{lhv-uniform-random/run/AliceA1/1}{0.500}
\setval{lhv-uniform-random/run/AliceA1/2}{0.496}
\setval{lhv-uniform-random/run/BobB0/1}{0.508}
\setval{lhv-uniform-random/run/BobB0/2}{0.498}
\setval{lhv-uniform-random/run/BobB1/1}{0.497}
\setval{lhv-uniform-random/run/BobB1/2}{0.497}
\setval{lhv-uniform-random/sep/AliceA0/1}{0.500}
\setval{lhv-uniform-random/sep/AliceA0/2}{0.500}
\setval{lhv-uniform-random/sep/AliceA1/1}{0.500}
\setval{lhv-uniform-random/sep/AliceA1/2}{0.500}
\setval{lhv-uniform-random/sep/BobB0/1}{0.500}
\setval{lhv-uniform-random/sep/BobB0/2}{0.500}
\setval{lhv-uniform-random/sep/BobB1/1}{0.500}
\setval{lhv-uniform-random/sep/BobB1/2}{0.500}
\setval{lhv-loophole/chsh}{3.000}
\setval{lhv-loophole/ns}{\no\ Bob at \texttt{B1}}
\setval{lhv-loophole/sep}{\yes}
\setval{lhv-loophole/run/AliceA0/1}{1.000}
\setval{lhv-loophole/run/AliceA0/2}{1.000}
\setval{lhv-loophole/run/AliceA1/1}{1.000}
\setval{lhv-loophole/run/AliceA1/2}{1.000}
\setval{lhv-loophole/run/BobB0/1}{1.000}
\setval{lhv-loophole/run/BobB0/2}{1.000}
\setval{lhv-loophole/run/BobB1/1}{1.000}
\setval{lhv-loophole/run/BobB1/2}{0.500}
\setval{lhv-loophole/sep/AliceA0/1}{1.000}
\setval{lhv-loophole/sep/AliceA0/2}{1.000}
\setval{lhv-loophole/sep/AliceA1/1}{1.000}
\setval{lhv-loophole/sep/AliceA1/2}{1.000}
\setval{lhv-loophole/sep/BobB0/1}{0.875}
\setval{lhv-loophole/sep/BobB0/2}{0.875}
\setval{lhv-loophole/sep/BobB1/1}{0.875}
\setval{lhv-loophole/sep/BobB1/2}{0.875}
\setval{lhv-blocks/chsh}{1.500}
\setval{lhv-blocks/ns}{\yes}
\setval{lhv-blocks/sep}{\yes}
\setval{lhv-blocks/run/AliceA0/1}{1.000}
\setval{lhv-blocks/run/AliceA0/2}{1.000}
\setval{lhv-blocks/run/AliceA1/1}{1.000}
\setval{lhv-blocks/run/AliceA1/2}{1.000}
\setval{lhv-blocks/run/BobB0/1}{0.875}
\setval{lhv-blocks/run/BobB0/2}{0.875}
\setval{lhv-blocks/run/BobB1/1}{0.875}
\setval{lhv-blocks/run/BobB1/2}{0.875}
\setval{lhv-blocks/sep/AliceA0/1}{1.000}
\setval{lhv-blocks/sep/AliceA0/2}{1.000}
\setval{lhv-blocks/sep/AliceA1/1}{1.000}
\setval{lhv-blocks/sep/AliceA1/2}{1.000}
\setval{lhv-blocks/sep/BobB0/1}{0.875}
\setval{lhv-blocks/sep/BobB0/2}{0.875}
\setval{lhv-blocks/sep/BobB1/1}{0.875}
\setval{lhv-blocks/sep/BobB1/2}{0.875}
\setval{lhv-fresh/chsh}{0.024}
\setval{lhv-fresh/ns}{\yes}
\setval{lhv-fresh/sep}{\yes}
\setval{lhv-fresh/run/AliceA0/1}{0.501}
\setval{lhv-fresh/run/AliceA0/2}{0.505}
\setval{lhv-fresh/run/AliceA1/1}{0.503}
\setval{lhv-fresh/run/AliceA1/2}{0.497}
\setval{lhv-fresh/run/BobB0/1}{0.504}
\setval{lhv-fresh/run/BobB0/2}{0.493}
\setval{lhv-fresh/run/BobB1/1}{0.500}
\setval{lhv-fresh/run/BobB1/2}{0.498}
\setval{lhv-fresh/sep/AliceA0/1}{0.498}
\setval{lhv-fresh/sep/AliceA0/2}{0.504}
\setval{lhv-fresh/sep/AliceA1/1}{0.501}
\setval{lhv-fresh/sep/AliceA1/2}{0.502}
\setval{lhv-fresh/sep/BobB0/1}{0.499}
\setval{lhv-fresh/sep/BobB0/2}{0.501}
\setval{lhv-fresh/sep/BobB1/1}{0.500}
\setval{lhv-fresh/sep/BobB1/2}{0.497}
\setval{superdet-cheat-fixed/chsh}{4.000}
\setval{superdet-cheat-fixed/ns}{\no\ Bob at \texttt{B1}}
\setval{superdet-cheat-fixed/sep}{--}
\setval{superdet-cheat-fixed/run/AliceA0/1}{1.000}
\setval{superdet-cheat-fixed/run/AliceA0/2}{1.000}
\setval{superdet-cheat-fixed/run/AliceA1/1}{1.000}
\setval{superdet-cheat-fixed/run/AliceA1/2}{1.000}
\setval{superdet-cheat-fixed/run/BobB0/1}{1.000}
\setval{superdet-cheat-fixed/run/BobB0/2}{1.000}
\setval{superdet-cheat-fixed/run/BobB1/1}{1.000}
\setval{superdet-cheat-fixed/run/BobB1/2}{0.000}
\setval{superdet-cheat-random/chsh}{4.000}
\setval{superdet-cheat-random/ns}{\no\ Bob at \texttt{B1}}
\setval{superdet-cheat-random/sep}{--}
\setval{superdet-cheat-random/run/AliceA0/1}{1.000}
\setval{superdet-cheat-random/run/AliceA0/2}{1.000}
\setval{superdet-cheat-random/run/AliceA1/1}{1.000}
\setval{superdet-cheat-random/run/AliceA1/2}{1.000}
\setval{superdet-cheat-random/run/BobB0/1}{1.000}
\setval{superdet-cheat-random/run/BobB0/2}{1.000}
\setval{superdet-cheat-random/run/BobB1/1}{1.000}
\setval{superdet-cheat-random/run/BobB1/2}{0.000}
\setval{superdet-legal/chsh}{2.000}
\setval{superdet-legal/ns}{\yes}
\setval{superdet-legal/sep}{--}
\setval{superdet-legal/run/AliceA0/1}{1.000}
\setval{superdet-legal/run/AliceA0/2}{1.000}
\setval{superdet-legal/run/AliceA1/1}{1.000}
\setval{superdet-legal/run/AliceA1/2}{1.000}
\setval{superdet-legal/run/BobB0/1}{1.000}
\setval{superdet-legal/run/BobB0/2}{1.000}
\setval{superdet-legal/run/BobB1/1}{1.000}
\setval{superdet-legal/run/BobB1/2}{1.000}
\setval{superdet-cheatns/chsh}{4.000}
\setval{superdet-cheatns/ns}{\yes}
\setval{superdet-cheatns/sep}{--}
\setval{superdet-cheatns/run/AliceA0/1}{0.498}
\setval{superdet-cheatns/run/AliceA0/2}{0.495}
\setval{superdet-cheatns/run/AliceA1/1}{0.500}
\setval{superdet-cheatns/run/AliceA1/2}{0.500}
\setval{superdet-cheatns/run/BobB0/1}{0.498}
\setval{superdet-cheatns/run/BobB0/2}{0.500}
\setval{superdet-cheatns/run/BobB1/1}{0.495}
\setval{superdet-cheatns/run/BobB1/2}{0.500}
\setval{peek-fixed/chsh}{4.000}
\setval{peek-fixed/ns}{\no\ Bob at \texttt{B1}}
\setval{peek-fixed/sep}{\yes}
\setval{peek-fixed/run/AliceA0/1}{1.000}
\setval{peek-fixed/run/AliceA0/2}{1.000}
\setval{peek-fixed/run/AliceA1/1}{1.000}
\setval{peek-fixed/run/AliceA1/2}{1.000}
\setval{peek-fixed/run/BobB0/1}{1.000}
\setval{peek-fixed/run/BobB0/2}{1.000}
\setval{peek-fixed/run/BobB1/1}{1.000}
\setval{peek-fixed/run/BobB1/2}{0.000}
\setval{peek-fixed/sep/AliceA0/1}{1.000}
\setval{peek-fixed/sep/AliceA0/2}{1.000}
\setval{peek-fixed/sep/AliceA1/1}{1.000}
\setval{peek-fixed/sep/AliceA1/2}{1.000}
\setval{peek-fixed/sep/BobB0/1}{1.000}
\setval{peek-fixed/sep/BobB0/2}{1.000}
\setval{peek-fixed/sep/BobB1/1}{0.750}
\setval{peek-fixed/sep/BobB1/2}{0.750}
\setval{peek-random/chsh}{1.990}
\setval{peek-random/ns}{\yes}
\setval{peek-random/sep}{\yes}
\setval{peek-random/run/AliceA0/1}{1.000}
\setval{peek-random/run/AliceA0/2}{1.000}
\setval{peek-random/run/AliceA1/1}{1.000}
\setval{peek-random/run/AliceA1/2}{1.000}
\setval{peek-random/run/BobB0/1}{1.000}
\setval{peek-random/run/BobB0/2}{1.000}
\setval{peek-random/run/BobB1/1}{0.747}
\setval{peek-random/run/BobB1/2}{0.752}
\setval{peek-random/sep/AliceA0/1}{1.000}
\setval{peek-random/sep/AliceA0/2}{1.000}
\setval{peek-random/sep/AliceA1/1}{1.000}
\setval{peek-random/sep/AliceA1/2}{1.000}
\setval{peek-random/sep/BobB0/1}{1.000}
\setval{peek-random/sep/BobB0/2}{1.000}
\setval{peek-random/sep/BobB1/1}{0.752}
\setval{peek-random/sep/BobB1/2}{0.746}
\setval{quantum-fixed/chsh}{2.820}
\setval{quantum-fixed/ns}{\yes}
\setval{quantum-fixed/sep}{\yes}
\setval{quantum-fixed/run/AliceA0/1}{0.496}
\setval{quantum-fixed/run/AliceA0/2}{0.497}
\setval{quantum-fixed/run/AliceA1/1}{0.503}
\setval{quantum-fixed/run/AliceA1/2}{0.504}
\setval{quantum-fixed/run/BobB0/1}{0.495}
\setval{quantum-fixed/run/BobB0/2}{0.499}
\setval{quantum-fixed/run/BobB1/1}{0.499}
\setval{quantum-fixed/run/BobB1/2}{0.497}
\setval{quantum-fixed/sep/AliceA0/1}{0.499}
\setval{quantum-fixed/sep/AliceA0/2}{0.499}
\setval{quantum-fixed/sep/AliceA1/1}{0.500}
\setval{quantum-fixed/sep/AliceA1/2}{0.505}
\setval{quantum-fixed/sep/BobB0/1}{0.502}
\setval{quantum-fixed/sep/BobB0/2}{0.501}
\setval{quantum-fixed/sep/BobB1/1}{0.508}
\setval{quantum-fixed/sep/BobB1/2}{0.499}
\setval{quantum-random/chsh}{2.818}
\setval{quantum-random/ns}{\yes}
\setval{quantum-random/sep}{\yes}
\setval{quantum-random/run/AliceA0/1}{0.501}
\setval{quantum-random/run/AliceA0/2}{0.499}
\setval{quantum-random/run/AliceA1/1}{0.500}
\setval{quantum-random/run/AliceA1/2}{0.503}
\setval{quantum-random/run/BobB0/1}{0.500}
\setval{quantum-random/run/BobB0/2}{0.498}
\setval{quantum-random/run/BobB1/1}{0.504}
\setval{quantum-random/run/BobB1/2}{0.497}
\setval{quantum-random/sep/AliceA0/1}{0.499}
\setval{quantum-random/sep/AliceA0/2}{0.499}
\setval{quantum-random/sep/AliceA1/1}{0.500}
\setval{quantum-random/sep/AliceA1/2}{0.505}
\setval{quantum-random/sep/BobB0/1}{0.502}
\setval{quantum-random/sep/BobB0/2}{0.501}
\setval{quantum-random/sep/BobB1/1}{0.508}
\setval{quantum-random/sep/BobB1/2}{0.499}
\setval{quantum-partial/chsh}{2.421}
\setval{quantum-partial/ns}{\yes}
\setval{quantum-partial/sep}{\yes}
\setval{quantum-partial/run/AliceA0/1}{0.852}
\setval{quantum-partial/run/AliceA0/2}{0.852}
\setval{quantum-partial/run/AliceA1/1}{0.497}
\setval{quantum-partial/run/AliceA1/2}{0.498}
\setval{quantum-partial/run/BobB0/1}{0.750}
\setval{quantum-partial/run/BobB0/2}{0.746}
\setval{quantum-partial/run/BobB1/1}{0.748}
\setval{quantum-partial/run/BobB1/2}{0.751}
\setval{quantum-partial/sep/AliceA0/1}{0.852}
\setval{quantum-partial/sep/AliceA0/2}{0.850}
\setval{quantum-partial/sep/AliceA1/1}{0.502}
\setval{quantum-partial/sep/AliceA1/2}{0.499}
\setval{quantum-partial/sep/BobB0/1}{0.749}
\setval{quantum-partial/sep/BobB0/2}{0.751}
\setval{quantum-partial/sep/BobB1/1}{0.747}
\setval{quantum-partial/sep/BobB1/2}{0.750}
\setval{pr-fixed/chsh}{4.000}
\setval{pr-fixed/ns}{\yes}
\setval{pr-fixed/sep}{\yes}
\setval{pr-fixed/run/AliceA0/1}{0.503}
\setval{pr-fixed/run/AliceA0/2}{0.498}
\setval{pr-fixed/run/AliceA1/1}{0.495}
\setval{pr-fixed/run/AliceA1/2}{0.495}
\setval{pr-fixed/run/BobB0/1}{0.503}
\setval{pr-fixed/run/BobB0/2}{0.495}
\setval{pr-fixed/run/BobB1/1}{0.498}
\setval{pr-fixed/run/BobB1/2}{0.505}
\setval{pr-fixed/sep/AliceA0/1}{0.497}
\setval{pr-fixed/sep/AliceA0/2}{0.500}
\setval{pr-fixed/sep/AliceA1/1}{0.498}
\setval{pr-fixed/sep/AliceA1/2}{0.500}
\setval{pr-fixed/sep/BobB0/1}{0.500}
\setval{pr-fixed/sep/BobB0/2}{0.504}
\setval{pr-fixed/sep/BobB1/1}{0.503}
\setval{pr-fixed/sep/BobB1/2}{0.497}
\setval{pr-random/chsh}{4.000}
\setval{pr-random/ns}{\yes}
\setval{pr-random/sep}{\yes}
\setval{pr-random/run/AliceA0/1}{0.497}
\setval{pr-random/run/AliceA0/2}{0.498}
\setval{pr-random/run/AliceA1/1}{0.500}
\setval{pr-random/run/AliceA1/2}{0.503}
\setval{pr-random/run/BobB0/1}{0.497}
\setval{pr-random/run/BobB0/2}{0.500}
\setval{pr-random/run/BobB1/1}{0.498}
\setval{pr-random/run/BobB1/2}{0.497}
\setval{pr-random/sep/AliceA0/1}{0.497}
\setval{pr-random/sep/AliceA0/2}{0.500}
\setval{pr-random/sep/AliceA1/1}{0.498}
\setval{pr-random/sep/AliceA1/2}{0.500}
\setval{pr-random/sep/BobB0/1}{0.500}
\setval{pr-random/sep/BobB0/2}{0.504}
\setval{pr-random/sep/BobB1/1}{0.503}
\setval{pr-random/sep/BobB1/2}{0.497}
\setval{prstate-uniform-fixed/chsh}{4.000}
\setval{prstate-uniform-fixed/ns}{\yes}
\setval{prstate-uniform-fixed/sep}{\yes}
\setval{prstate-uniform-fixed/run/AliceA0/1}{0.500}
\setval{prstate-uniform-fixed/run/AliceA0/2}{0.500}
\setval{prstate-uniform-fixed/run/AliceA1/1}{0.500}
\setval{prstate-uniform-fixed/run/AliceA1/2}{0.500}
\setval{prstate-uniform-fixed/run/BobB0/1}{0.500}
\setval{prstate-uniform-fixed/run/BobB0/2}{0.500}
\setval{prstate-uniform-fixed/run/BobB1/1}{0.500}
\setval{prstate-uniform-fixed/run/BobB1/2}{0.500}
\setval{prstate-uniform-fixed/sep/AliceA0/1}{0.500}
\setval{prstate-uniform-fixed/sep/AliceA0/2}{0.500}
\setval{prstate-uniform-fixed/sep/AliceA1/1}{0.500}
\setval{prstate-uniform-fixed/sep/AliceA1/2}{0.500}
\setval{prstate-uniform-fixed/sep/BobB0/1}{0.500}
\setval{prstate-uniform-fixed/sep/BobB0/2}{0.500}
\setval{prstate-uniform-fixed/sep/BobB1/1}{0.500}
\setval{prstate-uniform-fixed/sep/BobB1/2}{0.500}
\setval{prstate-uniform-random/chsh}{4.000}
\setval{prstate-uniform-random/ns}{\yes}
\setval{prstate-uniform-random/sep}{\yes}
\setval{prstate-uniform-random/run/AliceA0/1}{0.498}
\setval{prstate-uniform-random/run/AliceA0/2}{0.499}
\setval{prstate-uniform-random/run/AliceA1/1}{0.499}
\setval{prstate-uniform-random/run/AliceA1/2}{0.499}
\setval{prstate-uniform-random/run/BobB0/1}{0.498}
\setval{prstate-uniform-random/run/BobB0/2}{0.499}
\setval{prstate-uniform-random/run/BobB1/1}{0.499}
\setval{prstate-uniform-random/run/BobB1/2}{0.501}
\setval{prstate-uniform-random/sep/AliceA0/1}{0.500}
\setval{prstate-uniform-random/sep/AliceA0/2}{0.500}
\setval{prstate-uniform-random/sep/AliceA1/1}{0.500}
\setval{prstate-uniform-random/sep/AliceA1/2}{0.500}
\setval{prstate-uniform-random/sep/BobB0/1}{0.500}
\setval{prstate-uniform-random/sep/BobB0/2}{0.500}
\setval{prstate-uniform-random/sep/BobB1/1}{0.500}
\setval{prstate-uniform-random/sep/BobB1/2}{0.500}
\setval{prstate-plus2/chsh}{4.000}
\setval{prstate-plus2/ns}{\no\ Bob at \texttt{B1}}
\setval{prstate-plus2/sep}{\no}
\setval{prstate-plus2/run/AliceA0/1}{1.000}
\setval{prstate-plus2/run/AliceA0/2}{1.000}
\setval{prstate-plus2/run/AliceA1/1}{1.000}
\setval{prstate-plus2/run/AliceA1/2}{1.000}
\setval{prstate-plus2/run/BobB0/1}{1.000}
\setval{prstate-plus2/run/BobB0/2}{1.000}
\setval{prstate-plus2/run/BobB1/1}{1.000}
\setval{prstate-plus2/run/BobB1/2}{0.000}
\setval{prstate-plus2/sep/AliceA0/1}{1.000}
\setval{prstate-plus2/sep/AliceA0/2}{1.000}
\setval{prstate-plus2/sep/AliceA1/1}{1.000}
\setval{prstate-plus2/sep/AliceA1/2}{1.000}
\setval{prstate-plus2/sep/BobB0/1}{1.000}
\setval{prstate-plus2/sep/BobB0/2}{1.000}
\setval{prstate-plus2/sep/BobB1/1}{1.000}
\setval{prstate-plus2/sep/BobB1/2}{0.000}
\setval{bell-blocks-fixed/chsh}{2.500}
\setval{bell-blocks-fixed/ns}{\yes}
\setval{bell-blocks-fixed/sep}{\yes}
\setval{bell-blocks-fixed/run/AliceA0/1}{0.500}
\setval{bell-blocks-fixed/run/AliceA0/2}{0.500}
\setval{bell-blocks-fixed/run/AliceA1/1}{0.500}
\setval{bell-blocks-fixed/run/AliceA1/2}{0.500}
\setval{bell-blocks-fixed/run/BobB0/1}{0.500}
\setval{bell-blocks-fixed/run/BobB0/2}{0.500}
\setval{bell-blocks-fixed/run/BobB1/1}{0.500}
\setval{bell-blocks-fixed/run/BobB1/2}{0.500}
\setval{bell-blocks-fixed/sep/AliceA0/1}{0.499}
\setval{bell-blocks-fixed/sep/AliceA0/2}{0.499}
\setval{bell-blocks-fixed/sep/AliceA1/1}{0.500}
\setval{bell-blocks-fixed/sep/AliceA1/2}{0.500}
\setval{bell-blocks-fixed/sep/BobB0/1}{0.499}
\setval{bell-blocks-fixed/sep/BobB0/2}{0.500}
\setval{bell-blocks-fixed/sep/BobB1/1}{0.500}
\setval{bell-blocks-fixed/sep/BobB1/2}{0.500}
\setval{bell-uniform-random/chsh}{2.508}
\setval{bell-uniform-random/ns}{\yes}
\setval{bell-uniform-random/sep}{\yes}
\setval{bell-uniform-random/run/AliceA0/1}{0.501}
\setval{bell-uniform-random/run/AliceA0/2}{0.503}
\setval{bell-uniform-random/run/AliceA1/1}{0.501}
\setval{bell-uniform-random/run/AliceA1/2}{0.500}
\setval{bell-uniform-random/run/BobB0/1}{0.501}
\setval{bell-uniform-random/run/BobB0/2}{0.500}
\setval{bell-uniform-random/run/BobB1/1}{0.498}
\setval{bell-uniform-random/run/BobB1/2}{0.502}
\setval{bell-uniform-random/sep/AliceA0/1}{0.500}
\setval{bell-uniform-random/sep/AliceA0/2}{0.500}
\setval{bell-uniform-random/sep/AliceA1/1}{0.500}
\setval{bell-uniform-random/sep/AliceA1/2}{0.500}
\setval{bell-uniform-random/sep/BobB0/1}{0.500}
\setval{bell-uniform-random/sep/BobB0/2}{0.500}
\setval{bell-uniform-random/sep/BobB1/1}{0.500}
\setval{bell-uniform-random/sep/BobB1/2}{0.500}
\setval{bell-uniform-fixed/chsh}{4.000}
\setval{bell-uniform-fixed/ns}{\yes}
\setval{bell-uniform-fixed/sep}{\yes}
\setval{bell-uniform-fixed/run/AliceA0/1}{0.500}
\setval{bell-uniform-fixed/run/AliceA0/2}{0.500}
\setval{bell-uniform-fixed/run/AliceA1/1}{0.500}
\setval{bell-uniform-fixed/run/AliceA1/2}{0.500}
\setval{bell-uniform-fixed/run/BobB0/1}{0.500}
\setval{bell-uniform-fixed/run/BobB0/2}{0.500}
\setval{bell-uniform-fixed/run/BobB1/1}{0.500}
\setval{bell-uniform-fixed/run/BobB1/2}{0.500}
\setval{bell-uniform-fixed/sep/AliceA0/1}{0.500}
\setval{bell-uniform-fixed/sep/AliceA0/2}{0.500}
\setval{bell-uniform-fixed/sep/AliceA1/1}{0.500}
\setval{bell-uniform-fixed/sep/AliceA1/2}{0.500}
\setval{bell-uniform-fixed/sep/BobB0/1}{0.500}
\setval{bell-uniform-fixed/sep/BobB0/2}{0.500}
\setval{bell-uniform-fixed/sep/BobB1/1}{0.500}
\setval{bell-uniform-fixed/sep/BobB1/2}{0.500}

\newcommand{\yes}{\ding{51}}
\newcommand{\no}{\ding{55}}

\newcommand{\missedrow}{\rowcolor{BlockRedBody}}
\newcommand{\CHSH}{\mathrm{CHSH}}
\newcommand{\pr}[2]{[\,#1 \vdash #2\,]}
\newcommand{\fr}[1]{[\,#1\,]}
\newcommand{\tA}[1]{\texttt{A#1}}
\newcommand{\tB}[1]{\texttt{B#1}}
\newcommand{\xp}{x{=}{+}}
\newcommand{\yp}{y{=}{+}}
\newcommand{\pausehere}{\begin{center}\small\textcolor{TextGray}{\ding{167}\quad\ding{167}}\end{center}}

\newcommand{\vty}[1]{\textcolor{HsType}{\texttt{#1}}}
\newcommand{\vfn}[1]{\textcolor{HsFunction}{\texttt{#1}}}

\newcommand{\runtable}[1]{{\small\renewcommand{\arraystretch}{1.12}%
  \begin{tabular}{@{}cc|cccc|ccr@{}}
  $a$ & $b$ & $(+,+)$ & $(+,-)$ & $(-,+)$ & $(-,-)$ & $\fr{\xp}$ & $\fr{\yp}$ & trials\\ \hline
  \input{generated/table-#1.tex} 
  \end{tabular}}}
\newcommand{\idealtable}[4]{{\small\renewcommand{\arraystretch}{1.12}%
  \begin{tabular}{@{}cc|cccc@{}}
  $a$ & $b$ & $(+,+)$ & $(+,-)$ & $(-,+)$ & $(-,-)$\\ \hline
  \tA0 & \tB0 & #1\\ \tA0 & \tB1 & #2\\ \tA1 & \tB0 & #3\\ \tA1 & \tB1 & #4
  \end{tabular}}}

\tikzset{
  bell/.style={font=\small, >={Stealth[length=2mm]}, line width=0.6pt},
  fbox/.style={fill=FBoxFill, minimum width=13mm, minimum height=8mm},
  gbox/.style={fill=GBoxFill, minimum width=13mm, minimum height=8mm},
  lam/.style={circle, fill=black!14, minimum size=7.5mm, inner sep=0pt},
  comp/.style={draw=Navy, fill=PortFill, rounded corners=2pt, inner sep=3pt,
               font=\footnotesize, align=center},
  recnode/.style={comp, draw=BlockTeal, fill=BlockTealBody},
  chshnode/.style={comp, draw=BlockBlue, fill=BlockBlue, text=white},
  nsnode/.style={comp, draw=BlockRed, fill=BlockRed, text=white},
  membrane/.style={draw=BlockTeal, densely dashed, rounded corners=4pt, line width=0.8pt, inner sep=4pt},
  sfbox/.style={draw=black!45, fill=FBoxFill, minimum width=6.5mm, minimum height=5.5mm, inner sep=1pt},
  sgbox/.style={draw=black!45, fill=GBoxFill, minimum width=6.5mm, minimum height=5.5mm, inner sep=1pt},
  wire/.style={->, draw=black},
  lbl/.style={font=\scriptsize},
  note/.style={font=\scriptsize, text=TextGray, align=right},
  band/.style={draw=Navy, rounded corners=3pt, line width=0.8pt, inner sep=6pt},
  bandlbl/.style={font=\footnotesize\bfseries, text=Navy},
  fn/.style={font=\scriptsize\ttfamily, text=Navy},
  mm/.style={comp, minimum height=8.5mm, text width=25mm, inner sep=2pt, font=\footnotesize},
}

\newcommand{\twoboxes}{%
  \node[fbox] (f) at (0,0) {$f$};
  \node[gbox] (g) at (5.4,0) {$g$};
  \node (a) at (0,1.3) {$a \in \{0,1\}$};
  \node (b) at (5.4,1.3) {$b \in \{0,1\}$};
  \node (x) at (0,-1.3) {$x \in \{\pm1\}$};
  \node (y) at (5.4,-1.3) {$y \in \{\pm1\}$};
  \draw[->] (a) -- (f); \draw[->] (b) -- (g);
  \draw[->] (f) -- (x); \draw[->] (g) -- (y);
  \node[font=\bfseries, anchor=east] at (-1.25,0) {Alice};
  \node[font=\bfseries, anchor=west] at (6.65,0) {Bob};}

\tikzset{
  layer/.style={rounded corners=4pt, line width=0.5pt},
  elayer/.style={layer, draw=CExp!30, fill=BlockBlueBody},
  mlayer/.style={layer, draw=CMem!55, densely dashed, fill=BlockTealBody},
  tlayer/.style={layer, draw=CTrial!30, fill=CTrial!4},
  layername/.style={font=\footnotesize\bfseries, text=Navy, anchor=west},
  layertype/.style={font=\footnotesize\ttfamily, anchor=west},
  lnode/.style={comp, minimum height=8.6mm, text width=21mm},
  flow/.style={draw=black!65, line width=0.6pt},
  ehl/.style={}, mhl/.style={}, thl/.style={}, schedhl/.style={}, samphl/.style={},
  setpath/.style={flow}, lampath/.style={flow}, respath/.style={flow},
}
\newcommand{\layers}{%
  \path[elayer, ehl] (0,-0.55) rectangle (12.4,0.55);
  \path[mlayer, mhl] (0,-1.42) rectangle (12.4,-0.88);
  \path[tlayer, thl] (0,-3.42) rectangle (12.4,-1.74);
  \node[layername] at (0.18,0.18) {experiment layer};
  \node[layertype, text=CExp] at (0.18,-0.21) {Exp e};
  \node[layername] at (0.18,-1.15) {membrane\quad{\mdseries\ttfamily\color{CMem}RunTrial e t}};
  \node[layername] at (0.18,-2.39) {trial layer};
  \node[layertype, text=CTrial] at (0.18,-2.76) {TrialModel t};
  \node[lnode, schedhl] (sched) at (5.0,0) {scheduler\\[-1pt] {\scriptsize\ttfamily Scheduler e}};
  \node[lnode, samphl] (samp) at (8.0,0) {sampler\\[-1pt] {\scriptsize\ttfamily Sampler e}};
  \node[lnode] (run) at (10.9,0) {runner\\[-1pt] {\scriptsize $i = 0, \ldots, n{-}1$}};
  \node[sfbox] (f) at (4.4,-2.15) {$f$};
  \node[sgbox] (g) at (5.6,-2.15) {$g$};
  \coordinate (J) at (5.0,-1.15);
  \draw[setpath] (sched.south) -- (J);
  \draw[setpath, ->] (J) -- node[left, lbl, pos=0.78] {$a_i$} (f.north);
  \draw[setpath, ->] (J) -- node[right, lbl, pos=0.78] {$b_i$} (g.north);
  \node[lbl, anchor=west] at (5.07,-0.71) {$(a_i, b_i)$};
  \draw[lampath, ->] (samp.south) -- (8.0,-1.15);
  \draw[lampath] (8.0,-1.15) -- (J);
  \node[lbl, anchor=west] at (8.07,-0.71) {$\lambda_i$};
  \fill[black!65] (J) circle (1.3pt);
  \coordinate (K) at (5.0,-3.15);
  \draw[respath] (f.south) -- node[left, lbl, pos=0.3] {$x_i$} (K);
  \draw[respath] (g.south) -- node[right, lbl, pos=0.3] {$y_i$} (K);
  \draw[respath, ->] (K) -| (run.south);
  \fill[black!65] (K) circle (1.3pt);
  \node[lbl, anchor=south] at (8.3,-3.15) {$(x_i, y_i)$};}

\newcommand{\pipeline}{%
  \node[mm, fill=BlockTealBody, draw=BlockTeal] (trial) at (0,0) {page $i$: $a_i\ b_i\ x_i\ y_i$\\[-1pt] {\scriptsize\ttfamily Record, trial}};
  \node[mm] (experiment) at (3.35,0) {record: pages $0, \ldots, n{-}1$\\[-1pt] {\scriptsize\ttfamily experiment}};
  \node[mm, text width=27mm] (empirical) at (6.85,0) {table: $N(a, b, x, y)$\\[-1pt] {\scriptsize\ttfamily empirical, \mbox{numTrials}, freq}};
  \node[mm, fill=BlockBlueBody, draw=BlockBlue] (chsh) at (10.55,0.55) {CHSH value\\[-1pt] {\scriptsize\ttfamily correlator, chshOf}};
  \node[mm, fill=BlockRedBody, draw=BlockRed] (ns) at (10.55,-0.55) {no-signaling\\[-1pt] {\scriptsize\ttfamily comparisons, violated}};
  \draw[wire] (trial) -- (experiment);
  \draw[wire] (experiment) -- (empirical);
  \draw[wire] (empirical.east) -- (chsh.west);
  \draw[wire] (empirical.east) -- (ns.west);}

\newcommand{\runtop}{%
  \node[comp] (step) at (0,0) {step $i = 0, \ldots, n{-}1$};
  \node[comp] (sched) at (-1.55,-1.2) {scheduler};
  \node[sfbox] (f) at (0.95,-1.2) {$f$};
  \node[sgbox] (g) at (1.75,-1.2) {$g$};
  \node[membrane, fit=(f)(g)] (mem) {};
  \draw[wire] (step) -- (sched);
  \draw[wire] (step) -- (mem);
  \draw[wire] (sched) -- node[below, lbl] {$(a_i, b_i)$} (mem);}

\newcommand{\oldcheck}{%
  \node[comp] (fix) at (5.9,0) {fixed $(a,b)$,\\ step $i = 0, \ldots, n{-}1$};
  \foreach \k in {3,2,1}
    \draw[membrane, fill=white] (5.9-0.87+\k*0.1, -1.2-0.415-\k*0.1) rectangle ++(1.74,0.83);
  \draw[membrane, fill=white] (5.9-0.87, -1.2-0.415) rectangle ++(1.74,0.83);
  \node[sfbox] (f2) at (5.5,-1.2) {$f$};
  \node[sgbox] (g2) at (6.3,-1.2) {$g$};
  \coordinate (stack) at (5.9+0.87+0.3, -1.2-0.415-0.3);
  \node[lbl, text=AlertRed, anchor=east] at (4.98,-1.2) {$\times 4$};
  \draw[wire] (fix) -- node[right, lbl] {$(a,b)$} (5.9,-1.2+0.415);
  \node[comp] (count) at (5.9,-2.45) {count $\xp$, $\yp$};
  \draw[wire] (5.9,-1.2-0.415-0.3) -- node[right, lbl] {$(x,y)$} (count);
  \node[comp] (cmp) at (5.9,-3.55) {compare marginals};
  \draw[wire] (count) -- (cmp);
  \node[nsnode] (ns) at (5.9,-4.65) {no-signaling};
  \draw[wire] (cmp) -- (ns);
  \node[band, fit=(fix)(f2)(g2)(stack)(count)(cmp)(ns)] (checkband) {};
  \node[bandlbl, anchor=south west] at (checkband.north west) {separate experiments};
  \node[lbl, text=AlertRed, anchor=north east] at ($(checkband.south east)+(0,-0.05)$) {no scheduler};}

\usepackage{adjustbox}
\usepackage{newunicodechar}
\newunicodechar{^^^^2713}{\ding{51}}
\newunicodechar{^^^^2717}{\ding{55}}
\renewcommand{\hs}{\lstinline[style=hask,basicstyle=\ttfamily,frame=none,breaklines=true,breakatwhitespace=true]}

\def\orcidDing{}      % e.g. 0000-0002-1825-0097
\def\orcidSabry{}

\begin{document}

\title{The Bell experiment, monadically}

\author{Chenchao Ding}
\ifx\orcidDing\empty\else\orcid{\orcidDing}\fi
\affiliation{%
  \institution{Indiana University}
  \country{USA}
  \authoremail{cd17@iu.edu}
}
\author{Amr Sabry}
\ifx\orcidSabry\empty\else\orcid{\orcidSabry}\fi
\affiliation{%
  \institution{Indiana University}
  \country{USA}
  \authoremail{sabry@iu.edu}
}

\begin{abstract}
\noindent Bell's theorem says that no classical model satisfying certain independence assumptions can reproduce the correlations that quantum mechanics predicts and experiments observe. In this paper we write the two-party Bell experiment as a small Haskell program, and find those assumptions in its structure: in what each part of the program can see, and in how the parts are composed. The boxes of the two parties compute in one monad, and the experiment that schedules and repeats their trials computes in another; a \emph{membrane} runs the first inside the second. A run keeps the record of its trials, and both of its verdicts, the CHSH value and whether the boxes signal, are read off one table: the empirical model of that run.

In this setting, a model is classical when it has three properties. It is \emph{blind}: the hidden strategy of the boxes is drawn without seeing the measurement settings. It is \emph{local}: given that strategy, neither box's outcome depends on the other party's setting. It is \emph{factored}: given the strategy and the settings, the two outcomes are independent. Together, they give the classical bound of $2$. We give them up one at a time and run what results. Without blindness, we get experimental loopholes and superdeterminism. Without factoring, we get quantum boxes, which reach $2\sqrt2$, and the hypothetical PR boxes, which reach $4$, both without signaling. Without locality, we get boxes that rewrite a shared hidden state; whether they signal depends on how that state is sampled.

We also compare two ways to check no-signaling: reading it off the run's own table, and estimating it in separate experiments at fixed settings.
\end{abstract}

\maketitle

% written by jfpify.py from body.tex; do not edit
%% =====================================================================
\FloatBarrier
\section{Introduction}\label{sec:intro}

Bell's theorem draws one of the sharpest lines between classical and quantum models of information. \citet{Bell1964} showed that no classical theory satisfying a natural \emph{locality assumption} can reproduce all of the correlations that quantum mechanics predicts: systems whose behavior is determined by local information alone cannot account for certain statistical patterns.

Clauser, Horne, Shimony, and Holt~\citep{CHSH1969} recast Bell's argument in operational terms, as an inequality, now called the CHSH inequality, whose violation can be tested in the laboratory. Experiments of increasing precision followed, from those of \citet{Aspect1982} to the loophole-free tests of \citet{Hensen2015}, and they have consistently confirmed the quantum predictions. The 2022 Nobel Prize in Physics was awarded to Alain Aspect, John F. Clauser, and Anton Zeilinger ``\emph{for experiments with entangled photons, establishing the violation of Bell inequalities and pioneering quantum information science}.''

The scenario behind the CHSH inequality takes no physics to describe. We are given two black boxes. Each box accepts one of two inputs, a measurement \emph{setting}, and produces one of two outputs, an \emph{outcome}. We assume nothing about the inside of the boxes; only their input-output behavior is observable.

\begin{figure}[htbp]
\centering
\begin{adjustbox}{max width=\linewidth}\begin{tikzpicture}[bell]
  \twoboxes
\end{tikzpicture}\end{adjustbox}
\caption{Two black boxes $f$ and $g$.}
\label{fig:boxes}
\end{figure}
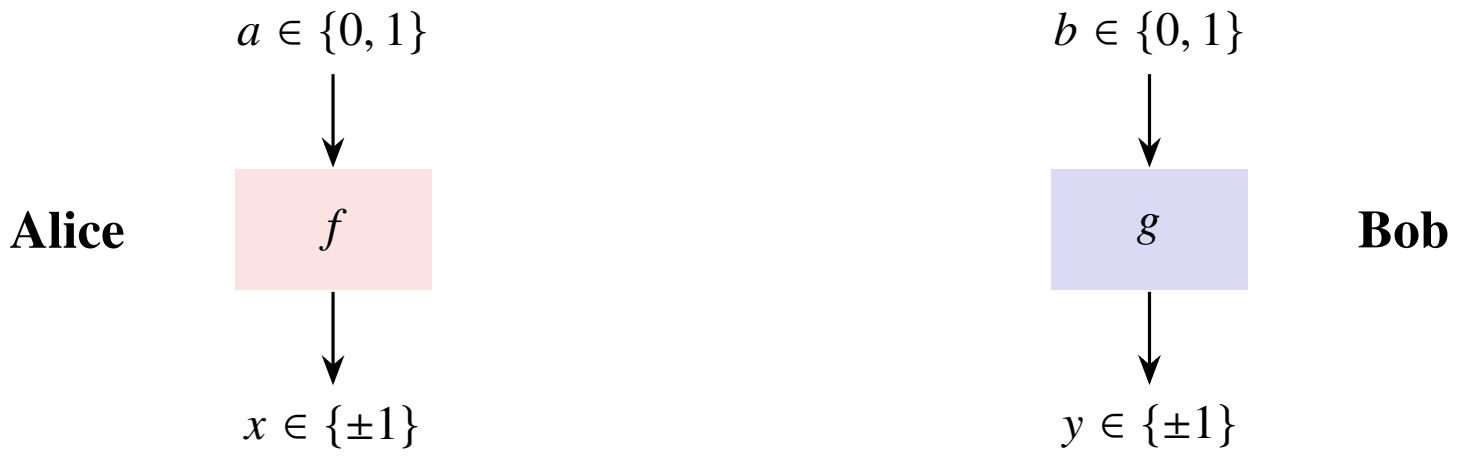

As a first approximation, we model the two boxes as deterministic functions
\[
  f, g : \{0, 1\} \longrightarrow \{\pm 1\}.
\]
Two space-like separated experimenters, Alice and Bob, each hold one box, $f$ and $g$ respectively (Figure~\ref{fig:boxes}). In one \emph{trial} of the experiment, Alice chooses a setting $a$ and Bob a setting $b$, and they observe the outcomes $(x, y) = \bigl(f(a), g(b)\bigr)$. We call the pair of settings $(a, b)$ the \emph{context} of the trial; there are four contexts.

To quantify the correlation between the two outcomes, we give each trial a score: the product $f(a)g(b)$, which is $+1$ when the outcomes agree and $-1$ when they disagree. The CHSH expression combines the scores of the four contexts:
\[
  \CHSH(f, g) = f(0)g(0) + f(0)g(1) + f(1)g(0) - f(1)g(1).
\]
The minus sign on the last term is essential. We invite the reader to pause here and work out which values this expression can take when $f$ and $g$ are fixed functions.

\pausehere

The four values $f(0), f(1), g(0), g(1)$ are fixed in advance, and each is $\pm1$. Regrouping the terms gives
\[
  \CHSH(f, g) = f(0)\bigl(g(0) + g(1)\bigr) \;+\; f(1)\bigl(g(0) - g(1)\bigr).
\]
If $g(0) = g(1)$, their sum is $\pm2$ and their difference vanishes; if $g(0) \neq g(1)$, the sum vanishes and the difference is $\pm2$. Either way exactly one term survives, and its value is $\pm2$. For deterministic boxes, then, the CHSH expression takes only the values $+2$ and $-2$. Its absolute value is bounded by $2$: the \emph{classical bound}.

\paragraph{Two quantum surprises.} The first surprise, and the one that history emphasizes, is that quantum mechanics exceeds this bound. Measuring a quantum system yields one of its possible outcomes at random, with probabilities determined by the state of the system and by the measurement chosen. Two systems are \emph{entangled} when their joint state is not made of a state of each; the outcomes of measurements on the two can then be correlated, however far apart the systems are. Boxes that answer at random give a context no fixed score: the score $xy$ varies from trial to trial, and the CHSH expression combines its expected values, the \emph{correlators} $E(a, b)$,
\[
  \CHSH = E(0, 0) + E(0, 1) + E(1, 0) - E(1, 1),
\]
which is $\CHSH(f, g)$ again for deterministic boxes, whose correlators are $E(a, b) = f(a)g(b)$. When Alice and Bob each measure one of two entangled systems, with well-chosen measurements, quantum mechanics predicts the values $\pm 2\sqrt2$, and experiments observe them, as statistical averages over many \emph{repeated trials}.

The second surprise is that $2\sqrt2$ is as far as quantum mechanics goes: this is the Tsirelson bound~\citep{Tsirelson1980}. Algebraically, the bound is unexpected. The CHSH expression combines four terms, each between $-1$ and $+1$, so its algebraic maximum is $4$.

\paragraph{Signaling or not?} Quantum mechanics cannot reach the value $4$, but boxes can, in two ways. The first is trivial: boxes that communicate during a trial can coordinate their outcomes so that all four terms contribute $+1$. Such boxes violate a condition of Bell experiments called \emph{no-signaling}~\citep{AbramskyBrandenburger2011,Bell1964,PopescuRohrlich1994}. We write $\pr{c}{e}$ for the probability of $e$ given $c$, and $\fr{e}$ for that of $e$. No-signaling requires that one party's choice of setting does not affect the distribution of the other party's outcome:
\begin{equation}\label{eq:ns}
  \pr{a, b}{x} = \pr{a}{x}, \qquad \pr{a, b}{y} = \pr{b}{y}.
\end{equation}
In physical terms, the boxes cannot carry a message between Alice and Bob, who are space-like separated: special relativity allows none.

The second way is subtle. \citet{PopescuRohrlich1994} described hypothetical correlations, now called \emph{PR boxes}, that satisfy no-signaling exactly and still reach $\CHSH = 4$: they are stronger than any correlations that quantum mechanics allows.

No-signaling alone therefore does not explain the Tsirelson bound. Quantum correlations exceed the classical bound $2$, and stop well short of the maximum $4$ that no-signaling permits. The three bounds form a hierarchy:
\[
  \text{Classical } (\CHSH \le 2) \;\subset\; \text{Quantum } (\CHSH \le 2\sqrt2) \;\subset\; \text{No-Signaling } (\CHSH \le 4).
\]
It corresponds to strict inclusions between sets of empirical models that are studied extensively in the literature on nonlocality and contextuality~\citep{Abramsky2013,AbramskyBrandenburger2011}.

\paragraph{This paper.} What must a model give up in order to climb this hierarchy? Bell's theorem answers with independence assumptions stated in probabilities. We answer with a program. We write the CHSH experiment in Haskell so that the boxes may use any computational effect, run it, and read the CHSH value and a no-signaling verdict off the record of the run. Bell's assumptions then become properties of the program, and each step up the hierarchy, like each experimental loophole, becomes a small, visible change to the code. That presentation is the contribution. Specifically, we give
\begin{itemize}
\item an executable model of a Bell experiment in two layers, in which the boxes compute in one monad and the experiment in another, joined by a \emph{membrane} (Sections~\ref{sec:boxes} and~\ref{sec:experiment});
\item Bell's assumptions as three properties of such programs, \emph{blind}, \emph{local} and \emph{factored}, two of which have the same shape at the two layers: a pair built by applicative composition, whose components come to depend on each other when the property is given up (Sections~\ref{sec:lambda} and~\ref{sec:conclusion});
\item a zoo of models that give the properties up one at a time: experimental loopholes and superdeterminism, which are not blind (Sections~\ref{sec:clocked} and~\ref{sec:lambda}); quantum boxes and PR boxes, which are not factored (Sections~\ref{sec:quantum} and~\ref{sec:pr}); and a shared location and hidden states that the boxes rewrite, which are not local, the latter giving the PR box and also a table that quantum mechanics realizes (Sections~\ref{sec:boxes} and~\ref{sec:hidden});
\item for every run, both verdicts read off one table, its empirical model, and a demonstration of why: a no-signaling check made in separate experiments passes runs that signal (Section~\ref{sec:checked}).
\end{itemize}

We assume a reading knowledge of Haskell, including type classes and the \hs|Reader| and \hs|State| monads~\citep{Hutton2016}; library functions are explained where they first appear. We assume no background in quantum mechanics: Section~\ref{sec:quantum} introduces the state vectors, projectors and Born rule that it uses, and Appendix~\ref{app:quantum} explains them from scratch. The complete implementation, with a test suite, is included as supplementary material. Every listing here is quoted from it, without its comments, and numbered in the margin. Every number, table and printed report of a run is written by a program that runs the models and stops if a claim of the text about them does not hold; runs that use random numbers fix the generator first, each with a seed of its own, so they are reproducible.

%% =====================================================================
\FloatBarrier
\section{Boxes as effectful computations}\label{sec:boxes}

Pure functions cannot exceed the classical bound (Section~\ref{sec:intro}), so any violation requires a computational effect. Our plan is to write the CHSH experiment once, parameterized by the effect that the boxes may use inside: randomness, shared state, or something more structured. Each choice of effect is an assumption about the resources of the boxes. Seeing which effects lead to which CHSH values will isolate the assumptions behind the classical bound, those that permit the quantum violation, and those that overshoot it.

Settings and outcomes are enumerations:
\code{17-18,23-23,35-37,41-42}{CHSH/Util.hs}[lst:util]
\hs|trialValue| is the score $xy$ of a trial, and \hs|chshOf| combines a number per context as the CHSH expression does: a score in this section, a correlator from Section~\ref{sec:run} on.

The boxes are the methods of a type class, in a monad \hs|t| that stands for whatever resources they may use:
\code{16-16,18-18,20-20,22-23}{CHSH/Experiment.hs}[lst:trialmodel]
The methods \hs|localA| and \hs|localB| are the two boxes, each a function of its own setting: the \emph{local responses}. The method \hs|jointAB| is one trial of both boxes: the \emph{joint response}. By default it is \hs|liftA2 (,) (localA a) (localB b)|, which runs Alice's box, then Bob's, and pairs their outcomes. An instance may override it, as the models of Sections~\ref{sec:quantum} and~\ref{sec:pr} do.

\paragraph{Boxes as pure functions.} The deterministic boxes of the introduction live in the \hs|Identity| monad. As an example, let each box answer \hs|Plus| at either setting. For the moment we compute the CHSH expression directly, from the score of one trial in each context:
\code{21-23,25-26}{CHSH/Explore.hs}
Evaluating \hs|chshId| gives $\val{chshId}$. The statistics of these boxes fill a table with a row for each context and a column for each pair of outcomes (Figure~\ref{fig:allplus}). Such tables may be viewed as \emph{empirical models} \`a la \citet{AbramskyBrandenburger2011}: they give the distribution of the joint outcome in each context. Here every entry is $0$ or $1$, because nothing in the boxes is random.

\begin{figure}[htbp]
\centering
\idealtable{1 & 0 & 0 & 0}{1 & 0 & 0 & 0}{1 & 0 & 0 & 0}{1 & 0 & 0 & 0}
\caption{The deterministic all-plus empirical model.}
\label{fig:allplus}
\end{figure}

\paragraph{Boxes with a shared location.} At the other extreme, a state monad makes the value $4$ easy to reach. Alice and Bob share a location. Alice's box writes her setting there and answers \hs|Minus|. Bob's box reads the location: if the settings are \hs|(A1, B1)|, it disagrees with Alice and answers \hs|Plus|; otherwise it answers \hs|Minus|. Below, \hs|modify'| updates the location, \hs|gets| reads it, and \hs|evalState| runs each trial from an empty location:
\code{39-39,41-49,51-52}{CHSH/Explore.hs}
Evaluating \hs|chshST| gives $\val{chshST}$, beyond even the quantum bound. It does so by signaling: Bob's outcome depends on Alice's setting. Nothing in the types of the methods prevented that. \hs|localB| is never handed Alice's setting; it finds it in the state.

These two instances mark out the range of the framework. Pure boxes are too weak to account for quantum correlations, and unrestricted effects trivialize the experiment by letting the boxes communicate. The question is what lies between the two.

%% =====================================================================
\FloatBarrier
\section{An experiment in two layers}\label{sec:experiment}

A CHSH experiment is not one trial but many, with varying settings. This section adds what repetition requires: a second layer of effects, a scheduler that chooses the settings, the record of a run with the table that is tallied from it, and the two verdicts that are read off that table.

%% ---------------------------------------------------------------------
\subsection{Two layers of effects}\label{sec:layers}

Repetition gives the boxes an opportunity that a single trial does not: they can coordinate \emph{across} trials, without communicating during any one of them. This is how the memory and timing loopholes of real Bell tests arise. To model them, we distinguish the effects inside one trial, in the monad \hs|t|, from effects that persist across the trials of an experiment, in a second monad:
\code{32-33}{CHSH/Experiment.hs}[lst:exp]
An experiment is a sequence of \emph{steps}, numbered from $0$, with one trial at each. The type \hs|Exp e| is the monad of instructions that may read the current step: an instruction of type \hs|Exp e a| wraps a function \hs|Int -> e a|, which, given the step $i$, computes an \hs|a| with the effect \hs|e| of the lab in which the experiment runs (\hs|Identity| or \hs|IO| in this paper). A value of type \hs|RunTrial e t| embeds a trial computation \hs|t a| into the experiment, for every result type \hs|a|. We call such an embedding a \emph{membrane}: it decides what passes from the experiment into a trial.

The membranes of the two models of Section~\ref{sec:boxes} let nothing through. \hs|ReaderT| makes an instruction from a function of the step; both functions ignore the step and run each trial on its own, the second from an empty location:
\code{18-19}{CHSH/Identity.hs}
\code{33-34}{CHSH/Shared.hs}
Most membranes in this paper ignore the step. It matters for boxes that adapt across trials (Section~\ref{sec:clocked}) and for strategies that vary from trial to trial (Section~\ref{sec:staging}).

\begin{figure}[htbp]
\centering
\begin{adjustbox}{max width=\linewidth}\begin{tikzpicture}[bell]
  \layers
\end{tikzpicture}\end{adjustbox}
\caption{The architecture: two layers, and the membrane between them. The \textcolor{CExp}{experiment layer} (\mbox{\texttt{Exp e}}) holds the scheduler, the sampler and the runner; the \textcolor{CTrial}{trial layer} (\mbox{\texttt{TrialModel t}}) holds Alice's box $f$ and Bob's box $g$; the \textcolor{CMem}{membrane} (\mbox{\texttt{RunTrial e t}}) runs the boxes for the experiment layer.}
\label{fig:layers}
\end{figure}
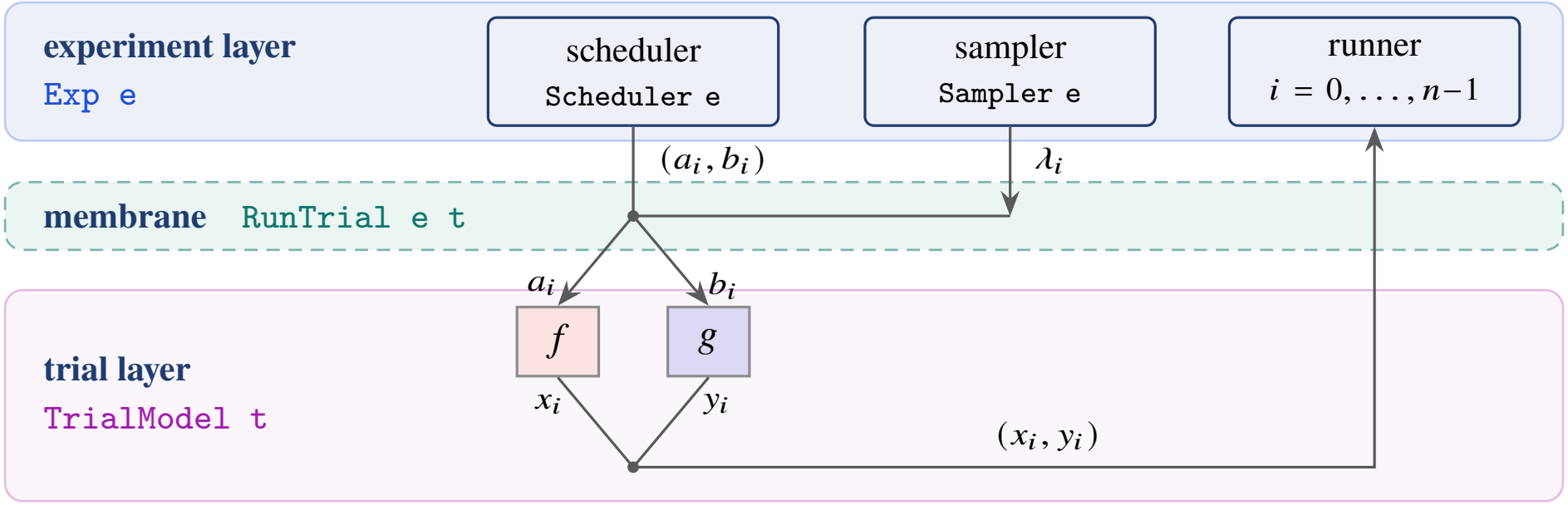

Figure~\ref{fig:layers} shows the architecture that results: two layers, and the membrane between them. The experiment layer chooses the settings with a \emph{scheduler}, may draw a strategy for the boxes with a \emph{sampler} (Section~\ref{sec:staging}), and runs the steps; the trial layer holds the boxes; the membrane runs the boxes for the experiment. At step $i$ the settings $(a_i, b_i)$ and a strategy $\lambda_i$ flow down, and the outcomes $(x_i, y_i)$ flow back up.

%% ---------------------------------------------------------------------
\subsection{Schedulers}\label{sec:schedules}

Every experiment now needs a \emph{scheduler}: an instruction of the experiment layer that returns the settings Alice and Bob use at the current step. Section~\ref{sec:boxes} went through the four contexts once, in a fixed order: \hs|(A0, B0)|, then \hs|(A0, B1)|, then \hs|(A1, B0)|, and finally \hs|(A1, B1)|. That order was arbitrary, and we do not expect it to affect the result of the experiment. In a real Bell test, Alice and Bob choose their settings at random, and so can a scheduler:
\code{36-37,44-51,53-57}{CHSH/Experiment.hs}[lst:schedules]
The fixed scheduler reads the step with \hs|ask| and cycles through the four contexts. The random scheduler never reads the step: it flips two coins with \hs|randomRIO|, lifted into the experiment monad by \hs|liftIO|. We write $\fr{a, b}_i$ for the probability that a scheduler picks $(a, b)$ at step $i$. For \hs|fixedSchedule| it is $1$ or $0$, and it changes with $i$; for \hs|randomScheduleIO| it is $1/4$ for every context, at every step. The step is an index here, not an event: a subscript $i$ on a probability always means ``at step $i$''.

The settings that a scheduler produces, step after step, are its \emph{schedule}, and we speak of runs under the fixed and under the random schedule. Experiments in \hs|IO| draw from the global generator of random numbers, and \hs|seeded s| sets it to the generator of the seed \hs|s| before it runs an experiment:
\code{61-62}{CHSH/Experiment.hs}[lst:seeded]
Every run in \hs|IO| in this paper is wrapped in a \hs|seeded s| of its own, with its own seed, so it starts from the same generator whatever ran before it; runs in \hs|Identity| draw nothing at random. Figure~\ref{fig:runs} lists the seed of every run. Within a run, the draws are made step after step and, within a step, in the order of the code: first the scheduler's coins (under the random schedule two, Alice's then Bob's; under the fixed one none), then whatever the membrane and the boxes draw. The generator is the \hs|StdGen| of Haskell's \texttt{random} library, version 1.3.1 on \texttt{splitmix} 0.1.2, which the build of this paper pins; another version may draw differently from the same seed.

%% ---------------------------------------------------------------------
\subsection{One experiment, one table}\label{sec:run}

We can now generalize the CHSH calculation of Section~\ref{sec:boxes} to a given scheduler and any number of steps.

\paragraph{A trial.} A trial runs the scheduler and the boxes at the same step, and keeps what an experimenter writes down: the two settings and the two outcomes.
\code{80-81,83-88}{CHSH/Experiment.hs}[lst:trial]
The scheduler and the membrane are both instructions of the experiment layer, so they read the same step $i$. The trial draws the settings once, names the pair \hs|s|, and hands it to the membrane together with the trial computation \hs|jointAB a b|. A trial thus takes a membrane for each context, of type \hs|Context -> RunTrial e t|, which may depend on the settings it is handed. \hs|blind runTrial| is one that does not: it ignores the settings and hands every trial computation to \hs|runTrial|:
\code{91-92}{CHSH/Experiment.hs}
Every model in this paper but one uses such a membrane. The exception is the subject of Section~\ref{sec:superdet}.

\paragraph{A run.} An experiment runs one trial at each of the steps $0, \ldots, n-1$ and returns the \emph{record} of the run, one \emph{page} per step:
\code{96-97}{CHSH/Experiment.hs}
The trial is an instruction of the experiment layer, so \hs|runReaderT oneTrial| unwraps its function of the step, \hs|Int -> e Record|, and \hs|mapM| applies that function to each step in order. In the \hs|IO| lab the coins are drawn in that order, so one seed fixes the whole record.

\paragraph{The table.} The empirical model of a run is a table: a row for each context $(a, b)$, a column for each pair of outcomes $(x, y)$, and in each cell the number of trials that recorded that context and those outcomes. A cell is thus a page that a trial can write, $(a, b, x, y)$, and the table gives every page its count. The rows, the columns and the cells all come from the values of the settings and the outcomes. Their types derive \hs|Enum| and \hs|Bounded| (\ref{lst:util}), so \hs|every| lists the values of each, in the order of its declaration. The list applicative builds the rest: \hs|liftA2 (,)|, which pairs the two outcomes in the default \hs|jointAB|, here pairs every value of one list with every value of the other, and \hs|<$>| with \hs|<*>| fills the four fields of a page with every combination of values:
\code{108-109,113-114,116-117,119-120}{CHSH/Experiment.hs}[lst:every]
The lists come out in the order of the tables in this paper, from \hs|(A0, B0)| to \hs|(A1, B1)| and from \hs|(Plus, Plus)| to \hs|(Minus, Minus)|. A table pairs each page with its count, and \hs|empirical| tallies a record into one: \hs|filter (== p)| keeps the pages of the record equal to \hs|p|, and \hs|length| counts them.
\code{124-125,127-128}{CHSH/Experiment.hs}
That is, $N(a, b, x, y) = \#\{\, i : \text{page } i = (a, b, x, y) \,\}$. With the all-plus boxes of Section~\ref{sec:boxes}, for instance, every page records the outcomes \hs|(Plus, Plus)|, so every row has all its trials in its first column; divided by the number of trials of its context, it is the row of Figure~\ref{fig:allplus}. Tables derive \hs|Eq|, so the accompanying code can compare two runs count for count.

\paragraph{Frequencies.} These sixteen counts are all that a run reports; everything else is read off them. Divided by $N(a, b)$, the number of trials of its context, a count estimates the conditional probability of its pair of outcomes, $\pr{a, b}{x, y} = N(a, b, x, y) / N(a, b)$. An \hs|Event| picks some pairs of outcomes, such as those in which Bob answers \hs|Plus|, or those in which the two outcomes agree. \hs|numTrials| adds up the counts of the pairs it picks in context $(a, b)$, and \hs|freq| divides the sum by $N(a, b)$:
\code{130-130,133-135,139-141}{CHSH/Experiment.hs}
So \hs|freq m a b (== (x, y))| estimates $\pr{a, b}{x, y}$, and an event that picks several pairs gets the sum of theirs, such as $\pr{a, b}{\xp} = \pr{a, b}{{+}, {+}} + \pr{a, b}{{+}, {-}}$ for Alice's \hs|Plus|.

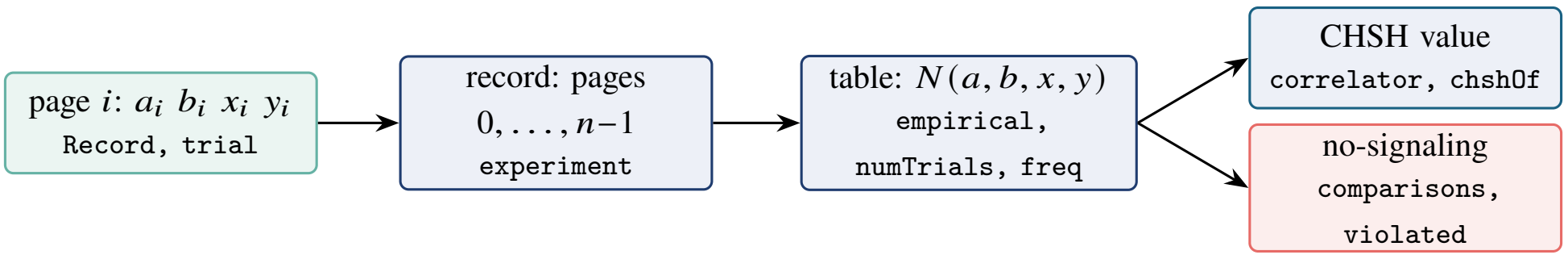
\begin{figure}[htbp]
\centering
\begin{adjustbox}{max width=\linewidth}\begin{tikzpicture}[bell]
  \pipeline
\end{tikzpicture}\end{adjustbox}
\caption{One run, one table. Each step writes a page; the record of the run is tallied into its empirical model, and both verdicts are read off that table.}
\label{fig:pipeline}
\end{figure}

\paragraph{The CHSH value.} The CHSH value is read off the same table. A context now holds many trials, so it contributes their average score, the correlator $E(a, b)$ of Section~\ref{sec:intro}: each pair of outcomes counts with its score $xy$, weighted by its frequency among the trials of that context:
\begin{gather*}
  E(a, b) = \sum_{x, y} xy\,\pr{a, b}{x, y}, \\
  \CHSH = E(\tA0, \tB0) + E(\tA0, \tB1) + E(\tA1, \tB0) - E(\tA1, \tB1).
\end{gather*}
In the code, \hs|correlator| is \hs|trialValue| in expectation, and the CHSH value is \hs|chshOf| of the four correlators, as \hs|chshId| was \hs|chshOf| of four scores:
\code{151-153,156-159}{CHSH/Experiment.hs}[lst:chsh]
Here \hs|<$>| applies a function to the result of a computation: \hs|chsh| runs the experiment with the settings that the scheduler provides, tallies its record, and computes the CHSH value from the table.\footnote{Averaging over all trials the scores with the signs of their CHSH terms, times $4$, gives the same value under the fixed schedule when $4$ divides $n$, but not under a random one, where the contexts occur unequally often.} Deterministic boxes give every trial of a context the same outcomes, $\bigl(f(a), g(b)\bigr)$, so each correlator is the score $f(a)g(b)$, and the table gives $\CHSH(f, g)$ again. Figure~\ref{fig:pipeline} shows the whole pipeline.

%% ---------------------------------------------------------------------
\subsection{Testing no-signaling}\label{sec:verdicts}

The same table decides no-signaling. With outcomes $\pm1$, Equation~\eqref{eq:ns} amounts to four equations, one per party and setting:
\begin{equation}\label{eq:four}
  \underbrace{\pr{a, \tB0}{\xp} = \pr{a, \tB1}{\xp}}_{\text{Alice at } a}
  \qquad\qquad
  \underbrace{\pr{\tA0, b}{\yp} = \pr{\tA1, b}{\yp}}_{\text{Bob at } b}
\end{equation}
Only two are displayed because each leaves its party's own setting free: Alice's stands for one equation at $a = \tA0$ and one at $a = \tA1$, and Bob's for one at each of $b = \tB0$ and $b = \tB1$. Each side is estimated from two counts of the run's table. A comparison names its condition, and holds for each of its two contexts the number of trials in which that party answered \hs|Plus| and the number of all trials there:
\code{26-30,32-39}{CHSH/NoSignaling.hs}
The frequencies of a finite run are rarely exactly equal, so \hs|violated| decides each equation with a two-proportion test. For frequencies $p_1 = k_1 / n_1$ and $p_2 = k_2 / n_2$,
\begin{gather*}
  p = \frac{k_1 + k_2}{n_1 + n_2}, \qquad
  \mathit{se} = \sqrt{p\,(1 - p) \left(\frac{1}{n_1} + \frac{1}{n_2}\right)}, \\
  \text{violated} \iff n_1 = 0 \,\lor\, n_2 = 0 \,\lor\, |p_1 - p_2| > 4\,\mathit{se}.
\end{gather*}
\code{51-58}{CHSH/NoSignaling.hs}
Frequencies that agree exactly are never flagged, and frequencies $1$ and $0$ are flagged as soon as the two contexts together have more than $16$ trials. At four standard errors, a large run of a model that does not signal is flagged in fewer than one run in three thousand.

A report holds the table of a run, its CHSH value, and the conditions that the run violates; we say that a run \emph{signals} when it violates at least one. \hs|report| runs the experiment once, builds its table, and reads both verdicts off it; \hs|chshReport| takes the same arguments as \hs|chsh|:
\code{63-67,69-73,76-78,80-81}{CHSH/NoSignaling.hs}[lst:report]
Every CHSH value and every verdict in the rest of this paper comes from such a report: one run of \val{nTrials} trials, one table, both verdicts. We run each model under the fixed schedule, the random one, or both; under the fixed schedule, each context gets \val{nPerContext} trials. Figure~\ref{fig:vocabulary} collects the vocabulary.

\begin{figure}[htbp]
\centering
{\small\renewcommand{\arraystretch}{1.12}%
\begin{adjustbox}{max width=\linewidth}\begin{tabular}{@{}l l l@{}}
\toprule
Term & Meaning & In the code\\
\midrule
setting & a party's input to its box & \vty{ASetting}, \vty{BSetting}\\
context & the two settings $(a, b)$ of a trial & \vty{Context}\\
outcome & what a box answers: $+1$ or $-1$ & \vty{Outcome}\\
trial & one use of both boxes, in one context & \vfn{jointAB}, \vfn{trial}\\
step & the position $i$ of a trial in an experiment & the \vty{Int} of \vty{Exp}\texttt{ e}\\
scheduler & chooses the settings of each step: its \emph{schedule} & \vty{Scheduler}\texttt{ e}\\
membrane & runs a trial computation for the experiment & \vty{RunTrial}\texttt{ e t}\\
run & one experiment: $n$ trials, at the steps $0, \ldots, n-1$ & \vfn{experiment}\\
record & what a run writes down: a \emph{page} $(a, b, x, y)$ per trial & \texttt{[}\vty{Record}\texttt{]}\\
table & the counts $N(a, b, x, y)$ of a record: its empirical model & \vty{Empirical}\\
verdicts & the CHSH value of a table, and whether it \emph{signals} & \vty{Report}\\
\bottomrule
\end{tabular}\end{adjustbox}}
\caption{The vocabulary of an experiment.}
\label{fig:vocabulary}
\end{figure}

The two models of Section~\ref{sec:boxes} already show the check at work. Pure boxes give CHSH \val{pure-fixed/chsh} and do not signal.\footnote{This run uses the pure boxes of the accompanying module \texttt{CHSH.Identity}: Alice answers \hs|Plus| at \hs|A0| and \hs|Minus| at \hs|A1|, and Bob the reverse. The all-plus boxes are a second instance for \hs|Identity|, in a module of its own.} The boxes with a shared location give CHSH \val{shared/chsh}, and Bob's frequency of \hs|Plus| at \hs|B1| is \val{shared/run/BobB1/1} next to \hs|A0| and \val{shared/run/BobB1/2} next to \hs|A1|: he signals, as intended.

%% ---------------------------------------------------------------------
\subsection{Scheduler-induced loopholes}\label{sec:clocked}

The two models so far cannot tell one schedule from another, because their outcomes depend on the settings alone. Here are \emph{clocked} boxes, which also consult the current step of the experiment:
\code{13-14,16-20}{CHSH/Clocked.hs}
Alice always answers \hs|Plus|, and Bob answers \hs|Minus| on every fourth step. Two membranes run these boxes. One passes the current step on to them; the other hides it, and runs every trial as if at step $0$:
\code{23-24,27-28}{CHSH/Clocked.hs}
Two membranes under two schedules make four runs. Their CHSH values concentrate around $1$, $2$, $2$, and $4$, in some order. We invite the reader to pause, to match each value with its run, and to guess which of the runs signal.

\pausehere

Under the fixed schedule, the settings follow a known pattern: \hs|(A0, B0)|, \hs|(A0, B1)|, \hs|(A1, B0)|, and finally \hs|(A1, B1)|. Bob follows a pattern too: he disagrees with Alice exactly when the step satisfies $i \bmod 4 = 3$. When the membrane shows the step to the boxes, the two patterns coincide. Every fourth step measures \hs|(A1, B1)|, the one context in which the CHSH expression rewards disagreement. All four terms then contribute $+1$, and the run reaches the algebraic maximum $4$. Its report, as \hs|printReport| prints it, shows how:
\printedreport{clock-visible-fixed}
Bob answers \hs|Minus| exactly in context \hs|(A1, B1)|, so his frequency of \hs|Plus| at \hs|B1| follows Alice's setting: \val{clock-visible-fixed/run/BobB1/1} next to \hs|A0|, \val{clock-visible-fixed/run/BobB1/2} next to \hs|A1|.

This coordination involves no communication during a trial. It comes from a correlation between the schedule of the experiment and the strategy of the boxes: boxes that can read the experiment's clock can anticipate its settings. The boxes do not signal to each other: the schedule signals to Bob, through the step.

The random schedule confirms this explanation. Bob still answers \hs|Minus| on every fourth step, but these steps now fall on all four contexts alike, so in each context the two outcomes agree on three trials in four. Every correlator is then $3/4 - 1/4 = 1/2$, and the CHSH value collapses to $1/2 + 1/2 + 1/2 - 1/2 = 1$ (\val{clock-visible-random/chsh} in our run); the table shows no signaling. Hiding the step from the boxes restores purely local behavior: the classical value $2$, without signaling, under either schedule.

The lesson is not that stronger effects yield larger CHSH values. It is that violations arise when the strategy of the boxes becomes correlated with the experimental protocol. A fixed, predictable schedule invites such a correlation; a random one destroys it. A model of a Bell experiment must therefore control not only the effects inside the boxes, but also how information about the experiment as a whole reaches them. The same boxes also show that no-signaling is a property of a run, not of the boxes alone: they signal under one schedule and not under the other. A check that ran them apart from the run's schedule could not tell the two runs apart (Section~\ref{sec:checked}).

%% =====================================================================
\FloatBarrier
\section{The \texorpdfstring{$\lambda$}{lambda}-model in code: Blind, local, factored}\label{sec:lambda}

The clocked boxes reached the algebraic maximum of $4$ without any communication during a trial. They did not refute Bell's theorem. They exposed a weakness of the experimental protocol: a schedule that the boxes could follow.

\begin{figure}[htbp]
\centering
\begin{adjustbox}{max width=\linewidth}\begin{tikzpicture}[bell]
  \twoboxes
  \node[lam] (l) at (2.7,0) {$\lambda$};
  \draw[->] (l) -- (f); \draw[->] (l) -- (g);
\end{tikzpicture}\end{adjustbox}
\caption{The $\lambda$-model: two boxes $f$ and $g$ with a hidden variable $\lambda$.}
\label{fig:lambda}
\end{figure}
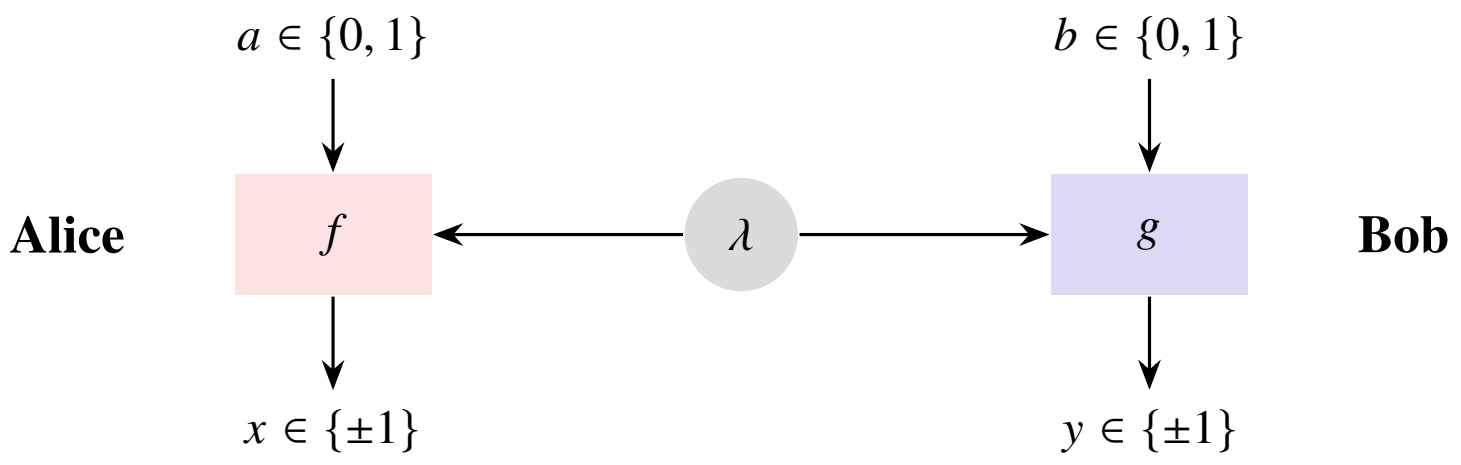

Bell's own analysis avoids such pitfalls by giving all variation from trial to trial a name: a hidden variable, traditionally written $\lambda$~\citep{Bell1964}. It stands for whatever the boxes may share before a trial, such as an agreed strategy or shared randomness, and it is assumed to be independent of the settings chosen in that trial. Figure~\ref{fig:lambda} shows this $\lambda$-model.

In our framework, $\lambda$ is not a new concept but a name for what state, clocks and other effects have been doing all along: varying the behavior of the boxes across trials. Bell's theorem rests on three assumptions about how they may do so. We state each as a property of a model, named by what it asks of the program, with the name of the assumption in the literature on Bell's theorem in parentheses (Section~\ref{sec:related}):
\begin{itemize}
\item \textbf{Blind} ($\lambda$-independence). The strategy is independent of the settings, $\pr{a, b}{\lambda} = \fr{\lambda}$: the code that draws $\lambda$ does not see them.
\item \textbf{Local} (parameter independence). Each box answers from its own setting and $\lambda$ alone, whatever the other setting: $\pr{a, b, \lambda}{x} = \pr{a, \lambda}{x}$ and $\pr{a, b, \lambda}{y} = \pr{b, \lambda}{y}$.
\item \textbf{Factored} (outcome independence). Given the settings and $\lambda$, the two outcomes are independent: $\pr{a, b, \lambda}{x, y} = \pr{a, b, \lambda}{x} \cdot \pr{a, b, \lambda}{y}$.
\end{itemize}
Together they bound the CHSH value by $2$~\citep{Bell1964,Brandenburger2008}. For a fixed $\lambda$, local and factored boxes answer, each from its own setting, independently of each other, so the regrouping of Section~\ref{sec:intro} applies to their average outcomes, which lie between $-1$ and $+1$, and gives the same bound. Blindness lets us average the bound over $\lambda$, because $\lambda$ has the same distribution in every context.

Each form of non-classical behavior in this paper comes from giving up one of the three properties. The loopholes of Sections~\ref{sec:clocked} and~\ref{sec:staging} and the superdeterminism of Section~\ref{sec:superdet} are not blind. The quantum boxes of Section~\ref{sec:quantum} and the PR boxes of Section~\ref{sec:pr} are not factored. The shared location of Section~\ref{sec:boxes} and the hidden states of Section~\ref{sec:hidden} are not local. This section first makes the three properties explicit in code: local and factored in the boxes (Section~\ref{sec:lhv}), blind in the sampler that draws $\lambda$ (Section~\ref{sec:staging}), whose type alone, we will see, does not make it so. It then gives up blindness outright (Section~\ref{sec:superdet}).

%% ---------------------------------------------------------------------
\subsection{Local hidden variables as immutable data}\label{sec:lhv}

In the CHSH scenario, a deterministic local strategy says in advance which outcome each box gives at each of its settings. It consists of four outcomes:
\code{18-23}{CHSH/LHV.hs}
A value of type \hs|Lambda| describes the complete behavior of both boxes in one trial. For example,
\code{36-37}{CHSH/LHV.hs}
\noindent always produces equal outcomes and yields a CHSH value of $+2$, while
\code{40-41}{CHSH/LHV.hs}
\noindent always produces opposite outcomes and yields a CHSH value of $-2$. There are sixteen such strategies, and the list \hs|allLambdas| enumerates them:
\code{44-52}{CHSH/LHV.hs}[lst:alllambdas]
The last generator varies fastest, and each runs from \hs|Minus| to \hs|Plus|, the reverse of the order of \hs|every| (\ref{lst:every}). So the entries \hs|a0|, \hs|a1|, \hs|b0|, \hs|b1| of the strategy at index $j$ are the binary digits of $j$, most significant first, with \hs|Minus| for $0$ and \hs|Plus| for $1$: \hs|b1| is \hs|Plus| when $j$ is odd, and \hs|b0| is \hs|Plus| when $j \bmod 4$ is $2$ or $3$. This order will matter for a sampler that goes through the list with the step (Section~\ref{sec:staging}).

The boxes read their strategy from a \hs|Reader| monad (\hs|LHV| stands for \emph{local hidden variables}); \hs|asks a0| reads the field \hs|a0| of $\lambda$:
\code{25-25,28-32}{CHSH/LHV.hs}
Note that $\lambda$ is a \emph{deterministic} strategy; randomness enters only through how $\lambda$ is sampled across trials.

\paragraph{Local and factored.} These boxes have both properties by construction. They are local: \hs|localA| receives an \hs|ASetting| and has access to $\lambda$, but none to a \hs|BSetting|, and likewise \hs|localB|; and a reader monad lets both boxes read $\lambda$, but lets neither write anything that the other could read. They are factored: \hs|jointAB| keeps its default (\ref{lst:trialmodel}), which runs the two reads and pairs their results. Locality needs both, the types and the effect. The shared location of Section~\ref{sec:boxes} has methods of the same types and is not local: \hs|State| lets Alice write her setting where Bob reads it.

%% ---------------------------------------------------------------------
\subsection{Blind sampling}\label{sec:staging}

It remains to say how $\lambda$ varies across trials, and what keeps it blind to the settings.

\paragraph{Sampling $\lambda$.} In our model, $\lambda$ is supplied by a \emph{sampler}, an instruction of the experiment layer, like the scheduler:
\code{65-65}{CHSH/LHV.hs}
A sampler may read the step, so the hidden variable can change from trial to trial, but it is not handed the settings. Here are two. \hs|pure| makes an instruction that ignores the step, and \hs|ReaderT|, as before, makes one from a function of the step:
\code{67-69}{CHSH/LHV.hs}
\noindent where \hs|uniformCycle xs i| picks \hs|xs !! (i `mod` length xs)|. So \hs|plus2| always draws the strategy \hs|lambdaAllPlus|, and \hs|uniformAll| goes through all sixteen strategies in turn. The accompanying code has a few more: \hs|minus2| always draws \hs|lambdaAntiAll|, and \hs|altExtrema| alternates between the two from step to step. The membrane wires the sampler into the trial; \hs|runReader r| runs the boxes' reader computation with the strategy that \hs|samp| draws:
\code{124-125}{CHSH/LHV.hs}[lst:runlhv]
The hidden variable is thus drawn outside the trial, by an instruction of the experiment layer, and supplied to the boxes as an immutable environment. Whatever happens inside the trial can observe the $\lambda$ that was chosen, and cannot influence how it was chosen.

Note that the membrane draws $\lambda$ \emph{after} the trial's settings (\ref{lst:trial}). What keeps $\lambda$ independent of the settings is not the order of the two draws but the flow of data: the membrane ignores the settings, so their values cannot flow into $\lambda$. In terms of the experiment monad, the settings and the strategy of one step are composed applicatively:
\code{17-18}{gen/Examples.hs}[lst:independent]
\hs|liftA2| runs \hs|sched| and then \hs|samp|, but \hs|samp| cannot see the value that \hs|sched| returns. (The accompanying code checks that feeding this pair to the boxes gives the record of \hs|trial| with the membrane \hs|blind (runTrialLHV samp)|.) We can now run several instructive samplers.

\paragraph{Extremal strategies.} Fixing $\lambda$ to one of the extremal strategies, with \hs|plus2| or \hs|minus2|, yields the classical bounds exactly: CHSH $\val{lhv-plus2/chsh}$ and $\val{lhv-minus2/chsh}$, with no signaling. No model that is blind, local and factored can exceed these values.

\paragraph{Uniform strategies.} If we cycle through all sixteen strategies with \hs|uniformAll|, positive and negative correlations cancel each other. Under the random schedule, the run's CHSH value is close to $0$ (\val{lhv-uniform-random/chsh}), and the run does not signal. Its table is the uniform empirical model, up to sampling fluctuations: in each context, every pair of outcomes occurs with frequency about $1/4$.
\begin{center}
\runtable{lhv-uniform-random}
\end{center}
Under the fixed schedule the CHSH value is exactly $0$, yet the run signals, and its table is far from uniform:
\begin{center}
\runtable{lhv-uniform-fixed}
\end{center}
The sampler cycles with the step, and so does the scheduler, so $i \bmod 4$ fixes the context and Bob's entries of $\lambda$ at once. In the order of \hs|allLambdas| (\ref{lst:alllambdas}), the steps with $i \bmod 4 = 0$ measure \hs|(A0, B0)| and draw a strategy whose \hs|b0| is \hs|Minus|, those with $i \bmod 4 = 2$ measure \hs|(A1, B0)| and draw one whose \hs|b0| is \hs|Plus|, and at the other two steps of every four, which measure \hs|(A0, B1)| and \hs|(A1, B1)|, \hs|b1| is \hs|Plus|. So in context \hs|(A0, B0)| Bob always answers \hs|Minus|, and in \hs|(A1, B0)| always \hs|Plus| (the marked frequencies). The sampler is never handed the settings, and yet, through the step, its $\lambda$ is not independent of them. The CHSH value does not show this dependence; the table does.

\paragraph{Exploiting the step.} A sampler that reads the step can do worse:
\code{85-91}{CHSH/LHV.hs}
\hs|biasPlus2Loophole 7 8| draws the all-plus strategy unless $i \bmod 8 = 7$, and the anti-correlated one when it is. Under the fixed schedule, a step with $i \bmod 8 = 7$ always measures \hs|(A1, B1)|, the context of the minus term in CHSH. The run reaches CHSH \val{lhv-loophole/chsh}, beyond the classical bound, with a sampler of type \hs|Sampler e| and local deterministic boxes; its table shows Bob at \hs|B1| signaling (\val{lhv-loophole/run/BobB1/1} next to \hs|A0|, \val{lhv-loophole/run/BobB1/2} next to \hs|A1|). Its variant \hs|biasPlus2Blocks 7 8| draws one strategy per block of four steps and so cannot single out a context: it gives $4 \cdot \frac78 - 2 = \val{lhv-blocks/chsh}$ and does not signal.

\paragraph{Why the type is not enough.} So the type of the sampler does not make a model blind. It keeps the settings themselves out of the sampler's hands, but not what determines them: the scheduler and the sampler read the same step, and through it a sampler can follow a scheduler that reads the step. In terms of probabilities, let $\fr{\lambda}_i$ be the probability that the sampler draws $\lambda$ at step $i$, and $\pr{a, b}{\lambda}_i$ that of drawing it there when the settings are $(a, b)$. The run's table pools the steps of each context: in context $(a, b)$ it sees $\lambda$ with probability
\begin{equation}\label{eq:pooled}
  \frac{\sum_i \fr{a, b}_i\, \pr{a, b}{\lambda}_i}{\sum_i \fr{a, b}_i}.
\end{equation}
A membrane that ignores the settings gives $\pr{a, b}{\lambda}_i = \fr{\lambda}_i$ at every step, as long as the lab's effect \hs|e| holds nothing that the scheduler wrote. The pooled distribution~\eqref{eq:pooled} is then the same in every context when $\fr{\lambda}_i$ does not change with $i$, or when $\fr{a, b}_i$ does not, as under the random schedule. The cycling sampler under the fixed schedule satisfies neither condition. A sampler that cannot read the step satisfies the first, provided that its action draws in the same way at every call. \hs|fresh| makes such a sampler from an action of the lab alone, and \hs|uniformIO| is an action that draws one of the sixteen strategies from coins:
\code{110-111,115-116}{CHSH/LHV.hs}
Under the fixed schedule, \hs|fresh uniformIO| gives CHSH \val{lhv-fresh/chsh} and does not signal, where the cycling \hs|uniformAll| signaled.

All of these samplers followed the settings at most indirectly, through the step. Section~\ref{sec:superdet} hands the settings to the sampler outright.

%% ---------------------------------------------------------------------
\subsection{Superdeterminism: A sampler that sees the settings}\label{sec:superdet}

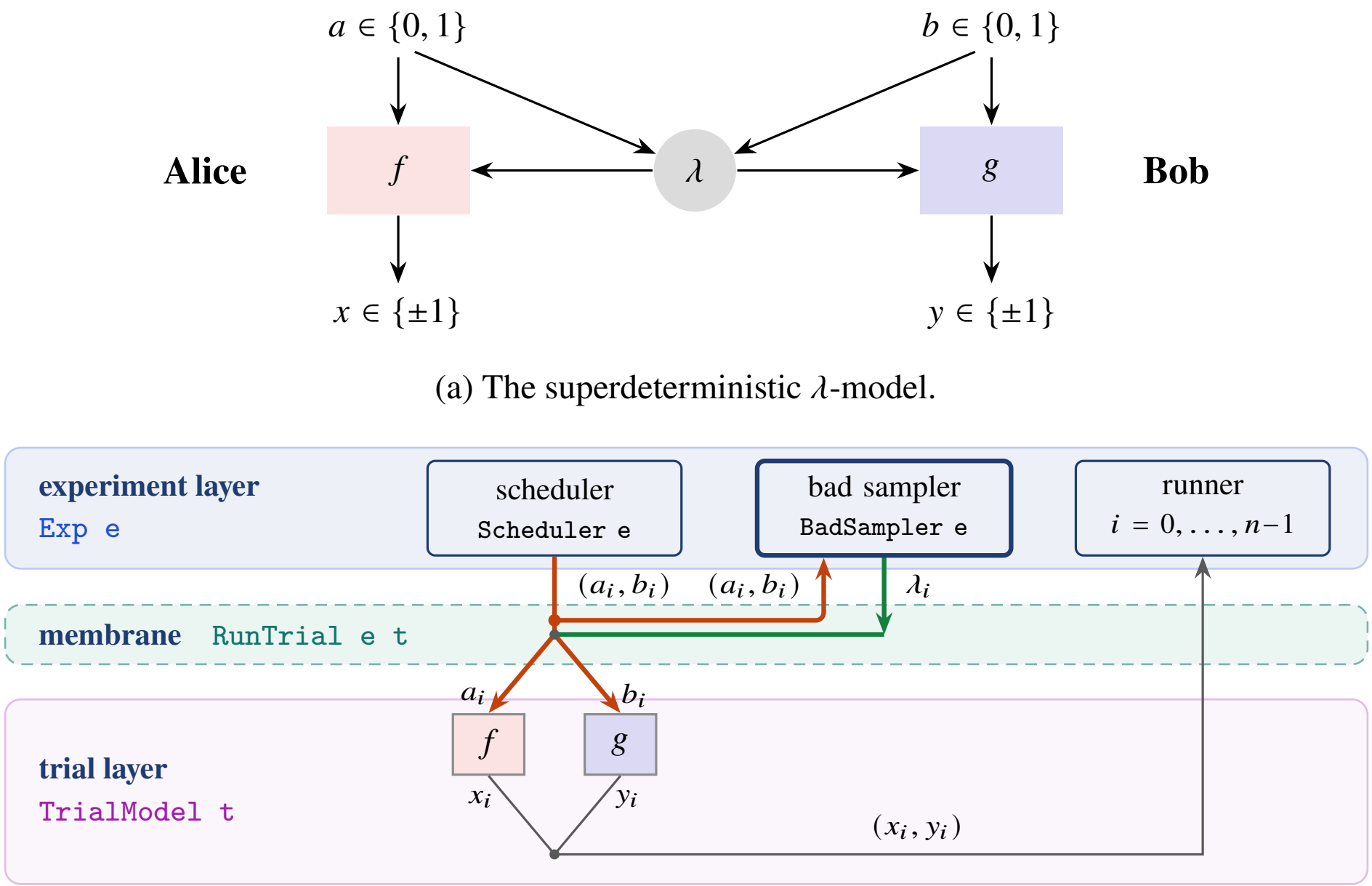
\begin{figure}[htbp]
\centering
\begin{subfigure}{\linewidth}
\centering
\begin{adjustbox}{max width=\linewidth}\begin{tikzpicture}[bell]
  \twoboxes
  \node[lam] (l) at (2.7,0) {$\lambda$};
  \draw[->] (l) -- (f); \draw[->] (l) -- (g);
  \draw[->] ($(a.south)+(0.15,0.05)$) -- (l);
  \draw[->] ($(b.south)+(-0.15,0.05)$) -- (l);
\end{tikzpicture}\end{adjustbox}
\caption{The superdeterministic $\lambda$-model.}
\end{subfigure}

\vspace{8pt}
\begin{subfigure}{\linewidth}
\centering
\begin{adjustbox}{max width=\linewidth}\begin{tikzpicture}[bell, setpath/.style={draw=CSet, line width=1.2pt},
    lampath/.style={draw=CLam, line width=1.2pt}]
  \layers
  \node[lnode, draw=Navy, line width=1.2pt] (bad) at (samp) {bad sampler\\[-1pt] {\scriptsize\ttfamily BadSampler e}};
  % the settings reach the membrane on their way down from the scheduler,
  % and the membrane passes them along to the bad sampler (runTrialSuperdet)
  \coordinate (up) at (7.45,0);
  \draw[->, draw=CSet, line width=1.2pt, rounded corners=2pt]
    (5.0,-1.02) -- (7.45,-1.02) -- (up |- bad.south);
  \fill[CSet] (5.0,-1.02) circle (1.6pt);
  \node[lbl, anchor=east] at (7.38,-0.71) {$(a_i, b_i)$};
\end{tikzpicture}\end{adjustbox}
\caption{In the architecture: the trial hands the settings of the step to the membrane (\ref{lst:trial}), and the membrane passes them on to a bad sampler (\ref{lst:superdet}), whose strategy $\lambda_i$ comes back down to the boxes.}
\end{subfigure}
\caption{Superdeterminism.}
\label{fig:superdet}
\end{figure}

The samplers of Section~\ref{sec:staging} are not handed the settings. Superdeterministic models give up exactly this: informally, superdeterminism~\citep{HossenfelderPalmer2020,Palmer2009,tHooft2016} allows the hidden variable $\lambda$ to depend on the measurement settings themselves, which some presentations describe as a failure of ``free choice''~\citep{ConwayKochen2006,ConwayKochen2009}. In our framework, it means letting the code that draws $\lambda$ see the settings $(a, b)$ (Figure~\ref{fig:superdet}). Nothing non-classical needs to happen inside the boxes: they may remain local, factored and deterministic.

\paragraph{A sampler that sees the settings.} Such a sampler maps the settings to an instruction of the experiment layer. We call it a \emph{bad sampler}, and a sampler of type \hs|Sampler e| a \emph{legal} one:
\code{24-24}{CHSH/Superdet.hs}
A legal sampler is a bad one that ignores its argument, \hs|const samp|. The trial already hands the settings it drew to the membrane (\ref{lst:trial}); the superdeterministic membrane passes them on to the bad sampler, and is otherwise that of local hidden variables:
\code{33-34}{CHSH/Superdet.hs}[lst:superdet]
A superdeterministic run thus goes through the same \hs|trial| as every other model, with the membrane \hs|runTrialSuperdet bad| where the others use \hs|blind runTrial|. In terms of the experiment monad, the applicative pair of \ref{lst:independent} becomes a bind, which passes the settings that \hs|sched| returns on to \hs|bad|:
\code{21-22}{gen/Examples.hs}[lst:dependent]
With \hs|bad = const samp| it is the draw of \ref{lst:independent} again. (The accompanying code checks that this pair, fed to the boxes, gives the record of \hs|trial sched (runTrialSuperdet bad)|.)

\paragraph{A maximally cheating sampler.} With the settings in hand, it is easy to make every CHSH term contribute $+1$. In context \hs|(A1, B1)| the sampler draws a strategy on which Alice and Bob disagree, so that the minus sign turns the term positive; in every other context, one on which they agree:
\code{60-64}{CHSH/Superdet.hs}
Its runs reach the algebraic maximum with local, factored, deterministic boxes: CHSH \val{superdet-cheat-fixed/chsh} under the fixed schedule, and \val{superdet-cheat-random/chsh} under the random one. Unlike for the loopholes of Sections~\ref{sec:clocked} and~\ref{sec:staging}, randomizing the settings does not help. This is the defining feature of superdeterminism, and in code the correlation is explicit: the sampler is handed the settings. A bad sampler that ignores them, \hs|legalAllPlus _ = plus2|, gives \val{superdet-legal/chsh}.

\paragraph{The table sees it, or not.} Both runs of \hs|cheatLam| signal: at \hs|B1|, Bob answers \hs|Plus| next to \hs|A0| and \hs|Minus| next to \hs|A1| (frequencies \val{superdet-cheat-random/run/BobB1/1} and \val{superdet-cheat-random/run/BobB1/2} in the random run). Local boxes cannot signal by themselves; it is the dependence of $\lambda$ on the settings that moves Bob's marginal, and the run's table records the move. The dependence need not move a marginal, though:
\code{75-79}{CHSH/Superdet.hs}
\hs|cheatNS| flips one fair coin with \hs|randomOutcome|\footnote{\hs|randomOutcome| draws an \hs|Int| between $0$ and $1$ with \hs|randomRIO|, and answers \hs|Minus| for $0$ and \hs|Plus| for $1$; \hs|flipOutcome| swaps the two outcomes.} and hands it to both boxes, turned over by \hs|flipOutcome| for Bob in context \hs|(A1, B1)|. Under the random schedule, its run reproduces the PR box of Section~\ref{sec:pr}, CHSH value $4$ included, and it does not signal:
\printedreport{superdet-cheatns}
No test on the table can detect this dependence: local deterministic boxes with a sampler that sees the settings are, as far as the record of a run can tell, a PR box.

\paragraph{A sampler that peeks.} The bind is open to a sampler of the legal type, too, because the sampler and the scheduler read the same step:
\code{86-87}{CHSH/Superdet.hs}
\noindent \hs|peek sched bad| runs the scheduler itself and hands the settings it gets to a bad sampler; nothing in the type of the sampler that it returns says so. Under \hs|fixedSchedule|, a function of the step, it finds the trial's own settings, and \hs|runTrialLHV (peek fixedSchedule cheatLam)| reproduces the run of \hs|cheatLam|: CHSH \val{peek-fixed/chsh}, with Bob at \hs|B1| signaling. Under \hs|randomScheduleIO| it flips coins of its own, and the cheat is gone: CHSH \val{peek-random/chsh}, with no signaling. The three types of sampler thus embed into one another, each reading more than the last:
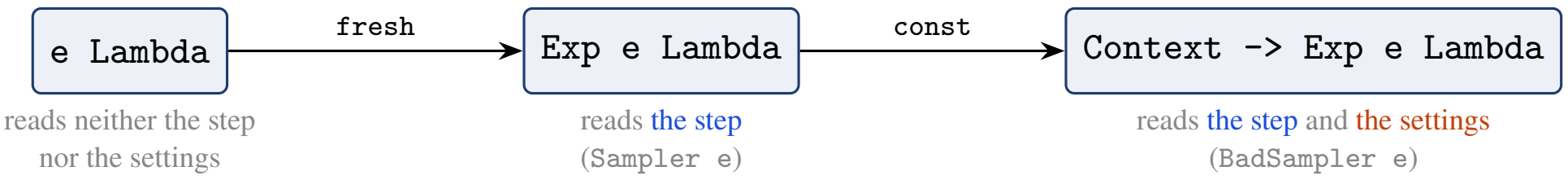
\begin{center}
\begin{adjustbox}{max width=\linewidth}\begin{tikzpicture}[bell]
  \tikzset{ty/.style={comp, font=\small, minimum height=7mm, inner sep=4pt}}
  \node[ty] (t0) at (0,0) {\texttt{e Lambda}};
  \node[ty] (t1) at (4.4,0) {\texttt{Exp e Lambda}};
  \node[ty] (t2) at (9.8,0) {\texttt{Context -> Exp e Lambda}};
  \draw[wire] (t0) -- node[above, lbl] {\texttt{fresh}} (t1);
  \draw[wire] (t1) -- node[above, lbl] {\texttt{const}} (t2);
  \node[lbl, text=TextGray, align=center] at (0,-0.75) {reads neither the step\\ nor the settings};
  \node[lbl, text=TextGray, align=center] at (4.4,-0.75) {reads \textcolor{CExp}{the step}\\ (\texttt{Sampler e})};
  \node[lbl, text=TextGray, align=center] at (9.8,-0.75) {reads \textcolor{CExp}{the step} and \textcolor{CSet}{the settings}\\ (\texttt{BadSampler e})};
\end{tikzpicture}\end{adjustbox}
\end{center}
Only the first rules out a dependence on the settings by itself, and only if the lab's effect \hs|e| carries nothing to it from the scheduler or from earlier steps; the second rules it out only together with a scheduler that does not read the step.

In executable terms, superdeterminism is the refactoring that turns the applicative pair of scheduler and sampler into a bind. It needs no signaling and no joint effects, which is why Bell's theorem must exclude it by assumption: without blindness, the CHSH inequality ceases to distinguish classical from non-classical correlations. In the next two sections the models stay blind and local, and give up only factoring.

%% =====================================================================
\FloatBarrier
\section{Quantum correlations in code}\label{sec:quantum}

So far, every model has left one method of the interface untouched. Every \hs|TrialModel| provides not only the local responses \hs|localA| and \hs|localB|, but also the joint response \hs|jointAB|, and we have relied on its default (\ref{lst:trialmodel}), which runs Alice's and Bob's local computations and pairs their results. With local hidden variables, that default is what made the boxes factored.

Quantum mechanics departs from classical behavior here. It does not let one box signal to the other, and it does not need a hidden variable that depends on the measurement settings. Each box, on its own, answers as its local response says; but once Alice's box has answered, Bob's box no longer answers as it would on its own. In our code, we express this departure by overriding \hs|jointAB|.

\paragraph{What this model is (and is not).} We write the quantum boxes as a quantum computing textbook writes them~\citep{NielsenChuang2010}: the state of two qubits is a vector of amplitudes, a measurement is a pair of projectors, and the Born rule gives the probability of each outcome and the state that the outcome leaves. Real amplitudes suffice for the CHSH experiment, so the linear algebra is that of lists of real numbers: vectors of four entries, and matrices of two by two or four by four. The model does not simulate a physical apparatus, and it is not a general simulator of quantum programs, as the Quantum IO monad of \citet{AltenkirchGreen2010} is: it computes only the statistics of measuring two qubits. Sections~\ref{sec:qubits} and~\ref{sec:pair} set up this linear algebra; a reader who knows it can go straight to Section~\ref{sec:override}.

%% ---------------------------------------------------------------------
\subsection{Qubits and measurements}\label{sec:qubits}

A qubit has two basis states, $|0\rangle$ and $|1\rangle$, and a pair of qubits has four, $|00\rangle$, $|01\rangle$, $|10\rangle$ and $|11\rangle$: the first digit for Alice's qubit, the second for Bob's. A state of the pair is a vector of four \emph{amplitudes}, one for each basis state; their squares are the probabilities of finding the pair in each basis state, and add up to $1$. An operator on the pair is a four-by-four matrix. Both are lists of numbers here.

We need three operations of linear algebra, and we write them as Dirac does: the inner product $\langle u|v\rangle$ of two vectors, a matrix applied to a vector, $M|v\rangle$, and the tensor (Kronecker) product $M \otimes N$, which makes an operator on the pair from one on each qubit. Haskell has no mixfix operators, but two infix ones come close: \hs!u <| v! is $\langle u|v\rangle$ and \hs!m |> v! is $M|v\rangle$, and since \hs!|>! binds tighter than \hs!<|!, the sandwich $\langle\psi|M|\psi\rangle$ reads \hs!psi <| m |> psi!. The tensor product, spelled \texttt{>{}<} in the source, is printed $\otimes$ here:
\code{23-24,31-33,35-36,38-39,41-42}{CHSH/Quantum.hs}
The module also defines \hs|add|, \hs|sub| and \hs|scale|, the entrywise sum and difference of two matrices and a multiple of one, and \hs|normalize|, which divides a vector $v$ by its length $\sqrt{\langle v|v\rangle}$.

Each setting measures an \emph{observable}: a two-by-two matrix whose eigenvalues, $+1$ and $-1$, are the outcomes \hs|Plus| and \hs|Minus|. Alice measures one of the Pauli matrices $Z$ and $X$, and Bob one of their normalized sum and difference:
\[
  Z = \begin{pmatrix} 1 & 0 \\ 0 & -1 \end{pmatrix}, \qquad
  X = \begin{pmatrix} 0 & 1 \\ 1 & 0 \end{pmatrix}, \qquad
  \frac{Z + X}{\sqrt2}, \qquad \frac{Z - X}{\sqrt2}.
\]
These give the largest CHSH value for the maximally entangled state below; the textbook's account of the CHSH inequality measures the same observables, up to sign, on another entangled state~\citep[Section~2.6]{NielsenChuang2010}:
\code{60-63,65-67,69-71}{CHSH/Quantum.hs}
Each outcome of an observable $O$ has a \emph{projector}, the matrix that keeps the part of a state in which $O$ has that outcome: $P_{+}$ for \hs|Plus| and $P_{-}$ for \hs|Minus|. On the pair, Alice's projectors act on her qubit and Bob's on his:
\[
  P_{\pm} = \frac{I \pm O}{2}, \qquad
  P \otimes I \ \text{for Alice}, \qquad I \otimes P \ \text{for Bob}.
\]
\hs|measA| and \hs|measB| build these projectors on the pair:
\code{75-77,81-82,84-85}{CHSH/Quantum.hs}
The Born rule gives the probability of the outcome with projector $P$ in the state $|\psi\rangle$, and, when that probability is positive, the state that the outcome leaves:
\[
  \langle\psi|P|\psi\rangle, \qquad
  \frac{P|\psi\rangle}{\sqrt{\langle\psi|P|\psi\rangle}}.
\]
In code, the projector \hs|p| plays $P$: \hs|probability p psi| is \hs!psi <| p |> psi!, and \hs|collapse| divides $P|\psi\rangle$ by its length, which is the square root of that probability. \hs|coin p| answers \hs|Plus| with probability \hs|p|:
\code{89-90,92-93,96-99}{CHSH/Quantum.hs}

%% ---------------------------------------------------------------------
\subsection{An entangled pair}\label{sec:pair}

A source prepares the pair in a state. The maximally entangled state $|\Phi^+\rangle$ gives equal amplitudes to $|00\rangle$ and $|11\rangle$, and none to the other two: the qubits are both $0$ or both $1$. A parameter $\theta$ weighs the two terms, and the states $|\psi_\theta\rangle$ include $|\Phi^+\rangle$ at $\theta = \pi/4$:
\[
  |\Phi^+\rangle = \frac{|00\rangle + |11\rangle}{\sqrt2} = \frac{1}{\sqrt2}\,[1, 0, 0, 1],
  \qquad
  |\psi_\theta\rangle = \cos\theta\,|00\rangle + \sin\theta\,|11\rangle.
\]
At $\theta = 0$ the pair is in $|00\rangle$, which is not entangled at all:
\code{104-104,107-108,112-113}{CHSH/Quantum.hs}
Either qubit of $|\Phi^+\rangle$, on its own, gives each outcome with probability $1/2$, whatever is measured. When Alice measures her qubit, the state collapses, and Bob's qubit is left in the state in which her observable is certain to give her outcome. Measuring $Z$ leaves $|00\rangle$ for \hs|Plus| and $|11\rangle$ for \hs|Minus|; measuring $X$ leaves $|{+}{+}\rangle$ for \hs|Plus| and $|{-}{-}\rangle$ for \hs|Minus|. Here $|{\pm}\rangle = (|0\rangle \pm |1\rangle)/\sqrt2$ are the states of one qubit in which $X$ is certain to give \hs|Plus| or \hs|Minus|, and $|00\rangle + |11\rangle$ is also $|{+}{+}\rangle + |{-}{-}\rangle$.

The correlator of a context, the average of $xy$, is
\[
  E(a, b) = \langle\psi|A \otimes B|\psi\rangle
\]
for Alice's observable $A$ and Bob's observable $B$, and \hs|chshOf (correlation psi)| is the exact CHSH value of a state:
\code{117-118}{CHSH/Quantum.hs}
For $|\Phi^+\rangle$, the averages $\langle\psi|Z \otimes Z|\psi\rangle$ and $\langle\psi|X \otimes X|\psi\rangle$ are $1$, and those of $Z \otimes X$ and $X \otimes Z$ are $0$. So the correlator is $1/\sqrt2$ in the three contexts other than \hs|(A1, B1)|, and $-1/\sqrt2$ in that one. The CHSH value is $4/\sqrt2 = 2\sqrt2$, the Tsirelson bound~\citep{Tsirelson1980}. For $|\psi_\theta\rangle$, the average of $X \otimes X$ is $\sin 2\theta$ and the others stay as they are, so
\[
  \CHSH(\theta) = \sqrt2\,(1 + \sin 2\theta).
\]
The parameter moves the CHSH value from $\sqrt2 \approx \val{theta/zero}$ at $\theta = 0$ up to the bound $2\sqrt2$ at $\theta = \pi/4$. On the way it exceeds the classical bound $2$ once $\sin 2\theta > \sqrt2 - 1$, that is, from $\theta \approx \val{theta/classical}$ on, about $\val{theta/classicalDeg}^\circ$. These observables are the best ones only at $\theta = \pi/4$; for any other $\theta$, other observables give a larger value.

%% ---------------------------------------------------------------------
\subsection{Overriding \texttt{jointAB}}\label{sec:override}

The trial monad holds the state of the pair, and flips coins in \hs|IO|. The local responses read the state with \hs|get|, measure each qubit on its own, and leave the state as it is. The joint response, however, pairs two computations that also write to the state:
\code{124-125,127-127,130-135,138-138}{CHSH/Quantum.hs}[lst:quantumjoint]
Alice's, \hs|nonlocalA|, answers as her local response does, and then collapses the state to what her outcome leaves. Bob's, \hs|nonlocalB|, is its dual:
\code{144-148,150-154}{CHSH/Quantum.hs}[lst:quantum]
The default pairs the local responses with \hs|liftA2 (,)|, and the override pairs these two computations in the same way. In a monad, \hs|liftA2| runs the first computation and then the second, so Bob's local response reads the state that Alice's outcome left, and his outcome depends on hers. The pair stays applicative: \hs|nonlocalB| never sees the value of her outcome, and the dependence travels through the state. In this order nothing reads the state after Bob's write, but either box could go first, as we see below.

The membrane supplies the state, as the membrane of local hidden variables supplies a strategy: \hs|evalStateT m psi| runs a trial from the state \hs|psi|, and drops the state in which the trial ends. The membrane ignores the step, so no information about the schedule can reach the boxes:
\code{160-161,163-164}{CHSH/Quantum.hs}

\paragraph{Local, but not factored.} One box writes and the other reads, as in the shared location of Section~\ref{sec:boxes}, and that model signals. What differs is what is written. Bob's outcome depends on Alice's \emph{outcome}, so the boxes are not factored. Averaged over her outcome, however, it does not depend on her \emph{setting}.\footnote{The names \hs|nonlocalA| and \hs|nonlocalB| say that these computations act on the state of the whole pair. The literature likewise calls correlations beyond the classical bound \emph{nonlocal}. That sense is wider than our property \emph{local}, which is parameter independence: the quantum boxes are nonlocal in the wide sense, and local in ours.}

To see this, write $P_x$ for Alice's projector of outcome $x$ on the pair, and $Q_y$ for Bob's of outcome $y$. Her outcome $x$ has probability $\langle\psi|P_x|\psi\rangle$. If that is positive, the outcome leaves $P_x|\psi\rangle$ normalized, in which Bob's outcome $y$ has probability $\langle\psi|P_x Q_y P_x|\psi\rangle / \langle\psi|P_x|\psi\rangle$. The projectors are symmetric matrices; Alice's and Bob's act on different qubits, so $P_x Q_y = Q_y P_x$; and $P_x P_x = P_x$. The joint probability is therefore
\[
  \pr{a, b}{x, y} = \langle\psi|P_x Q_y|\psi\rangle.
\]
If $\langle\psi|P_x|\psi\rangle = 0$, Alice never answers $x$, and the formula still holds. Since $\langle\psi|P_x|\psi\rangle = \langle\psi|P_x P_x|\psi\rangle$ is the squared length of $P_x|\psi\rangle$, that vector is $0$, and both sides are $0$: the left because $x$ never occurs, the right because it is the inner product of $P_x|\psi\rangle$ with $Q_y|\psi\rangle$. The formula is the same whichever box measures first: the two computations are dual in their effect as well as in their code. Each party's projectors add up to the identity, so summing over one party's outcome leaves the other's local response:
\[
  \sum_x \langle\psi|P_x Q_y|\psi\rangle = \langle\psi|Q_y|\psi\rangle, \qquad
  \sum_y \langle\psi|P_x Q_y|\psi\rangle = \langle\psi|P_x|\psi\rangle.
\]
So the marginals of the joint response are the local responses: \hs|fst <$> jointAB a b| has the distribution of \hs|localA a|, and \hs|snd <$> jointAB a b| that of \hs|localB b|. These two equations say that neither party's outcome depends on the other's setting: the boxes are local. (They are blind, since the state is the same at every trial.)

All this holds for every state. A state monad could carry a signal; that this write carries none is a fact about projectors, not about types. The test suite checks these identities for several states and every context, and runs the pair with Bob's box first. A computation in \hs|IO| does not expose its distribution, though, so we also check no-signaling on the outcomes of runs.

\paragraph{Where the collapse goes.} The argument uses what Alice's box writes, not where the code writes it. The membrane drops the state in which a trial ends, so a write that nothing reads afterwards changes nothing: run on its own, \hs|nonlocalA a| flips the same coin as \hs|localA a|, and gives the same outcome. We could therefore have put the collapse in the local responses, and kept the default:
\code{29-30,32-34}{CHSH/Staging.hs}
In a monad, \hs|liftA2| runs its two computations in turn, in the default as in an override, so the default \hs|jointAB| of \hs|Collapsing| is the override above. Its runs are those of Section~\ref{sec:quantum-ns}, draw for draw, as the program that writes our data checks. Conversely, the boxes of Section~\ref{sec:hidden} write in their local responses and keep the default, and local responses that only read, with an override that pairs the computations that write, give the same runs. So for these boxes, where a write sits, and whether \hs|jointAB| is overridden, is a choice of staging, and decides neither property. We keep the local responses of the quantum boxes free of writes, so that each is what one box does on its own, whatever runs after it, and the interaction of the two boxes stays in one place, the override.

What decides is what is written. Alice's box writes a collapse by an outcome drawn during the trial, so that for a given state there is an outcome to sum over; and the collapse acts through projectors that commute with Bob's and add up to the identity, so that the sum erases her setting. The boxes of Section~\ref{sec:hidden} lack the first: given $\lambda$, nothing in their trial is random, and what Alice's box writes into Bob's entries depends on her setting.

%% ---------------------------------------------------------------------
\subsection{CHSH behavior and no-signaling}\label{sec:quantum-ns}

We run the maximally entangled pair under each schedule. Both runs come close to $2\sqrt2 \approx \val{tsirelson}$: \val{quantum-fixed/chsh} under the fixed schedule and \val{quantum-random/chsh} under the random one. The membrane ignores the step, so the choice of schedule has no effect. Neither run signals, and the table of the first is the empirical model of the quantum boxes:
\begin{center}
\runtable{quantum-fixed}
\end{center}
Every marginal is close to $1/2$, as it must be for $|\Phi^+\rangle$, so here the no-signaling check compares fair coins with fair coins. A partially entangled pair puts it to a stronger test. At $\theta = \pi/8$, each qubit on its own is more often $0$ than $1$, and its marginal depends on the observable measured:
\begin{center}
\runtable{quantum-partial}
\end{center}
Alice answers \hs|Plus| with frequency \val{quantum-partial/run/AliceA0/1} at \hs|A0| and \val{quantum-partial/run/AliceA1/1} at \hs|A1|, where the Born rule predicts $\cos^2\theta = \val{partial/AliceA0}$ and $\val{partial/AliceA1}$, and neither frequency moves with Bob's setting. Bob answers \hs|Plus| with a frequency close to \val{partial/Bob} at both of his settings, whatever Alice's. The run does not signal. Its CHSH value is \val{quantum-partial/chsh}, close to $\CHSH(\pi/8) = 1 + \sqrt2 \approx \val{partial/chsh}$: a less entangled pair gives weaker correlations. The next section goes the other way, and replaces the quantum correlations with the strongest that no-signaling allows.

%% =====================================================================
\FloatBarrier
\section{PR boxes: The algebraic maximum without signaling}\label{sec:pr}

The quantum boxes of Section~\ref{sec:quantum} exceed the classical bound without signaling and without a hidden variable: they give up factoring, and stay blind and local. PR boxes~\citep{PopescuRohrlich1994} take this departure to its extreme. A PR box satisfies no-signaling exactly, and yet reaches the algebraic maximum $4$ of the CHSH expression. Its correlations are stronger than any that quantum mechanics allows, although they permit no faster-than-light communication between the parties; for this reason PR boxes are often described as ``super-quantum'' or ``post-quantum''. In our code they are written as the quantum boxes are, by overriding \hs|jointAB|, and for them the override is needed: their trial monad has no state through which one box could depend on the other.

%% ---------------------------------------------------------------------
\subsection{The PR constraint}

The defining property of a PR box is easiest to state in bits. Let the settings be bits $a, b \in \{0, 1\}$, and let the outcomes be bits $x, y \in \{0, 1\}$. The PR constraint is
\[
  x \oplus y = a \land b,
\]
together with the requirement that $x$ and $y$ are each uniform. Translated back into outcomes $\pm1$, the constraint makes the outcomes agree in three contexts and disagree in the fourth, so all four CHSH terms contribute $+1$, for a total of $4$. The two outcomes are perfectly correlated or anti-correlated, depending on the context, and yet neither party's marginal distribution depends on the other party's setting. The model is therefore exactly no-signaling.

%% ---------------------------------------------------------------------
\subsection{A PR trial monad}

Like the quantum boxes, PR boxes are most naturally written with genuine randomness. Their trial monad wraps \hs|IO|:
\code{21-22}{CHSH/PR.hs}
PR trials, like quantum trials, have no memory and do not depend on the step, which the membrane discards:
\code{54-55}{CHSH/PR.hs}
The \hs|TrialModel| instance says what the definition says. Each box, on its own, is a fair coin at either setting. The joint response runs the two boxes as the default does, and keeps only the pairs that meet the constraint: \hs|m `suchThat` p| runs \hs|m| again until its result satisfies \hs|p|:
\code{24-24,27-28,35-35,41-42,45-48}{CHSH/PR.hs}[lst:pr]
The outcomes must differ exactly in context \hs|(A1, B1)|, where $a \land b = 1$: that is the constraint $x \oplus y = a \land b$. The two boxes are dual, and neither sees the other's setting or outcome. All the correlation is in the constraint, which only the joint response can state, since only it has both settings.

\paragraph{Local, and as far from factored as can be.} In each context two pairs of outcomes meet the constraint, equally likely since the coins are fair, and in those two each party answers \hs|Plus| once. So each outcome is still a fair coin, at either of the other party's settings: the marginals of the joint response are the local responses, and the boxes are local. Given the settings, each outcome determines the other, so the boxes are as far from factored as they can be.

Keeping only some pairs is post-selection, which in a real Bell test is the detection loophole~\citep{Brunner2014}. A PR box is hypothetical, and here the selection is its definition; the retries stay inside one trial, so the record of the run has no gaps.

\paragraph{The runs.} Under each schedule, the run gives CHSH \val{pr-fixed/chsh}. The schedule cannot matter, because the correlations are defined context by context. And the reports confirm that the boxes do not signal: in both runs, every marginal is close to $1/2$, whatever the other party's setting. Figure~\ref{fig:pr} tabulates the empirical model of the PR box; the table of each run is that model up to sampling fluctuations, and so was the table that \hs|cheatNS| produced in Section~\ref{sec:superdet}.

\begin{figure}[htbp]
\centering
\idealtable{$1/2$ & $0$ & $0$ & $1/2$}{$1/2$ & $0$ & $0$ & $1/2$}{$1/2$ & $0$ & $0$ & $1/2$}{$0$ & $1/2$ & $1/2$ & $0$}
\caption{The empirical model of the PR box.}
\label{fig:pr}
\end{figure}

\paragraph{Why fair coins?} Unlike the quantum boxes, the PR box cannot have local responses that depend on the setting. The uniform marginals in its definition are no choice: any box that reaches CHSH $4$ without signaling has fair coins for parts. Write $p_a$ for the probability that Alice answers \hs|Plus| at setting $a$, and $q_b$ for Bob at $b$; without signaling, neither depends on the other party's setting. CHSH $4$ needs agreement in the contexts \hs|(A0, B0)|, \hs|(A0, B1)| and \hs|(A1, B0)|, and disagreement in \hs|(A1, B1)|. The three agreements give $p_0 = q_0$, $p_0 = q_1$ and $p_1 = q_0$, and the disagreement gives $p_1 = 1 - q_1$. Hence $p_1 = 1 - p_1$, and all four marginals are $1/2$.

The constraint does not supply this by itself: on coins biased to $3/4$, the same \hs|jointAB| still reaches CHSH $4$, but Alice's probability of \hs|Plus| at \hs|A1| is \val{pr/biasedB0} next to \hs|B0| and \val{pr/biasedB1} next to \hs|B1|. Local responses that read, and rewrite, a hidden variable do appear in Section~\ref{sec:prstate}, at the price of locality.

%% ---------------------------------------------------------------------
\subsection{Quantum versus PR: Global assignments}\label{sec:contextuality}

The quantum boxes and the PR box give up the same property in the same method. What, then, separates their tables, apart from the CHSH value? A value of type \hs|Lambda| fixes an outcome for each of the four settings at once: it is a \emph{global assignment}. The statistics of a run come from blind, local, factored boxes exactly when they come from a probability distribution over such values~\citep{Fine1982,AbramskyBrandenburger2011}. Neither the quantum boxes nor the PR box has such a distribution, since both exceed the classical bound. They differ in how far they are from having one.

With the maximally entangled pair, every one of the sixteen joint outcomes of the quantum boxes is possible (the rarest have frequency about $0.07$ in the first table of Section~\ref{sec:quantum-ns}), so each possible outcome lies on some global assignment; only the probabilities cannot be spread over the assignments. The PR box admits not even that: the outcomes it allows (Figure~\ref{fig:pr}) fit no global assignment at all.

\begin{figure}[htbp]
\centering
\begin{adjustbox}{max width=\linewidth}\begin{tikzpicture}[bell, scale=1.1,
    pt/.style={circle, fill=black, inner sep=0pt, minimum size=4.2pt},
    obs/.style={font=\footnotesize\ttfamily, text=HsType}]
  % the base: the four observables, and the four contexts as edges
  \coordinate (A0) at (0,0);
  \coordinate (B0) at (2.2,-0.6);
  \coordinate (A1) at (3.5,0.25);
  \coordinate (B1) at (1.3,0.85);
  \draw (A0) -- (B0) -- (A1) -- (B1) -- cycle;
  \foreach \o in {A0,B0,A1,B1} {
    \node[pt] at (\o) {};
    \coordinate (\o+) at ($(\o)+(0,1.7)$);
    \coordinate (\o-) at ($(\o)+(0,2.35)$);
    \draw[densely dotted] (\o) -- (\o-);
  }
  \node[obs, anchor=north east] at (A0) {A0};
  \node[obs, anchor=north] at ($(B0)+(0,-0.05)$) {B0};
  \node[obs, anchor=west] at ($(A1)+(0.08,0)$) {A1};
  \node[obs, anchor=east] at ($(B1)+(-0.08,0.05)$) {B1};
  % the supported joint outcomes of the PR box
  \draw (A0+) -- (B0+);  \draw (A0-) -- (B0-);
  \draw (A0+) -- (B1+);  \draw (A0-) -- (B1-);
  \draw (A1+) -- (B0+);  \draw (A1-) -- (B0-);
  \draw (A1+) -- (B1-);  \draw (A1-) -- (B1+);
  \foreach \o in {A0,B0,A1,B1} {
    \node[pt] at (\o+) {};
    \node[pt] at (\o-) {};
  }
  \node[font=\footnotesize, anchor=east] at ($(A0+)+(-0.08,0)$) {$+$};
  \node[font=\footnotesize, anchor=east] at ($(A0-)+(-0.08,0)$) {$-$};
  \node[font=\footnotesize, anchor=north west] at ($(B0+)+(0.04,-0.02)$) {$+$};
  \node[font=\footnotesize, anchor=north west] at ($(B0-)+(0.04,-0.02)$) {$-$};
  \node[font=\footnotesize, anchor=west] at ($(A1+)+(0.08,0)$) {$+$};
  \node[font=\footnotesize, anchor=west] at ($(A1-)+(0.08,0)$) {$-$};
  \node[font=\footnotesize, anchor=south east] at ($(B1+)+(-0.03,0.03)$) {$+$};
  \node[font=\footnotesize, anchor=south east] at ($(B1-)+(-0.03,0.03)$) {$-$};
\end{tikzpicture}\end{adjustbox}
\caption{The PR box as a bundle. Over each setting stands a fiber of its two outcomes, and each edge between fibers is a pair of outcomes that the PR box allows in a context; the context \texttt{(A1, B1)} is joined crosswise.}
\label{fig:bundle}
\end{figure}
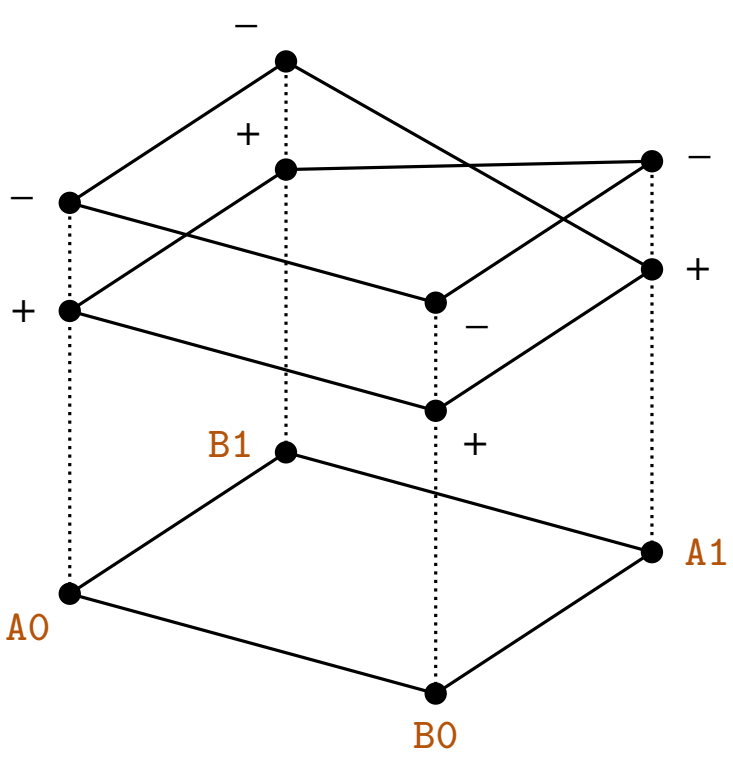

Figure~\ref{fig:bundle} draws the outcomes of the PR box as a bundle, with a fiber of two outcomes over each setting and an edge for each pair of outcomes that the PR box allows in a context. A global assignment would be a closed path that visits each fiber once: one outcome chosen in each fiber, such that all four chosen pairs are edges. The edges of the PR box form a discrete M\"obius strip, with the context \hs|(A1, B1)| joined crosswise, and no such path exists. \citet{AbramskyBrandenburger2011} call this \emph{strong contextuality}~\citep[see also][]{Abramsky2015}; quantum mechanics shows it only with more parties, as for GHZ states~\citep{GHZ1990}, or with more measurements, as in the Kochen--Specker theorem~\citep{KochenSpecker1967}.

The quantum boxes and the PR box thus differ only in how the two outcomes come to depend on each other in \hs|jointAB|: through the state that a measurement leaves, by the Born rule, or by the PR constraint on the pair. Both are blind and local and do not signal, but the quantum boxes are bounded by $2\sqrt2$ and, at that bound, leave every outcome possible, while the PR box attains $4$ and admits no global assignment.

%% =====================================================================
\FloatBarrier
\section{Hidden states that the boxes rewrite}\label{sec:hidden}

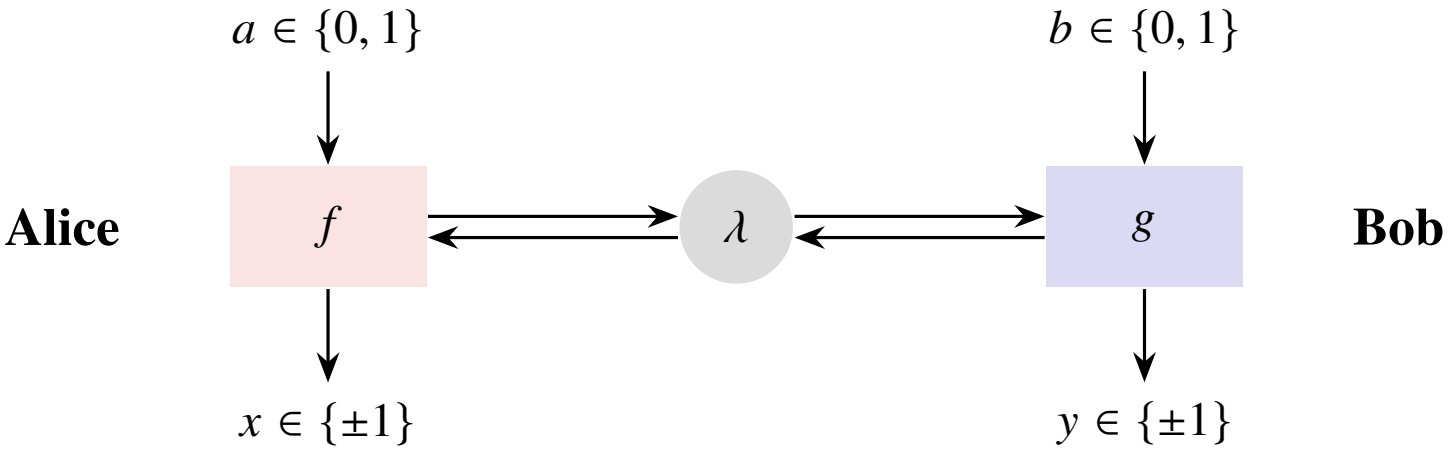
\begin{figure}[htbp]
\centering
\begin{adjustbox}{max width=\linewidth}\begin{tikzpicture}[bell]
  \twoboxes
  \node[lam] (l) at (2.7,0) {$\lambda$};
  \draw[->] ($(f.east)+(0,0.07)$) -- ($(l.west)+(0,0.07)$);
  \draw[<-] ($(f.east)+(0,-0.07)$) -- ($(l.west)+(0,-0.07)$);
  \draw[->] ($(l.east)+(0,0.07)$) -- ($(g.west)+(0,0.07)$);
  \draw[<-] ($(l.east)+(0,-0.07)$) -- ($(g.west)+(0,-0.07)$);
\end{tikzpicture}\end{adjustbox}
\caption{The nonlocal $\lambda$-model: the boxes read the hidden variable, and write to it.}
\label{fig:nonlocal}
\end{figure}

The quantum boxes and the PR box kept no hidden variable. The table of the PR box, and a table that quantum mechanics realizes, can also be produced with one, by boxes that stay factored and give up locality instead. The shared location of Section~\ref{sec:boxes} did so crudely, and signaled. In this section the boxes share $\lambda$ itself, and may update it during a trial (Figure~\ref{fig:nonlocal}): the definition of $\lambda$ stays the same, but \hs|Reader| becomes \hs|State|. Each box's local response reads its entry of $\lambda$, as in Section~\ref{sec:lhv}, and then rewrites the other party's entries; the joint response keeps its default (\ref{lst:trialmodel}). Whether such boxes signal will turn out to depend on their sampler.

%% ---------------------------------------------------------------------
\subsection{The PR box from a hidden state}\label{sec:prstate}

The trial monad is a state monad over $\lambda$:
\code{16-17}{CHSH/PRState.hs}
Each box reads its own entry of $\lambda$, as in Section~\ref{sec:lhv}, and then overwrites both of the other party's entries: with its outcome if it measured its first setting, and with its outcome and its flip if it measured its second. \hs|gets a0| reads the field \hs|a0| of the state, and \hs|modify| rewrites the state:
\code{28-36}{CHSH/PRState.hs}[lst:prstate]
Bob's box is its dual:\par\nopagebreak
\code{37-44}{CHSH/PRState.hs}
The joint response keeps its default (\ref{lst:trialmodel}), which runs Alice's box and then Bob's. Alice reads her outcome $x$ from $\lambda$; Bob then reads an entry that Alice's box overwrote, so his outcome is $y = x$, except in context \hs|(A1, B1)|, where it is the flip of $x$: the PR constraint. Bob's box rewrites Alice's entries in the same way, but she has already answered. The program has the shape of the shared location of Section~\ref{sec:boxes}, whose local responses also write. It could as well have the shape of the quantum boxes (\ref{lst:quantumjoint}), with local responses that only read and an override that pairs the two computations above; the runs would be the same (Section~\ref{sec:override}). What makes these boxes not local is what they write.

The membrane draws the initial $\lambda$ with a sampler, as for local hidden variables; \hs|evalState s| runs the trial from the state that \hs|samp| draws:
\code{50-51}{CHSH/PRState.hs}
We run the model with the cycling sampler under each schedule, and with the constant sampler \hs|plus2| under the fixed schedule. With the cycling sampler, both runs give CHSH \val{prstate-uniform-fixed/chsh} and do not signal. (Under the fixed schedule the cycling sampler correlates Bob's entries of $\lambda$ with the settings, as in Section~\ref{sec:staging}, but Alice's box overwrites them before Bob reads them.) With \hs|plus2|, the run still gives CHSH \val{prstate-plus2/chsh}, but Bob at \hs|B1| signals: his frequency of \hs|Plus| there is \val{prstate-plus2/run/BobB1/1} next to \hs|A0| and \val{prstate-plus2/run/BobB1/2} next to \hs|A1|.

\paragraph{Why a uniform $\lambda$ does not signal.} The model is not local: given $\lambda$, Bob's outcome depends on Alice's setting, through the entries that Alice's box overwrote. (It is trivially factored, since the settings and $\lambda$ determine both outcomes.) In bits, Bob's outcome is $y = x \oplus (a \land b)$, where $x$ is Alice's entry of $\lambda$. When $\lambda$ is uniform, $x$ is a fair coin, and so is $y$, whatever $a$ is: Alice's setting reaches Bob only masked by a uniformly random bit, as in a one-time pad, and Bob's marginal cannot depend on it. With \hs|plus2|, $x$ is always \hs|Plus|, the mask is gone, and Bob's outcome at \hs|B1| reveals Alice's setting. The shared location of Section~\ref{sec:boxes} is the same channel without the mask.

No-signaling here is a property of the sampler, not of the boxes. The quantum boxes also write to a state during a trial, and have the opposite properties: they are local and not factored. The difference is not where the writes sit, which is a choice (Section~\ref{sec:override}), but what they carry. Here the mask is an entry of $\lambda$, drawn by the sampler before the trial; for a given $\lambda$ there is no mask, and Bob's outcome depends on Alice's setting. There, what Alice's box writes depends on her outcome, drawn in the measurement during the trial, and averaged over that outcome it leaves Bob's distribution as it was, whatever the state. The quantum boxes do not signal from any state: there, no-signaling is a property of the boxes.

%% ---------------------------------------------------------------------
\subsection{The Bell table from a hidden state}\label{sec:belltable}

A hidden state that the boxes rewrite can also produce correlations weaker than the PR box's, among them correlations that quantum mechanics realizes. Figure~\ref{fig:belltable} shows the Bell table of \citet{Abramsky2017}: the outcomes always agree in context \hs|(A0, B0)|, agree three times in four in \hs|(A0, B1)| and \hs|(A1, B0)|, and agree once in four in \hs|(A1, B1)|. Its CHSH value is $1 + \frac12 + \frac12 + \frac12 = 2.5$, above the classical bound $2$ and below the Tsirelson bound $2\sqrt2$. The quantum boxes of Section~\ref{sec:quantum} realize it, with other observables: in the state $|\Phi^+\rangle$, with Alice measuring $Z$ or $(Z + \sqrt3\,X)/2$ and Bob measuring $Z$ or $(Z - \sqrt3\,X)/2$, the outcomes agree with probability $1$, $3/4$, $3/4$ and $1/4$. (The accompanying code computes these four probabilities by the Born rule on the pair.)

\begin{figure}[htbp]
\centering
\idealtable{$1/2$ & $0$ & $0$ & $1/2$}{$3/8$ & $1/8$ & $1/8$ & $3/8$}{$3/8$ & $1/8$ & $1/8$ & $3/8$}{$1/8$ & $3/8$ & $3/8$ & $1/8$}
\caption{The Bell table~\citep[Figure~3]{Abramsky2017}.}
\label{fig:belltable}
\end{figure}

\paragraph{The boxes.} The trial monad is again a state monad over $\lambda$:\par\nopagebreak
\code{27-28}{CHSH/BellTable.hs}
As for the PR box, each box reads its own entry of $\lambda$ and then rewrites the other party's entries, and the joint response keeps its default. To agree three times in four rather than always, a box copies its outcome into the other party's entries or flips it, by coins that it finds in $\lambda$ itself. Alice's box:
\code{40-49}{CHSH/BellTable.hs}
Bob's box is its dual:\par\nopagebreak
\code{50-58}{CHSH/BellTable.hs}
Once a box has read its entry, the other three entries of a uniform $\lambda$ are still fair coins, independent of its outcome, and two of them are both \hs|Minus| a quarter of the time. Such a pair is a coin, and we say that it \emph{comes up} when \hs|rare| holds of it:
\code{61-62,64-65}{CHSH/BellTable.hs}
In the order of the joint response, Bob's outcome is $x$ in context \hs|(A0, B0)|; $x$ flipped when the coin comes up in \hs|(A0, B1)| and \hs|(A1, B0)|; and $x$ flipped unless the coin comes up in \hs|(A1, B1)|. In bits, $y = x \oplus e$, where $e$ is $1$ with probability $0$, $1/4$, $1/4$ and $3/4$ in the four contexts. Nothing reads the state after Bob's write, but either box could go first: Alice's outcome would then be Bob's, flipped with the same probabilities, and a uniform $\lambda$ gives the same table.

As in Section~\ref{sec:prstate}, no-signaling is a property of the sampler: when $\lambda$ is uniform, the first outcome is a fair coin, the second is the first, flipped by an independent coin, and both marginals are $1/2$ whatever the other party's setting. The membrane is the one of that section:
\code{71-72}{CHSH/BellTable.hs}

\paragraph{The runs.} The coins are fair only if every context meets a uniform $\lambda$, which the cycling sampler under the fixed schedule does not provide (Section~\ref{sec:staging}). A sampler that keeps each strategy for a block of four steps, as \hs|biasPlus2Blocks| does, lets every context meet all sixteen equally often:
\code{97-98}{CHSH/LHV.hs}
With \hs|uniformBlocks| under the fixed schedule, the run's table is the Bell table, count for count, with CHSH \val{bell-blocks-fixed/chsh} and no signaling:
\begin{center}
\runtable{bell-blocks-fixed}
\end{center}
The test suite also runs the pair with Bob's box first, and gets the same table, count for count. With the cycling sampler under the random schedule, the run gives CHSH \val{bell-uniform-random/chsh} and does not signal.

Under the fixed schedule, the cycling sampler fixes Bob's entries of $\lambda$ with the context, as in Section~\ref{sec:staging}, and the coins with them: \hs|b1| is always \hs|Plus| in contexts \hs|(A0, B1)| and \hs|(A1, B1)|, and so is \hs|b0| in \hs|(A1, B0)|. No coin ever comes up, so Bob's outcome copies Alice's except in \hs|(A1, B1)|, where it flips it, and the run is the PR box: CHSH \val{bell-uniform-fixed/chsh}, without signaling. The boxes are the same in all three runs; the sampler and the schedule decide which table they produce, and the table shows it.

\paragraph{A logical Bell inequality.} In the terms of Section~\ref{sec:contextuality}, the Bell table behaves like the table of the quantum boxes. The outcomes it allows fit eight global assignments, those with equal entries for \hs|A0| and \hs|B0|, and each allowed outcome lies on one of them. Its CHSH value rules out a distribution over them. \citet{Abramsky2017} reads this as a logical Bell inequality: no global assignment makes all four events hold at once, the three agreements and the disagreement in \hs|(A1, B1)|, so their probabilities sum to at most $3$; in the Bell table they sum to $1 + \frac34 + \frac34 + \frac34 = 3.25$.

A hidden variable that the boxes may rewrite thus reaches a quantum table as easily as the PR box. What Bell's theorem rules out is a hidden variable that neither box can change, as in Section~\ref{sec:lhv}.

%% =====================================================================
\FloatBarrier
\section{No-signaling, checked in the run}\label{sec:checked}

Every verdict so far was read off the table of the run that produced the CHSH value. This section shows why that matters, by comparing it with the obvious alternative: checking no-signaling in experiments of its own.

%% ---------------------------------------------------------------------
\subsection{Two ways to check}

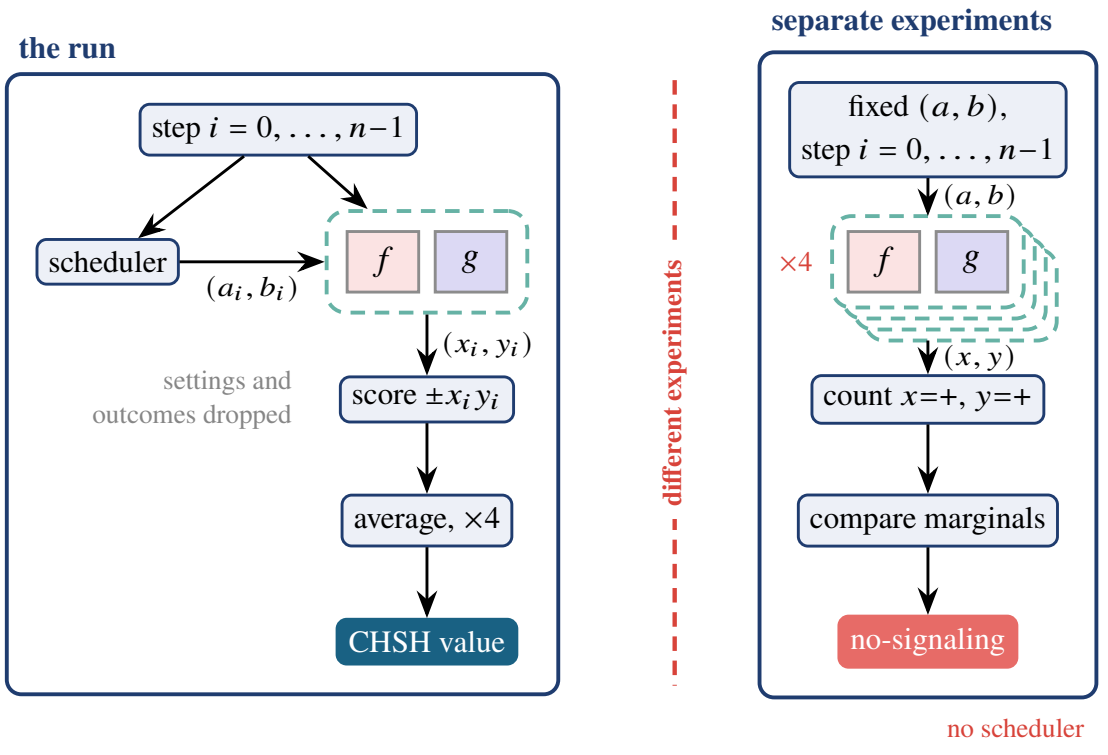
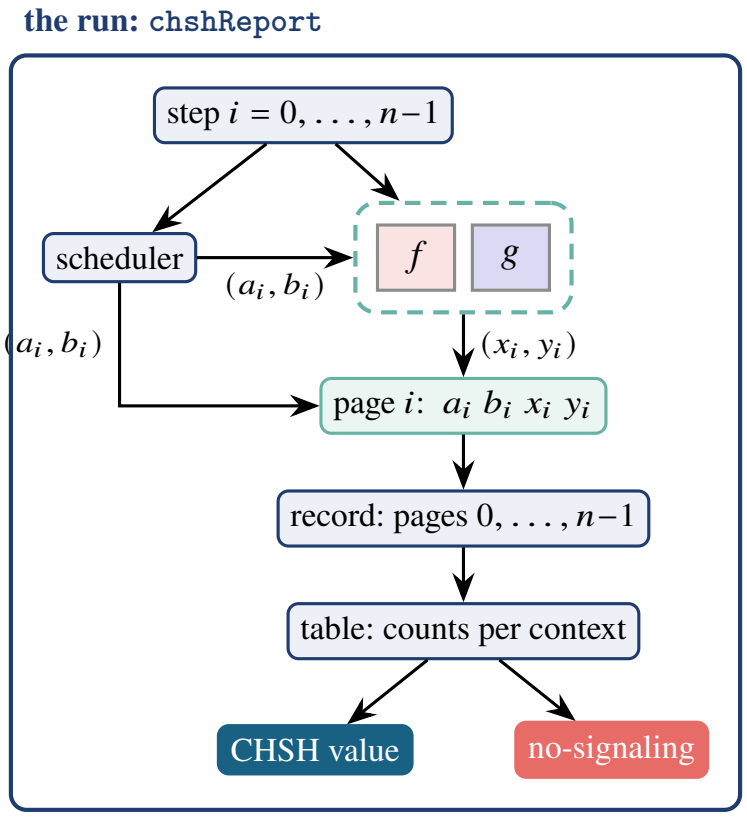
\begin{figure}[htbp]
\centering
\begin{subfigure}[b]{0.565\linewidth}
\centering
\begin{adjustbox}{max width=\linewidth}\begin{tikzpicture}[bell, scale=0.8, every node/.append style={transform shape}]
  \runtop
  \node[comp] (score) at (1.35,-2.45) {score $\pm x_i y_i$};
  \draw[wire] (mem) -- node[right, lbl] {$(x_i, y_i)$} (score);
  \node[note, anchor=east] at (0.25,-2.45) {settings and\\ outcomes dropped};
  \node[comp] (avg) at (1.35,-3.55) {average, $\times 4$};
  \draw[wire] (score) -- (avg);
  \node[chshnode] (chsh) at (1.35,-4.65) {CHSH value};
  \draw[wire] (avg) -- (chsh);
  \node[band, fit=(step)(sched)(mem)(score)(avg)(chsh)] (runband) {};
  \node[bandlbl, anchor=south west] at (runband.north west) {the run};
  \oldcheck
  \draw[AlertRed, densely dashed, line width=0.9pt] (3.6,0.45) -- (3.6,-5.05);
  \node[font=\scriptsize\bfseries, text=AlertRed, fill=white, rotate=90] at (3.6,-2.3) {different experiments};
\end{tikzpicture}\end{adjustbox}
\caption{A separate check: the verdict comes from other experiments.}
\end{subfigure}\hfill
\begin{subfigure}[b]{0.39\linewidth}
\centering
\begin{adjustbox}{max width=\linewidth}\begin{tikzpicture}[bell, scale=0.8, every node/.append style={transform shape}]
  \runtop
  \node[recnode] (rec) at (1.35,-2.45) {page $i$:\ \ $a_i\ b_i\ x_i\ y_i$};
  \draw[wire] (mem) -- node[right, lbl] {$(x_i, y_i)$} (rec);
  \draw[wire] (sched) |- node[pos=0.25, left, lbl] {$(a_i, b_i)$} (rec);
  \node[comp] (nb) at (1.35,-3.4) {record: pages $0, \ldots, n{-}1$};
  \draw[wire] (rec) -- (nb);
  \node[comp] (tab) at (1.35,-4.35) {table: counts per context};
  \draw[wire] (nb) -- (tab);
  \node[chshnode] (chsh) at (0.1,-5.35) {CHSH value};
  \node[nsnode] (nsrun) at (2.6,-5.35) {no-signaling};
  \draw[wire] (tab) -- (chsh);
  \draw[wire] (tab) -- (nsrun);
  \node[band, fit=(step)(sched)(mem)(rec)(nb)(tab)(chsh)(nsrun)] (runband) {};
  \node[bandlbl, anchor=south west] at (runband.north west) {the run: \texttt{chshReport}};
\end{tikzpicture}\end{adjustbox}
\caption{One run, one table, both verdicts.}
\end{subfigure}
\caption{Two ways to check no-signaling. A separate check fixes the settings in advance and never sees the schedule; in a run, each step writes a page, and both verdicts are read off the same table.}
\label{fig:twoways}
\end{figure}

Equation~\eqref{eq:four} compares conditional probabilities at fixed settings. It is tempting to estimate them in separate experiments at fixed settings: fix $(a, b)$, run the trial at the steps $0, \ldots, n-1$, and count the outcomes. No scheduler is involved, and the CHSH value comes from another run.\footnote{An earlier version of this paper checked no-signaling in this way, with $\val{nSep}$ trials per context and a tolerance of $\val{eps}$ on each difference of frequencies.} Figure~\ref{fig:twoways} sets this check beside the one of Section~\ref{sec:verdicts} (\ref{lst:report}), in which both verdicts come from one table.

%% ---------------------------------------------------------------------
\subsection{Why the two can differ}

% The table of every run, defined here so that it opens the next page:
% defined in Section 8.3, it is still pending at the end of the section,
% where the float barrier of placeins then clears the page.
\begin{figure}[t]
\centering
{\small\renewcommand{\arraystretch}{1.04}\setlength{\tabcolsep}{2.9pt}%
\begin{adjustbox}{max width=\linewidth}\begin{tabular}{@{}l l l r r l c@{}}
\toprule
Model & Sampler & Schedule & Seed & CHSH & In the run & Separate\\
\midrule
\input{generated/runs.tex} 
\bottomrule
\end{tabular}\end{adjustbox}}
\caption{Every run of this paper, \val{nTrials} trials each: the CHSH value and the no-signaling verdict of its own table, and the verdict of a separate check on the same membrane ($\val{nSep}$ trials per context). Seed: the seed of a run in \texttt{IO} (\ref{lst:seeded}); a run in \texttt{Identity} draws nothing at random. The separate checks in \texttt{IO} have seeds of their own, in the accompanying code. \no\ marks a run that signals, with the condition that its table violates. The shaded runs signal, and pass the separate check. Superdeterministic runs have no separate check.}
\label{fig:runs}
\end{figure}

Write $\pr{a, b}{y}_i$ for the probability of Bob's outcome $y$ at step $i$ when the settings are $(a, b)$, and recall that $\fr{a, b}_i$ is the probability that the scheduler picks $(a, b)$ at step $i$. The run's table averages over the steps that measured $(a, b)$; the separate check fixes $(a, b)$ and uses every step once:
\begin{equation}\label{eq:runcheck}
  \pr{a, b}{y}_{\mathrm{run}} = \frac{\sum_{i} \fr{a, b}_i\, \pr{a, b}{y}_i}{\sum_{i} \fr{a, b}_i},
  \qquad\qquad
  \pr{a, b}{y}_{\mathrm{check}} = \frac{1}{n} \sum_{i} \pr{a, b}{y}_i.
\end{equation}
The two agree when $\pr{a, b}{y}_i$ is the same at every step, as for the quantum and PR boxes, whose membranes ignore the step, or when $\fr{a, b}_i$ is, as under the random schedule. They can differ when both change with $i$: under the fixed schedule, with a membrane that passes the step on to the boxes, or with a sampler that reads it. The same holds for Alice's outcome $x$.

The clocked boxes with the clock visible are an example. For them, $\pr{a, b}{\yp}_i$ is $0$ when $i \bmod 4 = 3$ and $1$ otherwise, whatever $(a, b)$. The separate check therefore sees Bob answer \hs|Plus| at \hs|B1| with frequency $3/4$ next to either of Alice's settings (\val{clock-visible-fixed/sep/BobB1/1} and \val{clock-visible-fixed/sep/BobB1/2} in our run of it), and it passes. Under the fixed schedule, the run measures \hs|(A1, B1)| exactly at the steps where Bob answers \hs|Minus|, and finds \val{clock-visible-fixed/run/BobB1/1} next to \hs|A0| and \val{clock-visible-fixed/run/BobB1/2} next to \hs|A1|. Of the two experiments, only the run is the one whose CHSH value was reported.

%% ---------------------------------------------------------------------
\subsection{Every model, both checks}

Figure~\ref{fig:runs} lists every run of this paper, with the verdict of its own table and that of the separate check on the same membrane. The separate check passes four runs whose own table signals: the clocked boxes with the clock visible, the cycling sampler and the biased sampler of Section~\ref{sec:staging}, and the sampler that peeks, all under the fixed schedule. In all four, the step correlates the settings with the behavior of the boxes. The separate check never sees the schedule, so it never sees the correlation, and its verdict is the same under either schedule; the verdict of the run depends on the schedule, as it must.

The two checks agree on every other run that has both: on the shared location and the PR box from a hidden state with \hs|plus2|, which signal within a trial, and on the quantum and PR boxes, whose membranes ignore the step.

Superdeterministic runs have no entry in the last column. Their membrane depends on the settings, and a separate check must choose which settings to hand it. Our earlier version handed it the scheduler's settings at each step, by running the scheduler a second time inside the membrane. That membrane is \hs|runTrialLHV (peek fixedSchedule cheatLam)|, whose run signals while its separate check passes; and under a random schedule it loses the cheat altogether (CHSH \val{peek-random/chsh}), because the second run of the scheduler flips new coins. In a run, no such choice arises: the trial hands the membrane the settings it drew.

%% =====================================================================
% No float barrier before this section: the table of every run (Figure 13)
% may open the page on which Section 8 ends. With the barrier, a figure
% still pending here clears the page and strands the last line of Section 8.
\makeatletter\let\bf@secFB\@fb@secFB \let\@fb@secFB\relax\makeatother
\FloatBarrier
\section{Conclusion}\label{sec:conclusion}
\makeatletter\let\@fb@secFB\bf@secFB\makeatother

We conclude by collecting the models of this paper, in Figure~\ref{fig:picture}, by the three properties of Section~\ref{sec:lambda} and by the verdicts of their runs.

\begin{figure}[htbp]
\centering
{\small\renewcommand{\arraystretch}{1.15}%
\begin{adjustbox}{max width=\linewidth}\begin{tabular}{@{}l ccc c l@{}}
\toprule
Model & Blind & Local & Factored & No-signaling & CHSH\\
\midrule
Local hidden variables (\S\ref{sec:lhv}) & \yes & \yes & \yes & \yes & $\le 2$\\
\addlinespace
\multicolumn{6}{@{}l}{\emph{Not blind}}\\
\quad Scheduler-induced loopholes (\S\ref{sec:clocked}, \S\ref{sec:staging}) & \no & \yes & \yes & \no & $\le 4$\\
\quad Superdeterminism (\S\ref{sec:superdet}) & \no & \yes & \yes & \no\ / \yes & $\le 4$\\
\addlinespace
\multicolumn{6}{@{}l}{\emph{Not factored}}\\
\quad Quantum boxes (\S\ref{sec:quantum}) & \yes & \yes & \no & \yes & $\le 2\sqrt2$\\
\quad PR boxes (\S\ref{sec:pr}) & \yes & \yes & \no & \yes & $4$\\
\addlinespace
\multicolumn{6}{@{}l}{\emph{Not local}}\\
\quad Shared location (\S\ref{sec:boxes}) & \yes & \no & \yes & \no & $\le 4$\\
\quad PR box from a hidden state (\S\ref{sec:prstate}) & \yes & \no & \yes & \yes\ / \no & $4$\\
\quad Bell table from a hidden state (\S\ref{sec:belltable}) & \yes & \no & \yes & \yes & $2.5$ / $4$\\
\bottomrule
\end{tabular}\end{adjustbox}}
\caption{The models of this paper, by the property that each gives up. The no-signaling column is the verdict of each model's own runs (Figure~\ref{fig:runs}): the scheduler-induced loopholes signal under the fixed schedule, a superdeterministic sampler may move a marginal (\texttt{cheatLam}) or keep both fair (\texttt{cheatNS}), and the PR box from a hidden state does not signal with the cycling sampler but does with \texttt{plus2}. The Bell table from a hidden state gives $2.5$ when every context meets a uniform $\lambda$, and the PR box's $4$ with the cycling sampler under the fixed schedule, a run that is not blind either (Section~\ref{sec:staging}).}
\label{fig:picture}
\end{figure}

\paragraph{By rows: A hierarchy.} Local hidden variables have all three properties and stay within $2$. The scheduler-induced loopholes and superdeterminism give up blindness, and the shared location gives up locality; all three reach $4$ for reasons that have nothing to do with quantum theory. Quantum and PR boxes give up only factoring: they stay blind and local, do not signal, and reach $2\sqrt2$ and $4$. The hidden states of Section~\ref{sec:hidden}, like the shared location, give up locality, and do not signal only because their sampler masks the channel between the boxes. They reach the $4$ of the PR box and, with coins that the boxes find in $\lambda$ itself, a quantum value: the $2.5$ of the Bell table.

\paragraph{By columns: Three statements about a program.} In the code of local hidden variables, each property says something about one place.
\begin{itemize}
\item \textbf{Blind:} \hs|liftA2 (,) sched samp|, in the experiment layer (\ref{lst:independent}). The settings and the strategy of a step form an applicative pair, in which the sampler does not see what the scheduler returns.
\item \textbf{Factored:} \hs|liftA2 (,) (localA a) (localB b)|, in the trial layer (\ref{lst:trialmodel}). The two outcomes of a trial form an applicative pair, the default \hs|jointAB|, in which neither box sees what the other returns.
\item \textbf{Local:} \hs|localA :: ASetting -> t Outcome|, and likewise \hs|localB|. Each box is a function of its own setting, in an effect \hs|t| through which the other setting cannot reach it: a reader monad lets the boxes read $\lambda$ but not talk.
\end{itemize}
Blind and factored thus have the same shape at the two layers, and giving up either makes the two components of the pair depend on each other. Superdeterminism replaces the pair by a bind, which draws the strategy from the settings (\ref{lst:dependent}). The boxes that give up factoring keep \hs|liftA2|, and the dependence comes in another way. The quantum boxes collapse a shared state (\ref{lst:quantum}): in a monad, \hs|liftA2| runs one computation after the other, so the dependence can travel through the effect. The PR box keeps only the pairs that meet its constraint (\ref{lst:pr}), so the dependence comes from outside the pair. Local is what keeps that dependence from signaling: Bob's outcome may depend on Alice's outcome, but, averaged over it, not on her setting. The boxes with a hidden state keep \hs|liftA2| too, as the default, and their dependence travels through the same effect; but given $\lambda$ it carries Alice's setting, and they are not local. Where a dependence travels does not decide which property it breaks; what it carries does (Section~\ref{sec:override}).

\paragraph{What the code does not enforce.} None of the three properties is guaranteed by types. A sampler of the legal type can follow the settings through the step that it shares with the scheduler (Section~\ref{sec:staging}). A state monad hides a channel behind local types, in the shared location (Section~\ref{sec:boxes}) and in the boxes with a hidden state (Section~\ref{sec:hidden}), whose joint response is even the default. The quantum boxes write to a state too; that they write in their joint response is a choice, and that the write carries no signal is a fact about projectors, not about types (Section~\ref{sec:override}). The properties belong to what a program does, not to what it looks like. That is why the verdicts are read off its runs; and no-signaling in particular is a property of a run, not of the boxes alone (Section~\ref{sec:checked}).

\paragraph{Who goes first.} In a state monad, \hs|liftA2| runs Alice's box and then Bob's, and the second sees the effect of the first, though not its value. For the quantum boxes and the PR box the order does not show: their joint distribution is the same either way (Section~\ref{sec:override}; the PR box draws two independent coins, and its constraint treats both parties alike). For the boxes with a hidden state it shows, and must: the box that goes first answers from its own setting and $\lambda$ alone, so if the two orders always agreed, both boxes would be local, and with a blind sampler held to $2$. Their no-signaling holds in both orders when every context meets a uniform $\lambda$, as under the random schedule; otherwise it may depend on the order. The cycling sampler under the fixed schedule pins Bob's entries of $\lambda$ (Section~\ref{sec:staging}), which Alice's box overwrites when it goes first. With Bob's box first, both runs of Section~\ref{sec:hidden} with that sampler and schedule signal, and the Bell boxes fall from CHSH \val{bell-uniform-fixed/chsh} to \val{order/bell-uniform-fixed/chsh}. Their verdicts in Figure~\ref{fig:runs} hold for Alice's box first: the order of evaluation, which has no physical counterpart, is part of the run too.

That is the contribution. The Bell experiment is a small program whose behavior is determined by how information is allowed to flow---what each part sees, how the parts are composed, and what passes between them---and whose two verdicts come from the record of one run.

Bell's theorem and contextuality are often presented as metaphysical claims about the nature of reality. At a technical level, both concern which local behaviors admit a global explanation, and under which independence assumptions. The framework does not resolve the foundational questions, nor does it reduce quantum mechanics to classical computation; it makes Bell's assumptions visible, and testable, in the structure of a program.

%% =====================================================================
\FloatBarrier
\section{Related work}\label{sec:related}

\paragraph{Bell's assumptions.} \citet{Bell1964} introduced local hidden-variable theories, and \citet{CHSH1969} turned his argument into the inequality that experiments test, from \citet{Aspect1982} to the loophole-free test of \citet{Hensen2015}. Our three properties are the standard assumptions behind it. Blindness is $\lambda$-independence, also called measurement independence~\citep{Brunner2014}. Local and factored are the two conditions into which \citet{Jarrett1984} split Bell's locality, and which he called locality and completeness; \citet{Shimony1984} named them parameter independence and outcome independence. Quantum mechanics keeps parameter independence and violates outcome independence: in our code, Alice's measurement collapses the state of the pair by an outcome drawn during the trial, on which Bob's outcome then depends, and averaged over that outcome the collapse leaves his distribution as it was. The hidden states of Section~\ref{sec:hidden} violate parameter independence instead: the settings and $\lambda$ determine both outcomes, so outcome independence holds trivially, and Bob's outcome depends on Alice's setting. Whether a model overrides \hs|jointAB| does not decide which of the two it violates (Section~\ref{sec:override}). \citet{Brandenburger2008} classify these properties and the relations among them. Superdeterminism~\citep{HossenfelderPalmer2020,Palmer2009,tHooft2016} and the free will theorems~\citep{ConwayKochen2006,ConwayKochen2009} concern the first.

\paragraph{No-signaling and contextuality.} \citet{PopescuRohrlich1994} showed that no-signaling alone permits a CHSH value of $4$, which started a search for principles that single out the quantum correlations~\citep{Brunner2014}. \citet{AbramskyBrandenburger2011} treat nonlocality and contextuality alike: an empirical model is a family of distributions, one per context, that agree on overlaps---in a Bell scenario, this agreement is no-signaling---and contextuality is the absence of a global section, which here is a distribution over values of type \hs|Lambda| that gives every context its own \citep[see also][]{Abramsky2013,Abramsky2015}. The tables of our runs are empirical models in a weaker sense: nothing makes them agree on overlaps, and whether they do is one of the two verdicts. \citet{Specker1960} and \citet{KochenSpecker1967} found contextuality without spatial separation.

\paragraph{Causal models.} Causal analyses of Bell's theorem derive the inequalities from causal graphs and conditional independences, in the framework of \citet{Pearl2009}. We read the causal structure off the program instead: the step that the scheduler and the sampler share is a common cause, and superdeterminism is an applicative pair refactored into a bind.

\paragraph{Quantum programming.} The Quantum IO monad~\citep{AltenkirchGreen2010} computes with quantum states in Haskell. A trial model built on it would compute the statistics of any quantum experiment. Ours measures the state vector of two qubits as a textbook does~\citep{NielsenChuang2010}, which suffices for the CHSH experiment.

%% =====================================================================
\FloatBarrier
\section{Future directions}\label{sec:future}

\paragraph{Quantitative relaxations.} Partial measurement dependence or limited signaling can be modeled by controlled dependencies between the sampler, the scheduler and the boxes. Seeing how CHSH values degrade along such paths may clarify the boundary between classical and non-classical regimes in operational terms.

\paragraph{Other effects.} The present work uses \hs|Reader|, \hs|State| and \hs|IO|. Continuations and concurrency primitives offer other ways to express dependencies between local and joint behavior, and may reveal correspondences with interpretations of quantum mechanics such as pilot-wave theory~\citep{Bohm1952}, the transactional interpretation~\citep{Cramer1986} and many-worlds~\citep{Everett1957}.

\paragraph{Deriving the Tsirelson bound.} Our quantum model derives its correlations from the Born rule, and attains $2\sqrt2$ with observables chosen for it. That no quantum experiment does better is Tsirelson's theorem~\citep{Tsirelson1980}, which the code does not prove. A trial model built on the Quantum IO monad~\citep{AltenkirchGreen2010} would cover every quantum experiment, not only measurements of two qubits; better still would be a constraint on joint composition, perhaps reflected in a refined monadic structure, that entails the bound.

\paragraph{No-signaling by construction.} We check no-signaling on runs because the code does not enforce it. Pairing two computations with \hs|liftA2| does not stop them from communicating through the effect they share (Section~\ref{sec:boxes}), and a run can signal through the experiment layer even when the boxes cannot (Sections~\ref{sec:clocked} and~\ref{sec:staging}). Enforcing it requires resources that the parties cannot share, as in the tensor of linear logic~\citep{Girard1987} or the separating conjunction of separation logic~\citep{Reynolds2002}; no-signaling has been characterized equationally in categorical and linear-logic settings~\citep{KissingerUijlen2017,LindenhoviusZamdzhiev2025}. A type for it would have to constrain the boxes together with the experiment layer that runs them.

\paragraph{Qubits as computations, and beyond CHSH.} The type of a local response, $\mathrm{Setting} \longrightarrow t\ \mathrm{Outcome}$, records that a measurement does not reveal a value fixed in advance: an outcome is produced by interaction, inside an effect. Multipartite models such as GHZ experiments~\citep{GHZ1990} and multi-setting ones such as Mermin's device~\citep{Mermin1981} would add an applicative layer $f$ of separated parties, $f\,(\mathrm{Setting} \longrightarrow t\ \mathrm{Outcome})$, and the same construction could instantiate restricted toy theories such as that of \citet{Spekkens2007}.

%% =====================================================================
\FloatBarrier
\section*{AI use}\addcontentsline{toc}{section}{AI use}

An AI assistant, Claude (Anthropic), was used in preparing this paper and its artifact. It wrote and revised parts of the Haskell package, its test suite, and the program that generates the paper's numbers and tables. It checked claims of the text by running the code, and drafted and edited parts of the text, the figures and Appendix~\ref{app:quantum}. The authors directed this work and reviewed its results. The artifact guards what the paper reports: every listing is quoted from the code by line number, every number, table and printed report comes from the data program, which stops if a claim of the text about the runs does not hold, and the test suite passes.

\bibliographystyle{ACM-Reference-Format}
\bibliography{references}

\appendix
% written by jfpify.py from appendix.tex; do not edit
\FloatBarrier
\section{Quantum mechanics from scratch}\label{app:quantum}

This appendix introduces the quantum mechanics that Section~\ref{sec:quantum} uses, assuming none: one qubit and its measurements (Sections~\ref{app:qubit} and~\ref{app:measure}), two qubits and entanglement (Section~\ref{app:pair}), and the correlations that the CHSH value combines (Section~\ref{app:chsh}). It needs only vectors and matrices, and it keeps to real numbers, which suffice for every state and measurement in this paper.\footnote{In general, amplitudes are complex numbers, an observable is a Hermitian matrix, and a probability is the squared absolute value of an amplitude. With real numbers these become what we use here: real amplitudes, symmetric matrices, and squares.} Figure~\ref{fig:dictionary}, at the end, lists each notion with its name in the code. A textbook such as \citet{NielsenChuang2010} goes much further.

%% ---------------------------------------------------------------------
\subsection{One qubit}\label{app:qubit}

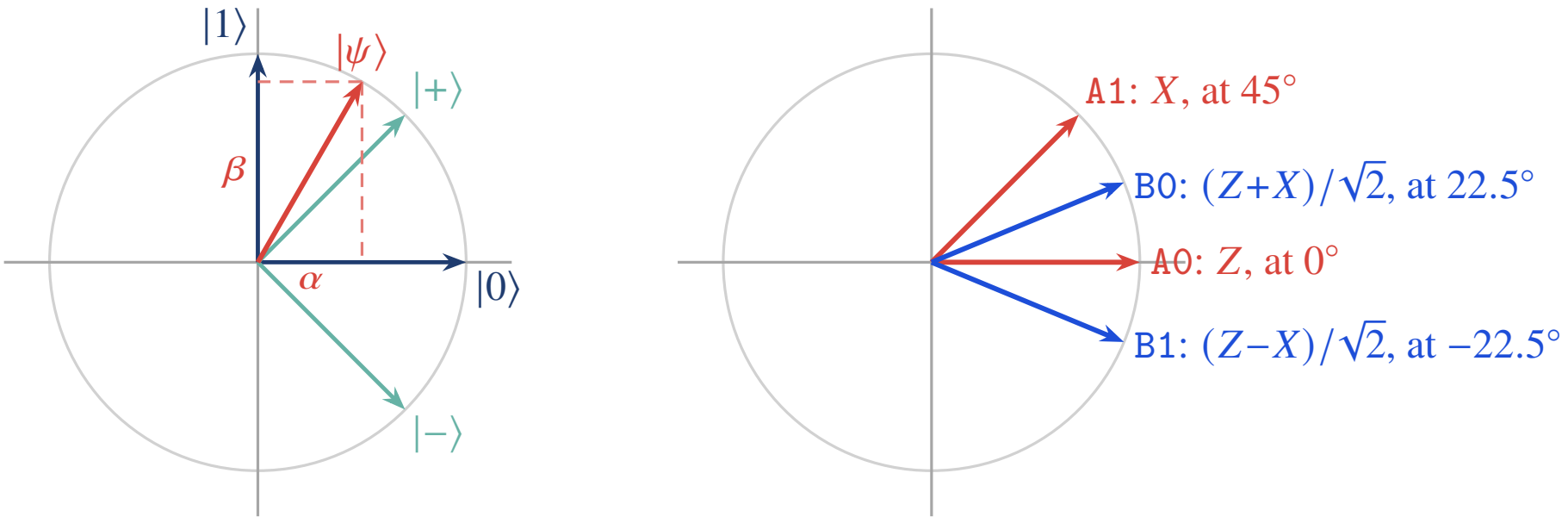
\begin{figure}[t]
\centering
\begin{subfigure}[b]{0.42\linewidth}
\centering
\begin{adjustbox}{max width=\linewidth}\begin{tikzpicture}[bell, >={Stealth[length=1.8mm]}]
  \def\R{1.6}
  \coordinate (O) at (0,0);
  \draw[black!18] (O) circle (\R);
  \draw[black!35] (-1.95,0) -- (1.95,0);
  \draw[black!35] (0,-1.95) -- (0,1.95);
  % the basis |0>, |1>
  \draw[->, Navy, line width=0.9pt] (O) -- (\R,0);
  \node[Navy, anchor=north west, inner sep=1.5pt] at (\R,0) {$|0\rangle$};
  \draw[->, Navy, line width=0.9pt] (O) -- (0,\R);
  \node[Navy, anchor=south east, inner sep=1.5pt] at (0,\R) {$|1\rangle$};
  % the basis |+>, |->
  \draw[->, BlockTeal, line width=0.9pt] (O) -- (45:\R);
  \node[BlockTeal, anchor=south west, inner sep=1pt] at (45:\R) {$|{+}\rangle$};
  \draw[->, BlockTeal, line width=0.9pt] (O) -- (-45:\R);
  \node[BlockTeal, anchor=north west, inner sep=1pt] at (-45:\R) {$|{-}\rangle$};
  % a state, and its two amplitudes
  \coordinate (psi) at (60:\R);
  \draw[densely dashed, AlertRed!70] (psi) -- (psi |- O);
  \draw[densely dashed, AlertRed!70] (psi) -- (O |- psi);
  \draw[->, AlertRed, line width=1.1pt] (O) -- (psi);
  \node[AlertRed, anchor=south, inner sep=2pt] at (psi) {$|\psi\rangle$};
  \node[AlertRed, anchor=north, font=\footnotesize, inner sep=2pt] at ($(O)!0.5!(psi |- O)$) {$\alpha$};
  \node[AlertRed, anchor=east, font=\footnotesize, inner sep=2pt] at ($(O)!0.5!(O |- psi)$) {$\beta$};
\end{tikzpicture}\end{adjustbox}
\caption{A state $|\psi\rangle = \alpha|0\rangle + \beta|1\rangle$, and the two measurements $Z$ (along $|0\rangle$ and $|1\rangle$) and $X$ (along $|{+}\rangle$ and $|{-}\rangle$).}
\end{subfigure}\hfill
\begin{subfigure}[b]{0.55\linewidth}
\centering
\begin{adjustbox}{max width=\linewidth}\begin{tikzpicture}[bell, >={Stealth[length=1.8mm]}]
  \def\R{1.6}
  \coordinate (O) at (0,0);
  \draw[black!18] (O) circle (\R);
  \draw[black!35] (-1.95,0) -- (1.95,0);
  \draw[black!35] (0,-1.95) -- (0,1.95);
  \draw[->, AlertRed, line width=1.1pt] (O) -- (0:\R);
  \node[AlertRed, anchor=west, inner sep=2pt] at (0:\R) {\tA0: $Z$, at $0^\circ$};
  \draw[->, AlertRed, line width=1.1pt] (O) -- (45:\R);
  \node[AlertRed, anchor=south west, inner sep=1pt] at (45:\R) {\tA1: $X$, at $45^\circ$};
  \draw[->, CExp, line width=1.1pt] (O) -- (22.5:\R);
  \node[CExp, anchor=west, inner sep=2pt] at (22.5:\R) {\tB0: $(Z{+}X)/\sqrt2$, at $22.5^\circ$};
  \draw[->, CExp, line width=1.1pt] (O) -- (-22.5:\R);
  \node[CExp, anchor=west, inner sep=2pt] at (-22.5:\R) {\tB1: $(Z{-}X)/\sqrt2$, at $-22.5^\circ$};
\end{tikzpicture}\end{adjustbox}
\caption{The four measurements of Section~\ref{sec:qubits}, each drawn as its direction for $+1$; the direction for $-1$ is perpendicular to it.}
\end{subfigure}
\caption{A qubit with real amplitudes, drawn in the plane.}
\label{fig:qubit}
\end{figure}

A bit is $0$ or $1$. A \emph{qubit} has a \emph{state}: a vector of two real numbers, its \emph{amplitudes}, whose squares add up to $1$,
\[
  |\psi\rangle = \alpha|0\rangle + \beta|1\rangle = [\alpha, \beta], \qquad \alpha^2 + \beta^2 = 1,
\]
where $|0\rangle = [1, 0]$ and $|1\rangle = [0, 1]$ are its two \emph{basis states}. The bracket $|\cdot\rangle$, a \emph{ket}, just marks a vector; the notation is Dirac's. A state is thus a unit vector in the plane (Figure~\ref{fig:qubit}a). Flipping the sign of the whole vector changes nothing: $-|\psi\rangle$ is the same state as $|\psi\rangle$, because every probability below is the square of an amplitude. Flipping the sign of one amplitude does change the state, as $|{+}\rangle$ and $|{-}\rangle$ below show. So a state is a line through the origin, which we call a \emph{direction}.

\emph{Measuring} a qubit gives one of two outcomes, at random. Measured in its basis, a qubit in the state $|\psi\rangle$ gives $0$ with probability $\alpha^2$ and $1$ with probability $\beta^2$, and afterwards it is in the basis state that was found: measuring it again in the same way gives the same outcome. The state fixes only the probabilities. A state with both amplitudes nonzero is a \emph{superposition} of $|0\rangle$ and $|1\rangle$. For example, each of
\[
  |{+}\rangle = \frac{|0\rangle + |1\rangle}{\sqrt2}, \qquad |{-}\rangle = \frac{|0\rangle - |1\rangle}{\sqrt2}
\]
gives $0$ and $1$ with probability $1/2$. They are nevertheless different states, and the measurement $X$ of Section~\ref{app:measure} tells them apart with certainty: a qubit is not a biased coin.

Two vectors have an \emph{inner product}, $\langle u|v\rangle = u_0 v_0 + u_1 v_1$, written \hs!u <| v! in the code. The length of $|v\rangle$ is $\sqrt{\langle v|v\rangle}$, two vectors are perpendicular when their inner product is $0$, and the amplitude $\alpha$ of $|\psi\rangle$ is $\langle 0|\psi\rangle$, its component along $|0\rangle$.

%% ---------------------------------------------------------------------
\subsection{Measurements: Observables and projectors}\label{app:measure}

Measuring in the basis $|0\rangle$, $|1\rangle$ is one measurement among many. Any two perpendicular unit vectors $|v_+\rangle$ and $|v_-\rangle$ define one, with the outcomes $+1$ and $-1$, by the same rule. Write the state in terms of them, $|\psi\rangle = c_+|v_+\rangle + c_-|v_-\rangle$: the outcome is $+1$ with probability $c_+^2$ and $-1$ with probability $c_-^2$, and afterwards the qubit is in $|v_+\rangle$ or $|v_-\rangle$ accordingly. Figure~\ref{fig:qubit}a shows two such pairs, $|0\rangle$, $|1\rangle$ and $|{+}\rangle$, $|{-}\rangle$.

\paragraph{Observables.} Physics names a measurement by a matrix, its \emph{observable}. An \emph{eigenvector} of a matrix $O$ is a nonzero vector that $O$ only scales, $O|v\rangle = \lambda|v\rangle$, and $\lambda$ is its \emph{eigenvalue}. The observable of the measurement above is the matrix with the eigenvectors $|v_+\rangle$ and $|v_-\rangle$, and the eigenvalues $+1$ and $-1$: the outcomes are its eigenvalues. Measuring in the basis $|0\rangle$, $|1\rangle$, with $+1$ for $0$ and $-1$ for $1$, is the observable $Z$, and measuring along $|{+}\rangle$ and $|{-}\rangle$ is $X$, two of the \emph{Pauli matrices}:
\begin{gather*}
  Z = \begin{pmatrix} 1 & 0 \\ 0 & -1 \end{pmatrix}, \quad Z|0\rangle = |0\rangle, \quad Z|1\rangle = -|1\rangle; \\
  X = \begin{pmatrix} 0 & 1 \\ 1 & 0 \end{pmatrix}, \quad X|{+}\rangle = |{+}\rangle, \quad X|{-}\rangle = -|{-}\rangle.
\end{gather*}
In the code, the outcomes $+1$ and $-1$ are \hs|Plus| and \hs|Minus|. The observable $-O$ is the same measurement as $O$, with the outcomes swapped. Every measurement in this paper has the form
\[
  \cos\varphi\, Z + \sin\varphi\, X
\]
for an angle $\varphi$. Its eigenvector for $+1$ is the direction at half that angle, $[\cos\frac\varphi2, \sin\frac\varphi2]$, and its eigenvector for $-1$ is the perpendicular direction. Written out, the matrix is $\bigl(\begin{smallmatrix} \cos\varphi & \sin\varphi \\ \sin\varphi & -\cos\varphi \end{smallmatrix}\bigr)$: the reflection of the plane in the line at angle $\varphi/2$, which leaves that line fixed (eigenvalue $+1$) and turns the perpendicular line around ($-1$). We name an observable by $\varphi$, and give its directions, at $\varphi/2$, in degrees. $Z$ is the case $\varphi = 0$ and points along $|0\rangle$; $X$ is $\varphi = \pi/2$ and points along $|{+}\rangle$, at $45^\circ$. Bob's observables in Section~\ref{sec:qubits}, $(Z \pm X)/\sqrt2$, are $\varphi = \pm\pi/4$ and point at $\pm 22.5^\circ$ (Figure~\ref{fig:qubit}b).\footnote{In the experiments with photons of Section~\ref{sec:intro}, such a measurement is a polarizer turned by the angle $\varphi/2$.}

\paragraph{Projectors and the Born rule.} The part of $|\psi\rangle$ along $|v_+\rangle$ is $c_+|v_+\rangle$. The matrix that keeps this part and drops the other is the \emph{projector} $P_+ = (I + O)/2$, where $I$ is the identity matrix, and $P_- = (I - O)/2$ keeps the other part:
\[
  P_+|\psi\rangle = c_+|v_+\rangle, \qquad P_-|\psi\rangle = c_-|v_-\rangle, \qquad P_+ + P_- = I.
\]
(Indeed $(I + O)/2$ maps $|v_+\rangle$ to $|v_+\rangle$, and $|v_-\rangle$ to $0$.) The probability of an outcome is the squared length of the part that its projector $P$ keeps, and the state that the outcome leaves is that part, divided by its length:
\[
  \langle\psi|P|\psi\rangle = c^2, \qquad \frac{P|\psi\rangle}{\sqrt{\langle\psi|P|\psi\rangle}},
\]
where $c$ is $c_+$ or $c_-$. Here $\langle\psi|P|\psi\rangle$ is the inner product of $|\psi\rangle$ with $P|\psi\rangle$, which is $c^2$ because $|v_+\rangle$ and $|v_-\rangle$ are perpendicular. The first formula is the \emph{Born rule}, and the second the rule for the state after a measurement; Section~\ref{sec:qubits} calls both the Born rule, and \hs|probability| and \hs|collapse| compute them. An outcome with $c = 0$ never occurs, and leaves no state. The state that an outcome leaves matters only to a later measurement of the same qubit, or of the same pair; for a measurement that nothing follows, only the probability matters. That is why the local responses of Section~\ref{sec:override} can leave the state alone: on its own, a box is followed by nothing. Whether the collapse is written there or in \hs|nonlocalA| and \hs|nonlocalB| changes no run (Section~\ref{sec:override}).

\paragraph{An example.} Measure $|0\rangle$ with $X$. Since $|0\rangle = (|{+}\rangle + |{-}\rangle)/\sqrt2$, the outcome is $+1$ or $-1$ with probability $1/2$ each, and the qubit is left in $|{+}\rangle$ or in $|{-}\rangle$. Measuring $Z$ now gives $+1$ or $-1$ with probability $1/2$ each, although before the measurement of $X$ it would have given $+1$ for sure. A state makes some measurements certain and leaves others to chance, and a measurement changes it.

\paragraph{Averages.} The average outcome, $(+1)\langle\psi|P_+|\psi\rangle + (-1)\langle\psi|P_-|\psi\rangle$, is $\langle\psi|O|\psi\rangle$, since $P_+ - P_- = O$. For $|0\rangle$ it is $1$ with $Z$, and $0$ with $X$.

%% ---------------------------------------------------------------------
\subsection{Two qubits and entanglement}\label{app:pair}

Two bits have four values, $00$, $01$, $10$ and $11$, and a state of two qubits has an amplitude for each:
\[
  |\psi\rangle = \psi_{00}|00\rangle + \psi_{01}|01\rangle + \psi_{10}|10\rangle + \psi_{11}|11\rangle = [\psi_{00}, \psi_{01}, \psi_{10}, \psi_{11}],
\]
with squares that add up to $1$. The first digit belongs to Alice's qubit, and the second to Bob's. Measured in their basis, the two qubits give $jk$ with probability $\psi_{jk}^2$. In the code, a state of the pair is this list of four numbers, a \hs|Vector|.

\paragraph{Product states.} If Alice's qubit is in the state $|u\rangle = [u_0, u_1]$ and Bob's, independently, in $|v\rangle = [v_0, v_1]$, the pair is in their \emph{tensor product}
\[
  |u\rangle \otimes |v\rangle = [u_0 v_0,\ u_0 v_1,\ u_1 v_0,\ u_1 v_1].
\]
Each amplitude is a product, so each probability is one too, $u_j^2\, v_k^2$: the two outcomes are independent. For example, $|00\rangle = |0\rangle \otimes |0\rangle$, and $|{+}{+}\rangle = |{+}\rangle \otimes |{+}\rangle = \frac12[1, 1, 1, 1]$.

\paragraph{Entangled states.} A state of the pair that is not a tensor product is \emph{entangled}. The state $|\Phi^+\rangle$ of Section~\ref{sec:pair},
\[
  |\Phi^+\rangle = \frac{|00\rangle + |11\rangle}{\sqrt2} = \frac{1}{\sqrt2}\,[1, 0, 0, 1],
\]
is one. A product with $u_0 v_0 \neq 0$ and $u_1 v_1 \neq 0$ has $u_0 \neq 0$ and $v_1 \neq 0$, so also $u_0 v_1 \neq 0$; but the second amplitude of $|\Phi^+\rangle$ is $0$. Neither qubit of $|\Phi^+\rangle$ has a state vector of its own: as we show below, each gives each outcome with probability $1/2$ whatever is measured, and no unit vector $|u\rangle$ does that, since measuring along $|u\rangle$ itself gives $+1$ for sure. Only the pair has a state. Measured in their basis, its two qubits give $00$ or $11$, with probability $1/2$ each: their outcomes are always equal, and random.

\paragraph{Operators on the pair.} A matrix that acts on the pair is four by four. The \emph{tensor product} $M \otimes N$ of two two-by-two matrices acts on each qubit of a product state, $(M \otimes N)(|u\rangle \otimes |v\rangle) = M|u\rangle \otimes N|v\rangle$, and is the four-by-four matrix of blocks
\[
  M \otimes N = \begin{pmatrix} m_{00}\, N & m_{01}\, N \\ m_{10}\, N & m_{11}\, N \end{pmatrix},
\]
which the code computes with the operator spelled \texttt{>{}<} and printed $\otimes$. A matrix $M \otimes I$ acts on Alice's qubit alone, and $I \otimes N$ on Bob's. For example, $Z \otimes I$ is the diagonal matrix with the entries $1, 1, -1, -1$: it maps $|jk\rangle$ to $\pm|jk\rangle$, with the sign set by Alice's digit $j$ alone. (Applying an observable to a state is not measuring it; a measurement uses the projectors.)

\paragraph{Four rules of algebra.} Section~\ref{sec:override} computes with projectors on the pair by four rules.
\begin{enumerate}[label=(\roman*)]
\item A matrix moves across an inner product as its transpose, $\langle Mu|v\rangle = \langle u|M^{\mathsf T}v\rangle$; a \emph{symmetric} matrix, $M = M^{\mathsf T}$, moves unchanged. We write $\langle u|M|v\rangle$ for $\langle u|Mv\rangle$.
\item Every observable $O = \cos\varphi\,Z + \sin\varphi\,X$ is symmetric, and $OO = I$: a reflection done twice. Hence its projectors are symmetric, $P_\pm P_\pm = P_\pm$, and $P_+ P_- = 0$.
\item The tensor product is linear in each argument, $(A \otimes B)(C \otimes D) = AC \otimes BD$, and $(A \otimes B)^{\mathsf T} = A^{\mathsf T} \otimes B^{\mathsf T}$. So $P \otimes I$ and $I \otimes Q$ are symmetric projectors too, they commute, and their product is $P \otimes Q$.
\item For a symmetric $P$ with $PP = P$, $\langle\psi|P|\psi\rangle = \langle P\psi|P\psi\rangle$, the squared length of $P|\psi\rangle$.
\end{enumerate}
By (iv), the rule of Section~\ref{app:measure} holds for every such projector, on one qubit or on the pair: an outcome has the squared length of the part that its projector keeps as its probability, and leaves that part, divided by its length.

\paragraph{Measuring one qubit of a pair.} Alice measures her qubit with the projectors $P \otimes I$ of her observable, and Bob his with $I \otimes Q$; by rule (iv), the rule of Section~\ref{app:measure} applies to these four-by-four projectors unchanged. They are \hs|measA| and \hs|measB| in Section~\ref{sec:qubits}. For example, let Alice measure $Z$ on $|\Phi^+\rangle$. Her projector for $+1$ is $P_+ = (I + Z)/2$, the diagonal matrix with the entries $1, 0$, so $P_+ \otimes I$ has the entries $1, 1, 0, 0$. It keeps the part $\frac{1}{\sqrt2}[1, 0, 0, 0]$ of $|\Phi^+\rangle$: her outcome is $+1$ with probability $1/2$, and it leaves the pair in $|00\rangle$. Bob's qubit is now $|0\rangle$, and his own $Z$ gives $+1$ for sure. Her outcome $-1$ leaves $|11\rangle$ instead.

In general, let $|d\rangle = [d_0, d_1]$ be the direction of Alice's outcome, so that its projector on her qubit maps $|u\rangle$ to $\langle d|u\rangle\,|d\rangle$. On the pair it keeps
\[
  (P_d \otimes I)\,|\psi\rangle = |d\rangle \otimes |w\rangle, \qquad w_k = d_0\,\psi_{0k} + d_1\,\psi_{1k},
\]
so her outcome has probability $\langle w|w\rangle$ and leaves Bob's qubit in $|w\rangle$, divided by its length. For $|\Phi^+\rangle$, $|w\rangle = |d\rangle/\sqrt2$: whatever Alice measures, each outcome has probability $1/2$. And her outcome leaves Bob's qubit in the direction that she found. (Equivalently, $|\Phi^+\rangle = (|d\rangle \otimes |d\rangle + |d_\perp\rangle \otimes |d_\perp\rangle)/\sqrt2$ for every direction $|d\rangle$ and its perpendicular $|d_\perp\rangle$, which needs real amplitudes.) This is the collapse of Section~\ref{sec:pair}, and the dependence that the joint response of the quantum boxes carries from \hs|nonlocalA| to \hs|nonlocalB| (Section~\ref{sec:override}).

\paragraph{Less entangled pairs.} The states $|\psi_\theta\rangle = \cos\theta\,|00\rangle + \sin\theta\,|11\rangle$ of Section~\ref{sec:pair} run from the product state $|00\rangle$ at $\theta = 0$ to $|\Phi^+\rangle$ at $\theta = \pi/4$. In between, the two qubits still give equal outcomes when both are measured in their basis, but $0$ is more likely than $1$. Here $|w\rangle = [d_0\cos\theta, d_1\sin\theta]$: when Alice measures $Z$, Bob's qubit is left in her direction, but when she measures $X$, it is left at the angle $\pm\theta$, not along $|{\pm}\rangle$. Only at $\theta = \pi/4$ is every measurement of either qubit on its own a fair coin; that is why $|\Phi^+\rangle$ is called \emph{maximally} entangled.

%% ---------------------------------------------------------------------
\subsection{Correlations, the CHSH value, and no-signaling}\label{app:chsh}

\paragraph{Joint probabilities.} When Alice measures with the projectors $P_x$ and Bob with $Q_y$, the pair of outcomes $(x, y)$ has the probability
\[
  \langle\psi|P_x \otimes Q_y|\psi\rangle,
\]
whichever of them measures first. Section~\ref{sec:override} derives this from the rule of Section~\ref{app:measure}; there $P_x$ and $Q_y$ stand for the projectors on the pair, $P_x \otimes I$ and $I \otimes Q_y$, whose product is $P_x \otimes Q_y$. The average of the product $xy$, the correlator $E$ of Section~\ref{sec:intro}, follows as the average outcome of Section~\ref{app:measure} did:
\[
  E = \sum_{x, y} xy\, \langle\psi|P_x \otimes Q_y|\psi\rangle = \langle\psi|(P_+ - P_-) \otimes (Q_+ - Q_-)|\psi\rangle = \langle\psi|A \otimes B|\psi\rangle
\]
for Alice's observable $A$ and Bob's observable $B$. This is the formula of Section~\ref{sec:pair}, which \hs|correlation| computes.

\paragraph{The CHSH value of $|\Phi^+\rangle$.} Let Alice measure along the direction at the angle $\varphi_A/2$, and Bob along the one at $\varphi_B/2$. Alice's outcome leaves Bob's qubit in the direction that she found, so his outcome agrees with hers with probability $\cos^2\frac{\varphi_A - \varphi_B}{2}$, the squared component of her direction along his. Hence
\[
  E = \cos^2\frac{\varphi_A - \varphi_B}{2} - \sin^2\frac{\varphi_A - \varphi_B}{2} = \cos(\varphi_A - \varphi_B).
\]
In the four contexts of Section~\ref{sec:qubits}, Alice measures along $0^\circ$ and $45^\circ$, and Bob along $22.5^\circ$ and $-22.5^\circ$ (Figure~\ref{fig:qubit}b). In three contexts their directions are $22.5^\circ$ apart, and the outcomes agree with probability $\cos^2 22.5^\circ \approx 0.854$; in \hs|(A1, B1)| they are $67.5^\circ$ apart, and agree with probability $\cos^2 67.5^\circ \approx 0.146$. These are close to the frequencies of agreement in the table of Section~\ref{sec:quantum-ns}. So $E$ is $1/\sqrt2$ in three contexts and $-1/\sqrt2$ in the fourth, and $\CHSH = 4/\sqrt2 = 2\sqrt2$. No other directions do better. Let $a_0, a_1, b_0, b_1$ be the unit vectors at the angles $\varphi$ of the four observables, so that $E = \cos(\varphi_A - \varphi_B)$ is the inner product of Alice's vector with Bob's. Regrouping as in Section~\ref{sec:intro}, $\CHSH = \langle a_0|b_0 + b_1\rangle + \langle a_1|b_0 - b_1\rangle$, which is at most the sum of the lengths of $b_0 + b_1$ and $b_0 - b_1$; their squared lengths add up to $4$, so the sum is at most $2\sqrt2$. By Tsirelson's theorem, no other quantum state or measurement does better either~\citep{Tsirelson1980}. The Bell table of Section~\ref{sec:belltable} measures along $0^\circ$ and $30^\circ$ for Alice, and $0^\circ$ and $-30^\circ$ for Bob: the directions are $0^\circ$, $30^\circ$, $30^\circ$ and $60^\circ$ apart, and the outcomes agree with probability $1$, $3/4$, $3/4$ and $1/4$.

\paragraph{No-signaling.} Bob's outcome depends on Alice's outcome, and, given her outcome, on what she measured: after her $Z$ gives $+1$ his $Z$ is certain, and after her $X$ gives $+1$ it is a fair coin. Averaged over her outcome, which Bob does not see, his qubit is left in her direction or in the perpendicular one, with probability $1/2$ each. Whatever he measures, at an angle $\gamma$ to her direction, he then answers $+1$ with probability $\frac12\cos^2\gamma + \frac12\sin^2\gamma = \frac12$. In general, Alice's two projectors add up to the identity, so summing the joint probability over her outcome leaves
\[
  \sum_x \langle\psi|P_x \otimes Q_y|\psi\rangle = \langle\psi|I \otimes Q_y|\psi\rangle,
\]
in which her observable does not appear. This is the argument of Section~\ref{sec:override}.

\paragraph{Less entangled pairs.} For $|\psi_\theta\rangle$, the averages of $Z \otimes Z$, $X \otimes X$, $Z \otimes X$ and $X \otimes Z$ are $1$, $\sin 2\theta$, $0$ and $0$, so
\[
  E = \cos\varphi_A \cos\varphi_B + \sin 2\theta\, \sin\varphi_A \sin\varphi_B,
\]
which gives the CHSH value $\sqrt2\,(1 + \sin 2\theta)$ of Section~\ref{sec:pair} for the paper's observables. The argument for $|\Phi^+\rangle$, with Bob's vectors scaled by $1$ along the first axis and by $\sin 2\theta$ along the second, bounds the CHSH value by $2\sqrt{1 + \sin^2 2\theta}$, and other observables reach that bound; it exceeds $\sqrt2\,(1 + \sin 2\theta)$ except at $\theta = \pi/4$. On its own, Alice's qubit gives $+1$ with probability $\cos^2\theta$ for $Z$ and $1/2$ for $X$, and Bob's gives $+1$ with probability $\frac12(1 + \cos 2\theta/\sqrt2)$ for either of his observables. At $\theta = \pi/8$ these are $\val{partial/AliceA0}$, $\val{partial/AliceA1}$ and $\val{partial/Bob}$, the probabilities that the run of Section~\ref{sec:quantum-ns} matches. Its table also has two cells at $0$: after Alice's $X$ gives $\pm1$, Bob's qubit is left at $\pm 22.5^\circ$, exactly the direction for $+1$ of \tB0 or of \tB1, so there Bob answers \hs|Plus| for sure.

\begin{figure}[htbp]
\centering
{\small\renewcommand{\arraystretch}{1.14}%
\begin{adjustbox}{max width=\linewidth}\begin{tabular}{@{}l l l@{}}
\toprule
Notion & In mathematics & In the code (Section~\ref{sec:quantum})\\
\midrule
a state of the pair & $[\psi_{00}, \psi_{01}, \psi_{10}, \psi_{11}]$ & \vty{Vector}, \vfn{maximallyEntangled}, \vfn{entangled}\\
inner product & $\langle u|v\rangle$ & {\outputfont u \opc{<|} v}\\
a matrix applied to a vector & $M|v\rangle$ & {\outputfont m \opc{|>} v}\\
tensor product & $M \otimes N$ & {\outputfont m \opc{$\otimes$} n}\\
observables & $Z$, $X$, $(Z \pm X)/\sqrt2$ & \vfn{pauliZ}, \vfn{pauliX}, \vfn{observableA}, \vfn{observableB}\\
projectors & $P_\pm = (I \pm O)/2$ & \vfn{projector}\\
projectors on the pair & $P \otimes I$, $I \otimes Q$ & \vfn{measA}, \vfn{measB}\\
probability of an outcome & $\langle\psi|P|\psi\rangle$ & \vfn{probability}\\
the state that it leaves & $P|\psi\rangle / \sqrt{\langle\psi|P|\psi\rangle}$ & \vfn{collapse}\\
an outcome, drawn & $+1$ with probability $p$ & \vfn{coin}\\
correlator & $\langle\psi|A \otimes B|\psi\rangle$ & \vfn{correlation}\\
\bottomrule
\end{tabular}\end{adjustbox}}
\caption{The notions of this appendix, and their names in the code of Section~\ref{sec:quantum}.}
\label{fig:dictionary}
\end{figure}

\end{document}